\documentclass[a4paper,fleqn,usenatbib]{mnras}

\usepackage{newtxtext,newtxmath}

\usepackage[T1]{fontenc}
\usepackage{ae,aecompl}

\usepackage{graphicx}	
\usepackage{amsmath}	
\usepackage{amssymb}	
\newcommand{\angstrom}{\text{\normalfont\AA}}

\usepackage{enumitem}
\usepackage{times}
\usepackage{url}
\usepackage[usenames]{color}

\def \simless {\mathbin{\lower 3pt\hbox{$\rlap{\raise 4pt
              \hbox{$\char'074$}}\mathchar"7218$}}}
\def \simgreat {\mathbin{\lower 3pt\hbox{$\rlap{\raise 4pt
              \hbox{$\char'076$}}\mathchar"7218$}}}
\def\ie {{\it i.e.}}
\def\cf {{\it c.f.}}
\def\eg {{\it e.g.}}

\title[Dust temperatures of high redshift galaxies]{On the dust temperatures of high redshift galaxies}

\author[L. Liang et al.]{Lichen Liang$^{1}$\thanks{Email: lliang@physik.uzh.ch}, Robert Feldmann$^{1}$, Du\v{s}an Kere\v{s}$^{2}$, Nick Z. Scoville$^{3}$,
\newauthor Christopher C. Hayward$^{4}$, Claude-Andr\'{e} Faucher-Gigu\`{e}re$^{5}$, Corentin Schreiber $^{6,7}$,
\newauthor Xiangcheng Ma$^{3,8}$, Philip F. Hopkins$^{3}$, Eliot Quataert$^{8}$  \\
$^{1}$Institute for Computational Science, University of Zurich, Zurich CH-8057, Switzerland\\
$^{2}$Department of Physics, Centre for Astrophysics and Space Sciences, University of California at San Diego, La Jolla, CA 92093, USA\\
$^{3}$TAPIR, California Institute of Technology, Pasadena, CA, USA\\
$^{4}$Centre for Computational Astrophysics, Flatiron Institute, 162 Fifth Avenue, New York, NY 10010, USA\\
$^{5}$Department of Physics and Astronomy and CIERA, Northwestern University, Evanston, IL 60208, USA\\
$^{6}$Astrophysics, University of Oxford, Denys Wilkinson Building, Keble Road, Oxford OX1 3RH, United Kingdom\\
$^{7}$Leiden Observatory, Leiden University, NL-2300 RA Leiden, The Netherlands\\
$^{8}$Department of Astronomy, 501 Campbell Hall, University of California, Berkeley, CA, 94720, USA}

\date{Accepted 2019. Received 2019; in original form 2019}

\pubyear{2019}

\begin{document}
\label{firstpage}
\pagerange{\pageref{firstpage}--\pageref{lastpage}}
\maketitle

\begin{abstract}
Dust temperature is an important property of the interstellar medium (ISM) of galaxies. It is required when converting (sub)millimeter broadband flux to total infrared luminosity ($L_{\rm IR}$), and hence star formation rate, in high-redshift galaxies. However, different definitions of dust temperatures have been used in the literature, leading to different physical interpretations of how ISM conditions change with, \eg, redshift and star formation rate. In this paper, we analyse the dust temperatures of massive ($M_{\rm star}>10^{10}\,M_{\odot}$) $z=2-6$ galaxies with the help of high-resolution cosmological simulations from the \textit{Feedback in Realistic Environments} (\textsc{\small FIRE}) project. At $z\sim2$,  our  simulations  successfully  predict  dust temperatures in good agreement with observations. We find that dust temperatures based on the peak emission wavelength increase  with redshift, in line with the higher star formation activity at higher redshift, and are strongly correlated with the specific star formation rate. In contrast, the \textit{mass-weighted} dust temperature, which is required to accurately estimate the total dust mass, does not strongly evolve with redshift over $z=2-6$ at fixed IR luminosity but is tightly correlated with $L_{\rm IR}$ at fixed $z$. We also analyse an `equivalent' dust temperature for converting (sub)millimeter flux density to total IR luminosity, and provide a fitting formula as a function of redshift and dust-to-metal ratio. We find that galaxies of higher equivalent (or higher peak) dust temperature (`warmer dust') do not necessarily have higher mass-weighted temperatures. A `two-phase' picture for interstellar dust can explain the different scaling relations of the various dust temperatures.
\end{abstract}


\section{Introduction}
\label{Sec:1}

\noindent Astrophysical dust, originating from the condensation of metals in stellar ejecta, is pervasive in the interstellar medium (ISM) of galaxies in both local and distant Universe \citep[\eg][and references therein]{S97,B99,C05,C11,R13,W13,C15,W15,I16,L17,V17,Z17,HL18,PR18,I19}. Dust scatters and absorbs UV-to-optical light, and therefore strongly impacts the observed flux densities and the detectability of galaxies at these wavelengths \citep[\eg][]{C94,K93,C00,K13,N18}. Despite that it accounts for no more than a few percent of the total ISM mass \citep{D07}, dust also plays a key role in star formation process of galaxies \citep[\eg][]{CT02,MQT05,M07,H12}. Constraining and understanding dust properties of galaxies is therefore essential for proper interpretation of the multi-wavelength data from observations and for facilitating our understanding of galaxy formation and evolution.

Much of the stellar emission of star-forming galaxies is absorbed by dust grains and re-emitted at infrared (IR)-to-millimeter (mm) wavelengths as thermal radiation, encoding important information about dust and galactic properties, such as dust mass, total IR luminosity\footnote{In this paper, $L_{\rm IR}$ is defined as the luminosity density integrated over the $8-1000\,\rm \mu m$ wavelength interval.} ($L_{\rm IR}$) and star formation rate (SFR) \citep[\eg][]{CE01,DH02,SK07,C08,I10,W11,C12,M12,C14,S16,S18}. The advent of the new facilities in the past two decades, such as the \textit{Spitzer Space Telescope} \citep{F04}, \textit{Herschel Space Observatory} \citep{PR10}, the Submillimetre Common-user Bolometer Array (SCUBA) camera on the James Clerk Maxwell Telescope (JCMT) \citep{H99, HB13}, the AzTEC millimeter camera on the Large Millimeter Telescope (LMT) \citep{W08}, the South Pole Telescope (SPT) \citep{CA13} and the Atacama Large Millimeter/sub-millimeter Array (ALMA) has triggered significant interests in the study of ISM dust. In particular, observations with the Photodetector Array Camera and Spectrometer \citep[PACS,][]{P10} and the Spectral and Photometric Imaging Receiver \citep[SPIRE,][]{G10} instruments aboard \textit{Herschel} made it possible to study the $70-500\;\rm \mu m$ wavelength range where most of the Universe's obscured radiation emerges, and many dust-enshrouded, previously unreported objects at distant space have been uncovered through the wide-area extra-galactic surveys \citep[\eg][]{E10, L11, O12}. Far-infrared (FIR)-to-mm spectral energy distribution (SED) modelling of dust emission has therefore become possible for objects at high redshift \citep[$z\sim4$,][]{W13,I16,S18} and various dust properties can be extracted using SED fitting techniques \citep{W11}. 

The Rayleigh-Jeans (RJ) side (\ie~$\lambda>\lambda_{\rm peak}$, where $\lambda_{\rm peak}$ is the wavelength of the peak emission) of the dust SED can empirically be well fit by a single-temperature ($T$) modified blackbody (MBB) function \citep{H83}. However, the shape of the Wien side (\ie~$\lambda<\lambda_{\rm peak}$) of the SED, which is tied to the warm dust component in vicinity of the young stars and active galactic nuclei (AGN), has a much larger variety \citep[\eg][]{K12, SV13}. Studies have shown that \textit{one} single-$T$ MBB function cannot well fit \textit{both} sides of the observed SEDs \citep{C12}. Motivated by this fact, fitting functions of multi-$T$ components have been adopted \citep[\eg][]{D01,B03,K10,D12,K12,G12,C12,C13,CW15}. Meanwhile, empirical SED templates have been developed based on an assumed distribution of interstellar radiation intensity ($U$) incident on dust grains \citep[\eg][]{DD01,DL07,GH11,S18}. Both approaches can produce functional shape of dust SED that better matches the observed photometry of galaxies compared to a single-$T$ MBB function.

At high redshift ($z\simgreat2$), however, it is more common that a galaxy has only a few (two or three) reliable photometric data points in its dust continuum so that SED fitting by multi-$T$ functions or more sophisticated SED templates is not possible. Therefore, the widely adopted approach is to simply fit the available data points with \textit{one} single-$T$ MBB function \citep[\eg][]{ML12,SV13,M14,TS17,SS17}. The $T$ parameter that yields the best-fit is then often referred to as the `dust temperature' of the galaxy in the literature. We specify this definition of dust temperature as the \textit{`effective'} temperature ($T_{\rm eff}$) in this paper. Another temperature also often used is the \textit{`peak'} temperature ($T_{\rm peak}$), which is defined based on the emission peak of the best-fit SED assuming Wien's displacement law \citep{C14}. These observationally-derived dust temperatures (both $T_{\rm eff}$ and $T_{\rm peak}$) can depend on the assumed functional form of SED as well as the adopted photometry \citep{C12,C14}. Despite that it is unclear how well these simplified fitting functions represent the true SED shape of high-redshift galaxies and the physical interpretation of the derived temperatures is not obvious, this approach is frequently used to analyse large statistical samples of data \citep[\eg][]{C05,H10,SV13,SS14,CW15,C18b}.

The scaling relations of $T_{\rm peak}$ ($T_{\rm eff}$) and other dust/galaxy properties, including the $L_{\rm IR}$-temperature and specific star formation rate (sSFR)-temperature relations, may be related to the physical conditions of the star-forming regions in distant galaxies and have attracted much attention \citep[\eg][]{M12, ML12, M14, L14,S18, C18a}. While observational studies derive dust temperatures in a variety of ways (using different fitting functions and/or different photometry), they generally infer that the temperature increases with $L_{\rm IR}$ and sSFR of galaxies. Accurate interpretation of these observed scaling relationships requires a knowledge of how different dust and galaxy properties (\eg~stellar mass, SFR and dust mass) shape the dust SED \citep{DL07,G08,S13,SH16}, and hence the derived dust temperatures. Radiative transfer (RT) analyses of galaxy models are important tools for understanding these temperatures since the intrinsic properties of the simulated galaxies are known \citep[\eg][]{NH10,HK11,HJ12,HS15,N15,C16,L18,M19}.

One important question is how the derived temperatures are related to the physical, \textit{mass-weighted} dust temperature ($T_{\rm mw}$). Observations of the local galaxies have shown that the bulk of the ISM dust remains at low temperature \citep{D01,HF13,LB14}. The cold dust component determines $T_{\rm mw}$ of the galaxy, which sets the shape of the RJ tail. For distant galaxies, it is very challenging to measure $T_{\rm mw}$ due to the limit of resolution. However, a good estimate of $T_{\rm mw}$ is important for deriving the ISM masses of high-redshift galaxies via the RJ method \citep[\eg][]{S14,S16,S17}. It is unclear whether, or how, one can infer $T_{\rm mw}$ from the observationally-derived temperatures. Alternatively, one can simply adopt a constant value if $T_{\rm mw}$ has relatively small variation among different galaxies, given that the mass estimates scale only \textit{linearly} with $T_{\rm mw}$ \citep{S16}. If that is the case, it can also be one major advantage of the RJ approach because the main difficulty of the traditional CO method is the large uncertainty of the CO-to-$\rm H_2$ conversion factor \citep[\eg][]{SG11,F12,CW13}. RT analyses are useful for understanding the relation between the derived temperatures and $T_{\rm mw}$ \citep{L18}.

Over the past two decades, many ground-based galaxy surveys at (sub)mm wavelengths (\eg~SCUBA, AzTEC, SPT and ALMA) that are complementary to \textit{Herschel} observations \citep[\eg][and references therein]{S97,D00,GC13,KS13,SS14,A16,B16,D16,W16,H16,G17,FE18} have been carried out. Deep (sub)mm surveys are capable of probing less actively star-forming ($\rm SFRs\simless100\,M_{\odot}\,yr^{-1}$) galaxies at $z\simless4$  \citep[\eg][]{HO13,cc14,O14,Z18}. Furthermore, they are effective at uncovering sources at $z>4$ thanks to the effect of ``negative-$K$ correction" \citep[\eg][]{C15,CM15,FO16,L17,C18a}.
The (sub)mm-detected sources do not necessarily have \textit{Herschel} counterparts. Deriving $L_{\rm IR}$ (and hence SFR) of these sources from a single (sub)mm flux density (S) requires adopting an assumed dust temperature, which we refer to as \textit{`equivalent'} temperature ($T_{\rm eqv}$) in this paper, along with an assumed (simplified) functional shape of the dust SED \citep{B16,C18a}. $T_{\rm eqv}$ is conceptually different from $T_{\rm eff}$ introduced above because the former is an \textit{assumed} quantity for extrapolating $L_{\rm IR}$ from a \textit{single} data point while the latter is a \textit{derived} quantity through SED fitting to \textit{multiple} data points.

A good estimate of $T_{\rm eqv}$ is important for translating the information (\eg~source number counts) extracted from the ALMA blind surveys to the \textit{obscured} cosmic star formation density at $z\simgreat4$ \citep{C18b,C18a,Z18b}, where currently only reliable data from rest-frame UV measurements are available \citep{MD14}. One common finding of the recent (sub)mm blind surveys is a dearth of faint sources at these early epochs, as noted by \citet{C18b} (see also the references therein). This can suggest that the early Universe is relatively dust-poor and only a small fraction of stellar emission is absorbed and re-emitted by dust \citep{C18b}. Alternatively, it could also be accounted for by a significantly higher $T_{\rm eqv}$ at high redshifts, meaning that galaxies of the same $L_{\rm IR}$ appear to be fainter in the (sub)mm bands \citep[\cf][]{C15,B16,FO17,FC17,ND18}.  Hence, understanding how $T_{\rm eqv}$ evolves with redshift and how it depends on different galaxy properties are crucial for constraining the total amount of dust and the amount of \textit{obscured} star formation density in the early Universe \citep{C18b,C18a}.

In this paper, we study in detail the observational and the physical (mass-weighted) dust temperatures with the aid of high-resolution cosmological galaxy simulations. In particular, we study a sample of massive ($M_{\rm star}>10^{10}\;M_{\odot}$) $z=2-6$ galaxies from the \textsc{\small FIRE} project\footnote{\url{fire.northwestern.edu}} \citep{H14} with dust RT modelling. This sample contains galaxies with $L_{\rm IR}$ ranging over two orders of magnitude, from $10^{10}$ to $10^{12}\,L_{\odot}$ and few dust-rich, ultra-luminous ($L_{\rm IR}\simgreat10^{12}\,L_{\odot}$) galaxies at $z\sim2$ that are candidates for both \textit{Herschel}- and submm-detected objects. A lot of them have $L_{\rm IR}\sim$ a few $\times10^{11}\,L_{\odot}$, which is accessible by $\textit{Herschel}$ using stacking techniques \citep[\eg][]{TS17,S18}. Our sample also contains fainter galaxies at $z=2-6$ with observed flux densities $S_{\rm 870\mu m}\,(S_{\rm 1.2 mm})\simgreat0.1$ mJy, which could be potentially detected with ALMA. We calculate and explicitly compare their $T_{\rm mw}$ with the observationally-derived temperatures ($T_{\rm peak}$ or $T_{\rm eff}$), as well as their scaling relationships with several galaxy properties. We also provide the prediction for $T_{\rm eqv}$ that is needed for deriving $L_{\rm IR}$ of galaxy from its observed single-band (sub)mm flux.

The paper is structured as follows. In Section~\ref{S2}, we introduce the simulation details and the methodology of radiative transfer modelling. In Section~\ref{S3}, we provide the various definitions of dust temperature in detail, discuss the impact of dust-temperature on SED shape, and compare the specific predictions of our simulations with observations. In Section~\ref{S4}, we focus on the conversion from single-band (sub)mm broadband flux to $L_{\rm IR}$ and provide useful fitting formulae. In Section~\ref{S5}, we discuss the observational implications of our findings. We summarise and conclude in Section~\ref{S6}.

Throughout this paper, we adopt cosmological parameters in agreement with the nine-year data from the Wilkinson Microwave Anisotropy Probe \citep{H13}, specifically $\Omega_{\rm m}=0.2821$, $\Omega_{\Lambda}=0.7179$, and $H_0=69.7\; \rm km\;s^{-1}\;Mpc^{-1}$.

\section{Simulation Methodology}
\label{S2}

In this section, we introduce our simulation methodology. In Section~\ref{S2a}, we briefly summarize the details of the cosmological hydrodynamic simulations from which our galaxy sample is extracted. In Section~\ref{S2b}, we introduce the methodology of our dust RT analysis and present mock images produced with \textsc{\small SKIRT}.
  
\subsection{Simulation suite and sample}
\label{S2a}

We extract our galaxy sample from the \textsc{\small MassiveFIRE} cosmological zoom-in suite \citep{F16,F17}, which is part of the Feedback in Realistic Environments (FIRE) project. 

The initial conditions for the \textsc{\small MassiveFIRE} suites are generated using the \textsc{MUSIC} code \citep{H11} within a $(100\;\rm Mpc/h)^3$ comoving periodic box with the WMAP cosmology. From a low-resolution (LR) dark matter (DM)-only run, isolated halos of a variety of halo masses, accretion history and environmental over-densities are selected. 
Initial conditions for the `zoom-in' runs use a convex hull surrounding all particles within $3R_{\rm vir}$ at $z=2$ of the chosen halo defining the Lagrangian high-resolution (HR) region. The mass resolution of the default HR runs are $m_{\rm DM}=1.7\times10^5\;M_{\odot}$ and $m_{\rm gas}=3.3\times10^4\;M_{\odot}$, respectively. The initial mass of the star particle is set to be the same as the parent gas particle from which it is spawned in the simulations. 

The simulations are run with the gravity-hydrodynamics code \textsc{GIZMO}\footnote{A public version of GIZMO is available at \url{http://www.tapir.caltech.edu/phopkins/Site/GIZMO.html}} (\textsc{\small FIRE-1} version) in the Pressure-energy Smoothed Particle Hydrodynamics (``P-SPH") mode \citep{H15}, which improves the treatment of fluid mixing instabilities and includes various other improvements to the artificial viscosity, artificial conductivity, higher-order kernels, and time-stepping algorithm designed to reduce the most significant known discrepancies between SPH and grid methods \citep{HP13}. Gas that is locally self-gravitating and has density over $5\; \rm cm^{-3}$ is assigned an SFR $\dot{\rho}=f_{\rm mol}\rho/t_{\rm ff}$, where $f_{\rm mol}$ is the self-shielding molecular mass fraction. The simulations explicitly incorporate several different stellar feedback channels (but not feedback from supermassive black holes) including 1) local and long-range momentum flux from radiative pressure, 2) energy, momentum, mass and metal injection from supernovae (Types Ia and II), 3) and stellar mass loss (both OB and AGB stars) and 4) photo-ionization and photo-electric heating processes. We refer the reader to \citet{H14} for details.

In the present study we analyse 18 massive ($10^{10}<M_{\rm star}<10^{11.3}\;M_{\odot}$ at $z=2$) central galaxies (from Series A, B and C in~\citealt{F17}) and their most massive progenitors (MMP) up to $z=6$, identified using the Amiga Halo Finder \citep{G04,K09}. These galaxies were extracted from the halos selected from the LR DM-only run. In order to better probe the dusty, IR-luminous galaxies at the extremely high-redshift ($z>4$) Universe, we also include another 11 massive ($10^{10}<M_{\rm star}<10^{11}\;M_{\odot}$ at $z=6$) galaxies extracted from a different set of \textsc{\small MassiveFIRE} simulations that stop at $z=6$, which are presented here for the first time. The latter were run with the same physics, initial conditions, numerics, and spatial and mass resolution, but were extracted from larger simulation boxes (400 $\rm Mpc/h$ and 762 $\rm Mpc/h$ on a side, respectively).

FIRE simulations successfully reproduce a variety of observed galaxy properties relevant for the present work, such as the stellar-to-halo-mass relation \citep{H14, F17},  the sSFRs of galaxies at the cosmic noon ($z\sim2$) \citep{H14, F16}, the stellar mass -- metallicity relation \citep{M16}, and the submm flux densities at 850 $\rm \mu m$ \citep{L18}.

\subsection{Predicting dust SED with SKIRT}
\label{S2b}

\begin{figure}
 \centering
 \includegraphics[width=75mm]{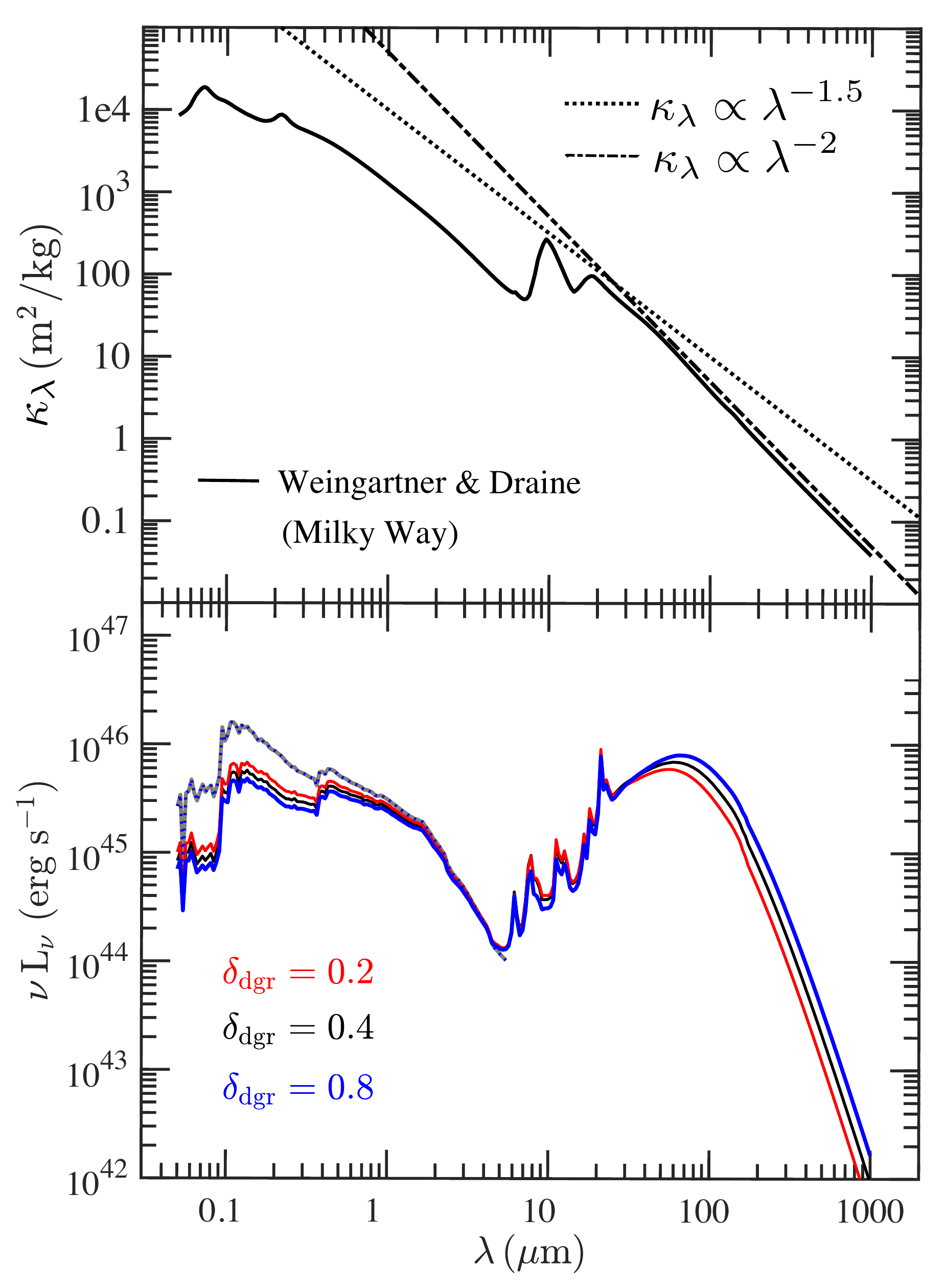}
  \caption{\textit{Upper panel:} Dust opacity curve of the \citet{WD01} dust model that is used for analysis in this work (solid line). The dashed and dash-dotted lines show the asymptotic power law $\kappa_\lambda \propto\lambda^{-1.5}$ and $\kappa_\lambda\propto\lambda^{-2}$, respectively. \textit{Lower panel:} SEDs of a selected $z=2$ {\sc\small MassiveFIRE} galaxy. The red, black and blue curves show results for $\delta_{\rm dzr}$ = 0.2, 0.4 and 0.8, respectively. The grey curve shows the intrinsic stellar emission. About half of the stellar radiative energy of this galaxy is absorbed and re-emits at IR.}
  \label{fig:Dust} 
\end{figure}

We generate the UV-to-mm spectral energy distribution (SED) using the open source\footnote{\textsc{\small SKIRT} code repository: \url{https://github.com/skirt}} 3D dust Monte Carlo RT code \textsc{\small SKIRT} \citep{B11,BC15}. \textsc{\small SKIRT} accounts for absorption and anisotropic scattering of dust and self-consistently calculates the dust temperature. We follow the approach by \citet{C16} (see also \citealt{T17}) to prepare our galaxy snapshots as RT input models.

Each star particle in the simulation is treated as a `single stellar population' (SSP). The spectrum of a star particle in the simulation is assigned using {\sc \small starburst99} SED libraries. In our default RT model, every star particle is assigned an SED according to the age and metallicity of the particle.

While our simulations have better resolution than many previous simulations modelling infrared and submm emission (e.g., \citealt{NH10,HK11,DF14}) and can directly incorporate various important stellar feedback processes, they are still unable to resolve the emission from HII and photo-dissociation regions (PDR) from some of the more compact birth-clouds surrounding star-forming cores. The time-average spatial scale of these HII+PDR regions typically vary from $\sim5$ pc to $\sim800$ pc depending on the local physical conditions \citep{J10}. Hence, in our alternative RT model, star particles are split into two sets based on their age. Star particles formed less than 10 Myrs ago are identified as `young star-forming' particles, while older star particles are treated as above. We follow \citet{C16} in assigning a source SED from the \textsc{\small MappingsIII} \citep{G08} family to young star-forming particles to account for the pre-processing of radiation by birth-clouds. Dust associated with the birth-clouds is removed from the neighbouring gas particles to avoid double-counting \citep[see also][]{C16}.

We present in Section~\ref{S3} and~\ref{S4} the results from our default (`no birth-cloud') model. In Section~\ref{S5} we will show that none of our results are qualitatively altered if we adopt the alternative RT model and account for unresolved birth-clouds.

\begin{figure}
 \begin{center}
 \includegraphics[height=120mm,width=88mm]{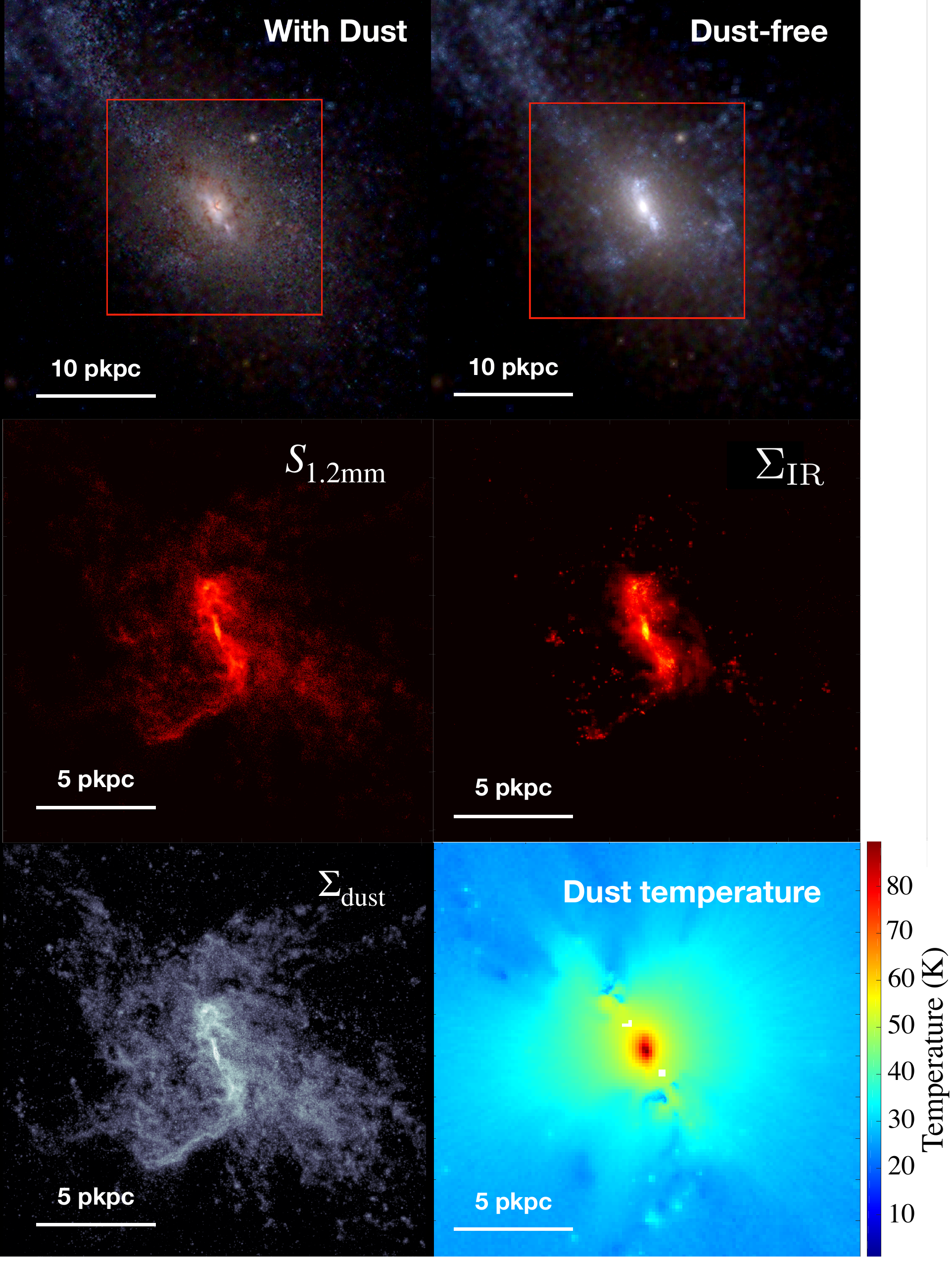}
 \caption{Example of the radiative transfer analysis applied to a $z=2$ \textsc{\small MassiveFIRE} galaxy. \textit{Upper panels}: UVJ image with (left) and without (right) the effect of dust extinction. {\it Middle panels}: Normalised $S_{\rm 1.2 mm}$ (left) and normalised $\Sigma_{\rm IR}$ (right). Compared with $S_{\rm 1.2 mm}$, $\Sigma_{\rm IR}$ traces more tightly to the star-forming regions. {\it Lower panels}: Dust surface density (left) and dust temperature weighted along the line-of-sight, weighted by mass (right). The middle and lower panels show the result for the zoomed-in region enclosed by the red box in the upper panels. }
    \label{fig:Image}
  \end{center}
  \vspace{-10pt}
\end{figure}

Our RT analysis uses $10^6$ photon packets for each stage. We use an octree for the dust grid and keep subdividing grid cells until the cell contains less than $f=3\times10^{-6}$ of the total dust mass and the $V$-band optical depth in each cell is less than unity. The highest grid level corresponds to a cell width of $\sim20$ pc,~\ie, about twice the minimal SPH smoothing length. For all the analysis in this paper, we adopt the \citet{WD01} dust model with Milky-Way size distribution for the case of $R_{\rm V}=3.1$. At FIR, the dust opacity can be well described by a power law, $\kappa_{\lambda}\propto 0.05\,(\lambda/{\rm 870\mu m})^{-\beta}\,\rm m^2/kg$, where $\beta\approx2.0$ (see the \textit{upper panel} of Figure~\ref{fig:Dust}) is the dust emissivity spectral index \citep[consistent with the observational constraints, \eg][]{D00,D07}. Gas hotter than $10^6$ K is assumed to be dust-free due to sputtering \citep{HN15}.  We self-consistently calculate the self-absorption of dust emission and include the transient heating function to calculate non-local thermal equilibrium (NLTE) dust emission by transiently heated small grains and PAH molecules \citep{B11}. Transient heating influences the rest-frame mid-infrared (MIR) emission ($\simless80\,\rm \mu m$) but has minor impact on the FIR and (sub)mm emission \citep{B18}. \textsc{\small SKIRT} outputs $T_{\rm mw}$ for each cell that is obtained by averaging the temperature over grains of different species (composition and size). A galaxy-wide dust temperature is calculated by mass-weighting $T_{\rm mw}$ of each cell in the galaxies. At high redshift ($z>4$), the radiation field from the cosmic microwave background (CMB) starts to affect the temperature of the cold ISM. We account for the CMB by adopting a corrected dust temperature \citep{C13}

\begin{equation}
    T^{\rm corr}_{\rm dust}(z)=\big[T^{4+\beta}_{\rm dust}+T^{4+\beta}_{\rm CMB}(z)-T^{4+\beta}_{\rm CMB}(z=0) \big]^{1/(4+\beta)},
    \label{eq.1}
\end{equation}

\noindent where $T_{\rm CMB}(z) = 2.73\,(1 + z)$ K is the CMB temperature
at $z$.

For this study, we assume that dust mass traces metal mass in the ISM, and adopt a constant dust-to-metal mass ratio $\delta_{\rm dzr}=0.4$ \citep{D98,D07,WD11} for our \textit{fiducial} analysis. We also try two different cases where $\delta_{\rm dzr}=0.2$ and $\delta_{\rm dzr}=0.8$, and throughout the paper, we refer to these two \textit{dust-poor} and \textit{dust-rich} cases, respectively. In the lower panel of Figure~\ref{fig:Dust}, we show the galaxy SED for the three models. $L_{\rm IR}$ increases when $\delta_{\rm dzr}$ increases because a higher optical depth leads to more absorption of stellar light and more re-emission at IR.

\textsc{\small SKIRT} produces spatially resolved, multi-wavelength rest-frame SEDs for each galaxy snapshot observed from multiple viewing angles. For the analysis in this paper, SEDs are calculated on an equally spaced logarithmic wavelength grid ranging from rest-frame 0.005 to 1000 $\rm \mu m$. We convolve the simulated SED output from \textsc{\small SKIRT} with the transmission functions of the PACS (70, 100, 160 $\rm \mu m$), SPIRE (250, 350, 500 $\rm \mu m$), SCUBA-2 (450, 850 $\rm \mu m$), ALMA band 6 (870 $\rm \mu m$) and 7 (1.2 mm) to yield the broadband flux density for each band.

We show in Figure~\ref{fig:Image} the result of running {\sc\small SKIRT} on one of our galaxies. In particular we show a composite U, V, J false-color image with and without accounting for dust absorption, scattering, and emission. We also show the image of ALMA 1.2 mm flux density, total IR luminosity, dust surface density and temperature. It can be seen that the 1.2 mm flux density traces the dust mass distribution, while IR luminosity appears to be more localised to the high-temperature region, since it is expected to be sensitive to temperature ($L\sim MT^{4+\beta}$). The local intensity of radiation, the dust temperatures, and the dust density all peak in the central region of the galaxy.

\vspace{-10pt}
\section{Understanding dust temperature and its scaling relations}
\label{S3}

In this section, we at first review the different ways of defining galaxy dust temperature that have been used in different observational and theoretical studies (Section~\ref{S3a}), and compute the different temperatures for the \textsc{\small MassiveFIRE} sample (Section~\ref{S3b}). We compare the calculated dust temperature(s) of the simulated galaxies with recent observational data (Section~\ref{S3c}). Finally, we reproduce several observed scaling relations (\eg~$L_{\rm IR}$ vs. temperature, sSFR vs. temperature) with the simulated galaxies and provide physical insights for these relations (Section~\ref{S3d}). 

\subsection{Defining dust temperature}
\label{S3a}

Dust temperature has been defined in different ways by observational and theoretical studies. Here, we focus on four different possibilities, which we call \emph{mass-weighted}, \emph{peak}, \emph{effective}, and \emph{equivalent} dust temperature.

\paragraph*{Mass-weighted dust temperature $T_{\rm mw}$\\}
$T_{\rm mw}$ is the physical, mass-weighted temperature of dust in the ISM.  $T_{\rm mw}$ is often explicitly discussed in theoretical studies where dust radiative transfer modelling is applied to the snapshots from the galaxy simulations, and dust temperature is calculated using LTE (for large grains) and non-LTE (for small grains and PAH molecules) approaches \citep[\eg][]{B18, L18}. 

\paragraph*{Peak dust temperature $T_{\rm peak}$\\}
The peak dust temperature is defined based on the wavelength $\lambda_{\rm peak}$ at which the far-infrared spectral flux density reaches a maximum \citep[\eg][]{C14}

\begin{equation}
T_{\rm peak}=\frac{2.90\times10^{3}\;{\rm \mu m\cdot K}}{\lambda_{\rm peak}}.
\end{equation}

\noindent The peak wavelength $\lambda_{\rm peak}$ is commonly derived from fitting the SED to a specific functional form, for instance, a modified blackbody (MBB), see below. $\lambda_{\rm peak}$ (and $T_{\rm peak}$) in practice depends on the adopted functional form as well as the broadband photometry used in the fit \citep{C12,C14}.

\paragraph*{Effective dust temperature $T_{\rm eff}$\\}
The effective temperature is obtained by fitting the SED with a parametrised function. The effective temperature is thus a fit parameter, and like $T_{\rm peak}$, depends on both the adopted functional form and the broadband photometry.

For most observed SEDs, the RJ side of the dust continuum can be well described by a generalised modified-blackbody function (\textbf{G-MBB}) of the form \citep{H83}  

\begin{align}
S_{\nu_0} (T) &= A\,\frac{(1+z)}{d^2_{\rm L}}\left ( 1-e^{-\tau_\nu}\right)B_\nu(T)\label{eq.3}\\
&=  \frac{1-e^{-\tau_\nu}}{\tau_\nu}\frac{(1+z)}{d^2_{\rm L}}\kappa_\nu M_{\rm dust}\,B_\nu(T)
\label{eq.3x}
\end{align}

\noindent where $\nu_{\rm o}$ is the observer's frequency, $\nu=\nu_{\rm o}\,(1+z)$ is the rest-frame frequency,  $\tau_\nu$ is the dust optical depth at $\nu$\footnote{Throughout this paper, all $\nu$ and $\lambda$ with no subscript stand for rest-frame quantities, while those with ``o" are the observed quantities.}, $\kappa_{\nu}$ is the dust opacity (per unit dust mass) at $\nu$, $B_{\nu}(T)$ is the Planck function, $A$ is the surface area of the emitting source and $d_{\rm L}$ is the luminosity distance from the source. $\tau_\nu$ is often fitted by a power law at FIR wavelengths, \ie~$\tau_{\nu}=(\nu/\nu_{\rm 1})^\beta$, where $\beta$ is the spectral emissivity index and $\nu_{\rm 1}$ is the frequency where optical depth is unity. Observational evidence has shown that the value of $\nu_{\rm 1}$ can differ between galaxies \citep{GE04,SS17,SM17}. In principle, $\nu_1$ can be determined from SED fitting given full FIR-to-mm coverage \citep{C12}. However, in practice, it is often taken to be a constant, $\sim$1.5-3 THz (\ie~$\lambda_1=c\,\nu^{-1}_1=100-200\;\rm \mu m$) \citep[\eg][]{D06,CC11,R13,SV13,C14,Z18,C18b,C18a}.

The Wien side of the dust emission is expected to be strongly affected by the warm dust component in the vicinity of the star-forming regions, which can significantly boost the luminosity of galaxy with only a small mass fraction \citep[\eg][]{D01,HF13}, knowing $L\sim M T^{4+\beta}$. Observations also show a variety of SED shape at MIR \citep[\eg][]{K12,SV13}. To better account for the emission at MIR, \citet{C12} introduced a simple (truncated) power-law component to Eq.~\ref{eq.3}, giving rise to a G-MBB with an additional power-law component (\textbf{GP-MBB})

\begin{equation}
S_{\nu_0} (T) = A\,\frac{(1+z)}{d^2_{\rm L}}\,\left[  \big(1-e^{-\tau_\nu} \big)B_\nu(T) + 
N_{\rm pl}\,\nu^{-\alpha}e^{-(\nu_{\rm c}(T)/\nu)^2}\right],
\label{eq.4}
\end{equation}

\noindent where $N_{\rm pl}$ is the normalisation factor, $\alpha$ is the power-law index, and $\nu_{\rm c}$ is a cutoff frequency where the power-law term turns over and no longer dominates the emission at MIR. We allow $N_{\rm pl}$ as a free parameter, fix $\alpha=2.5$, and adopt the functional form of $\nu_{\rm c}(T)$ provided by \citet{C12}. The latter were constrained by fitting the observational data of a sample of local IR-luminous galaxies from the Great-Origins All Sky LIRG Survey \citep[GOALS,][]{A09}.

In the optically-thin regime ($\tau\ll1$), Eq.\ref{eq.4} reduces to the optically-thin modified black body function (\textbf{OT-MBB}), \citep[see \eg][]{HK11}

\begin{equation}
\begin{split}
S_{\nu_0} &= \frac{(1+z)}{d^2_{\rm L}}\kappa_{\nu} M_{\rm dust} B_{\nu}(T)\\
&= \frac{(1+z)}{d^2_{\rm L}}\kappa_{870} \left(\frac{\nu}{\nu_{870}}\right)^\beta\,M_{\rm dust} B_{\nu}(T)\\
&=\mathcal{C}_\nu(z) M_{\rm dust} B_\nu(T)
\end{split}
\label{eq.5}
\end{equation}

\noindent where $\kappa_{870}$ is the opacity at $870\,\rm \mu m$ ($\kappa_{870}=0.05\,{\rm m^2\,kg^{-1}}$ for the dust model used in this work), $\nu_{870}=343\,\rm GHz$, and $\mathcal{C}_\nu(z)$ is a known constant for a given $\nu$, $\kappa_{870}$, $\beta$, and $z$ and $d_{\rm L}$.

The long-wavelength ($\lambda\simgreat200\rm\,\mu m$) RJ tail of the dust emission, where dust optical depth becomes low, can be well fit by the above equation. However,  Eq.~\ref{eq.5} is also frequently adopted to fit the \emph{full} dust SED, including both the Wien and the RJ sides, especially by the studies in the pre-\textit{Herschel} era, when not enough data is available to well cover both sides from the SED peak \citep{ML12}. The single-$T$ parameter in Eq.~\ref{eq.5} is then often referred to as the `dust temperature' of the galaxy. However, an effective temperature derived this way should be primarily understood as a fitting parameter and may not correspond to a \textit{physical} temperature \citep{SS17}. In particular, it differs in general from the mass-weighted temperature of dust in a galaxy. 

\begin{figure*}
\vspace{-10pt}
 \begin{center}
 \includegraphics[width=160mm]{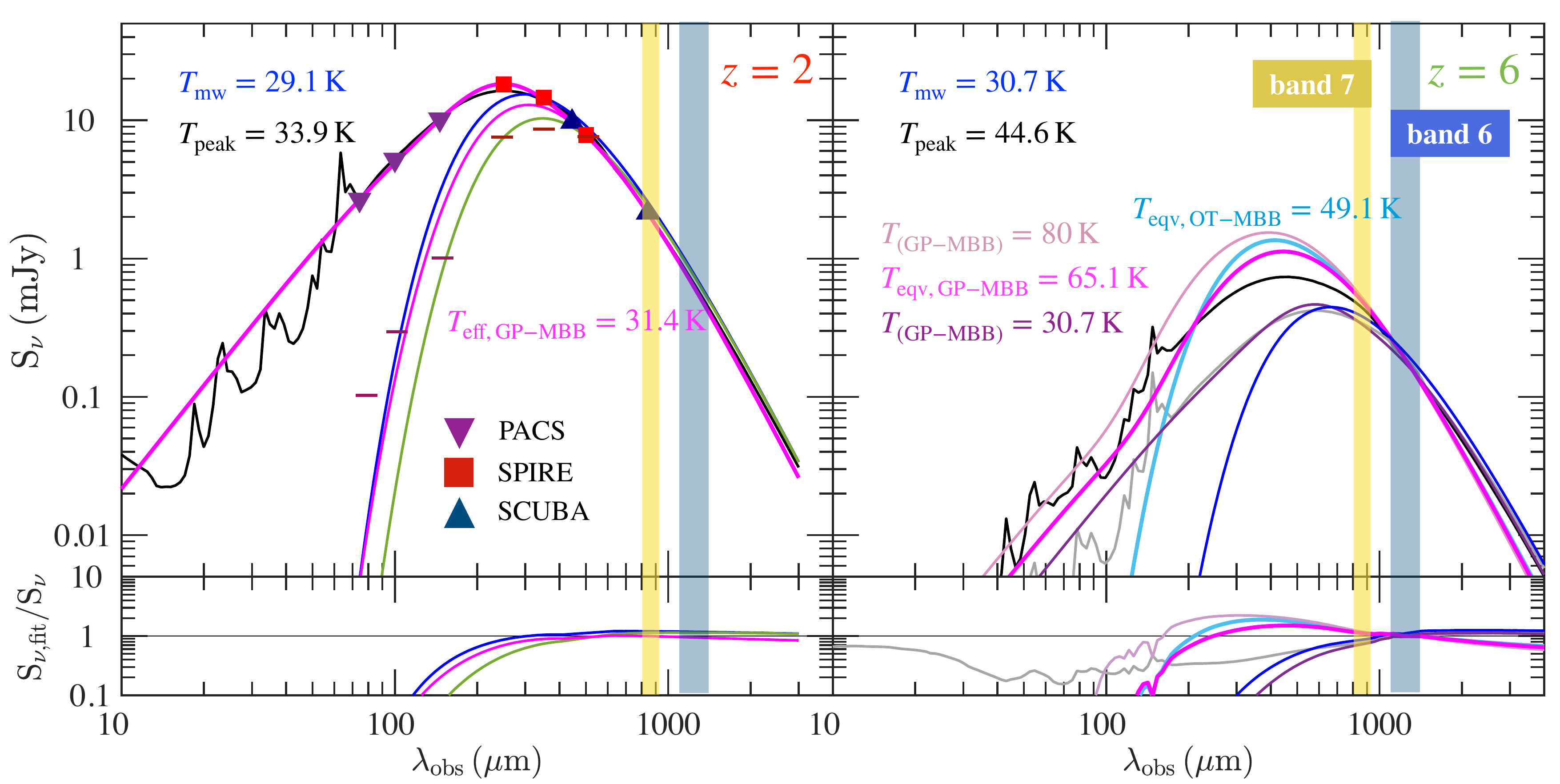}
 \caption{ \textsc{\small SKIRT} SED (black lines) of selected $z=2$ (\textit{upper left panel}) and $z=6$ (\textit{upper right panel}) \textsc{\small MassiveFIRE} galaxies and SED fitting functions (coloured lines) for the two galaxies. In the \textit{upper left panel}, the thick magenta line represents the GP-MBB function (Eq.~\ref{eq.4}, with $\alpha=2.5$, $\beta=2.0$ and $\lambda_1=100\rm \mu m$) that best fits PACS + SPIRE + SCUBA + ALMA photometry calculated from its \textsc{\small SKIRT} SED. The thin magenta line represents the MBB component of the GP-MBB function. The derived effective temperature $T_{\rm eff}$ of the GP-MBB function is 31.4 K. The blue line shows the OT-MBB function (Eq.~\ref{eq.5}, with $\beta=2.0$) with $T$ being equal to the mass-weighted temperature $T_{\rm mw}=29.1$ K of the galaxy. The green line shows the G-MBB function with the same $M_{\rm dust}$ and $T$ but $\lambda_1=100\rm \mu m$. \textbf{The optical depth in the G-MBB function results in a lower luminosity-to-mass ratio as well as a longer emission peak wavelength than the OT-MBB function with the same $M_{\rm dust}$ and $T$}. The calculated PACS, SPIRE and SCUBA flux densities of the galaxy are explicitly marked with the different symbols as labeled, and the horizontal ticks mark the confusion noise limit of the PACS/SPIRE bands. In the \textit{upper right panel}, we show the GP-MBB (thick salmon, magenta and purple lines) and OT-MBB (light blue line) functions that are normalised to match the observed flux density at ALMA band 6 (1.2 mm). The magenta and light blue lines correspond to the MBB functions with $T=T_{\rm eqv}$ that yield the $L_{\rm IR}$ of the $z=6$ galaxy. The salmon (purple) line corresponds to a GP-MBB function with $T>T_{\rm eqv}$ ($T<T_{\rm eqv}$), resulting in an over(under)-estimate of $L_{\rm IR}$. Like in the \textit{upper left panel}, we show with blue line the OT-MBB function with $T=T_{\rm mw}$ and $M_{\rm dust}$ of the selected galaxy. The grey line represents the SED of the $z=2$ galaxy that is redshifted to $z=6$ and rescaled to match the observed flux density of the $z=6$ galaxy at ALMA band 6. In the two \textit{upper panels}, the golden and grey shaded region mark ALMA band 7 and 6, respectively. In the \textit{lower panels}, the coloured lines show the ratio of the flux of the MBB fitting functions (excluding power-law component in Eq.~\ref{eq.4} for the GP cases) to the simulated flux calculated by \textsc{\small SKIRT} that are shown in the \textit{upper panels}. \textbf{An OT-MBB function with $T_{\rm mw}$ fits the RJ part of the dust SED quite well, while a GP-MBB function is able to also match the dust SED left of the peak. }} 
\label{fig:SED}
  \end{center}
    \vspace{-15pt}
\end{figure*}

\paragraph*{Equivalent dust temperature $T_{\rm eqv}$\\}

We define $T_{\rm eqv}$ as the temperature that reproduces the actual IR luminosity for a given broadband flux (e.g., at 870 $\mu$m) and adopted parametrised functional form of the SED (e.g., OT-MBB). The value of $T_{\rm eqv}$ typically depends on both the observing frequency band as well as the SED form (Section~\ref{S4}).

In the specific case of optically-thin dust emission, the specific luminosity, can be written as 

\begin{equation}
\begin{split}
L_{\rm \nu,\;OT}(T, M_{\rm dust}) &= 4\pi(1+z)^{-1}d^2_{\rm L} S_{\nu_{\rm o}}\\
&=4\pi \kappa_{\nu} M_{\rm dust} B_{\nu}(T)
\label{eq.6}    
\end{split}
\end{equation}

\noindent By directly integrating the above formula over $\nu$, one obtains the total IR luminosity \citep[\eg][]{HK11}

\begin{equation}
\begin{split}
L_{\rm IR, \;OT} (T, M_{\rm dust})& = \int^{\infty}_0 4\pi M_{\rm dust}\kappa_{\nu} B_{\nu}(T) \;d\nu\\
&= 4\pi M_{\rm dust} \kappa_{\nu_1}\nu_1^{-\beta}(\frac{k_B T}{h})^{4+\beta}(\frac{2h}{c^2})\\&\phantom{==}\Gamma (4+\beta) \zeta(4+\beta)\\
&= \mathcal{D} M_{\rm dust}T^{(4+\beta)},
\end{split}
\label{eq:OTMBB}
\end{equation}
where $\mathcal{D}(\kappa_{\nu_1}, \nu_1, \beta)$ is a constant and $\Gamma$ and $\zeta$ are Riemann functions.

Combining Eq.~\ref{eq:OTMBB} and Eq.~\ref{eq.5}, $T_{\rm eqv}$ can now be defined as the temperature satisfying
\begin{equation}
L_{\rm IR} / S_{\nu_0} = \frac{\mathcal{D} T_{\rm eqv}^{4+\beta}}{\mathcal{C}_\nu(z) B_\nu(T_{\rm eqv})}.
\label{eq:Tequiv}
\end{equation}

\noindent In the RJ regime, where $B_\nu(T_{\rm eqv}){}={}2\nu^2k_{\rm B}T_{\rm eqv}/c^2$,
\begin{equation}
    L_{\rm IR} / S_{\nu_0} \propto T^{3+\beta}_{\rm eqv}.
\label{eq:LSequiv}    
\end{equation}

$T_{\rm eqv}$ is therefore the temperature that one would need to adopt in order to obtain the correct IR luminosity and match the broadband flux density under the assumption that the SED has the shape of an OT-MBB function. Of course, the latter assumption is often a poor one and the actual SED shape can differ substantially from an OT-MBB curve. In this case, the equivalent temperature will be different from the mass-weighted dust temperature. Furthermore, the dust mass that is derived this way (via Eq.~\ref{eq.5} for a given $T_{\rm eqv}$ and $S_{\nu_0}$) will then differ from the actual physical dust mass.

In this paper, we compute $T_{\rm eqv}$ based on Eq.~\ref{eq:Tequiv} using the actual integrated IR luminosities and 870 $\rm \mu m$ (1.2 mm) flux densities unless explicitly noted otherwise. For equivalent temperatures based on G-MBB or GP-MBB spectral shapes, we numerically integrate Eq.~\ref{eq.3} and Eq.~\ref{eq.4} to obtain the IR luminosity for a given dust temperature and dust mass (analogous to Eq.~\ref{eq:OTMBB} for the OT-MBB case).

\subsection{The SEDs of simulated galaxies}
\label{S3b}

In Figure~\ref{fig:SED} we show example SEDs of a $z=2$ galaxy and a $z=6$ galaxy from the {\sc\small MassiveFIRE} sample. We separately discuss $z=2$ and the $z=6$ galaxies because the observational strategies for the two epochs are usually different. For $z=2$, an IR-luminous (\ie~$L_{\rm IR}\simgreat 10^{12}\,L_{\odot}$) galaxy may have both \textit{Herschel} coverage at FIR as well as (sub)mm coverage from ground-based facilities (\eg~SCUBA, ALMA and AzTEC). One can then \emph{derive} the dust temperature ($T_{\rm peak}$ or $T_{\rm eff}$) from the observed FIR-to-mm photometry via SED fitting. At $z>4$, the sources that have a good coverage of the SED peak (via \textit{Herschel} surveys) are currently limited to higher IR luminosity (\ie~$L_{\rm IR}\simgreat 10^{13}\,L_{\odot}$) and the majority are strongly lensed objects \citep[\eg][]{W13,I16,SW16,Z17,M18,R19}. Meanwhile, the unprecedented sensitivity of ALMA has allowed us to detect the dust continuum of a growing population of galaxies at these epochs \citep[\eg][]{C15,L17,H18,MSB19} that do not have detected \textit{Herschel} counterparts. Most of these observations cover only a single band (typically at ALMA band 6 or 7). Physical properties, such as $L_{\rm IR}$ and SFR, are thus often derived based on a single data point at (sub)mm \citep[\eg][]{FC17,S18}, by \emph{assuming} a dust temperature for the object \citep[\eg][]{B16,C18a}. This approach is sensible if the adopted dust temperature is close to $T_{\rm eqv}$ of the given galaxy (see section \ref{S3a}).

\subsubsection{Example: The SED of a galaxy at $z=2$}

Figure~\ref{fig:SED} shows the SED of a selected $z=2$ {\sc\small MassiveFIRE} galaxy (\textit{upper left panel}). This galaxy has $L_{\rm IR}{}={}3.0\times10^{12}\,L_{\odot}$, SFR${}={}210\,M_\odot\,\rm yr^{-1}$, $M_{\rm dust}{}={}5.4\times10^8\,M_{\odot}$ and $M_*{}={}5.3\times10^{11}\,M_{\odot}$ \footnote{Physical properties of the simulated galaxies reported in this paper are measured using the material within a 30 pkpc kernel around the DM halo centre, \ie~the minimum gravitational potential.}.

We calculate the PACS (70, 100 and 160 $\rm \mu m$) + SPIRE (250, 350 and 500 $\rm \mu m$) + SCUBA-2 (450 and 850 $\rm \mu m$) + ALMA (870 $\rm \mu m$ and 1.2 mm) broadband flux densities from the simulated SED. We fit its FIR-to-mm photometry --- assuming successful detection at every band, as we show in the left panel that the PACS/SPIRE fluxes of this galaxy are above the confusion noise limit (marked by the horizontal ticks) \citep{N10,MP13} and the submm fluxes are above the typical sensitivity limit of SCUBA-2 and ALMA --- by a GP-MBB function (with $\lambda_1=100\;\rm \mu m$, $\beta=2.0$ and $\alpha=2.5$) using least-$\chi^2$ method. $N_{\rm pl}$ and $T$ are left as two free parameters for the fitting. The best-fitting GP-MBB function is shown by the thick magenta line. The derived $T_{\rm eff}$ is $31.4$ K, which is similar to its mass-weighted temperature ($T_{\rm mw}=29.1$ K)\footnote{How well $T_{\rm eff}$ in the best-fitting GP-MBB function approximates $T_{\rm mw}$ depends on its parametrisation (see Section~\ref{S3a}). For instance, increasing $\lambda_1$ in Eq.~\ref{eq.4} from 100 to 200 $\rm \mu m$ changes $T_{\rm eff}$ from 29.1 K to 48.2 K (see also Figure 20 of \citealt{C14}).}. From the best-fitting GP-MBB function (and also the simulated SED), $T_{\rm peak}$ is found to be 33.9 K. 

For demonstration purpose, we also show with the blue line the exact solution of the OT-MBB function, with $T=T_{\rm mw}=29.1$ K, $M_{\rm dust} = 5.4\times10^8\,M_{\odot}$, $\kappa_{870}=0.05\,\rm m^2\,kg^{-1}$ and $\beta=2.0$. As expected, the OT-MBB function with a mass-weighted temperature is in very good agreement with the galaxy SED at long wavelength. For this galaxy, at $\lambda=100-650\,\rm \mu m$ ($\lambda_{\rm o}=300\,\rm \mu m-2\,\rm mm$), the difference between the flux of the OT-MBB function and the simulated flux is within $10\%$ (illustrated by the lower left panel). At shorter wavelength, the emission is more tied to the dense, warm dust component in the galaxy, which is poorly accounted for by this OT-MBB function with a mass-weighted temperature. Overall, the OT-MBB function accounts for $\sim50\%$ of $L_{\rm IR}$ of the galaxy, and the discrepancy is largely due to the MIR emission.

We also show the effect of optical depth. In the upper left panel, the green line shows the analytic solution from a G-MBB (Eq.~\ref{eq.3}) function with the same $M_{\rm dust}$ and $T$ ($T=T_{\rm mw}=29.1$ K), but with a power-law optical depth that equals unity at rest-frame $\nu_0=1.5$ THz, or $\lambda=100\,\rm \mu m$. While the emission looks identical to the optical-thin case (blue line) at long wavelength ($\lambda_{\rm o}>500\rm \,\mu m$), it appears to be lower at shorter wavelength when the effect of optical depth becomes important. The effect of increasing optical depth is that the overall light-to-mass ratio is lower and the emission peak wavelength is longer compared to the optically-thin case \citep[\cf][]{S13}.

\subsubsection{Example: The SED of a galaxy at $z=6$}

Figure~\ref{fig:SED} also shows the SED of a $z=6$ \textsc{\small MassiveFIRE} galaxy. This galaxy has lower $L_{\rm IR}$ ($3\times10^{11}\,L_{\odot}$) and $M_{\rm dust}$ ($8\times10^7\,M_{\odot}$) compared to the $z=2$ galaxy, but interestingly, it has similar $T_{\rm mw}$ (30.7 K). The calculated flux densities at ALMA band 7 ($S_{\rm 870\,\mu m}$) and 6 ($S_{\rm 1.2 mm}$) are 0.44 and 0.23 mJy, respectively. Like the $z=2$ galaxy, an OT-MBB function (blue line) with $M_{\rm dust}$ and $T=T_{\rm mw}$ can well describe the emission of the $z=6$ galaxy at long wavelength (for this case, $\lambda_{\rm o}>1.2$ mm, or rest-frame $\lambda>170\,\rm \mu m$), but it only accounts for $\sim30\%$ of $L_{\rm IR}$. A larger fraction of the total emission of this $z=6$ galaxy origins from the warm dust component.

To estimate $L_{\rm IR}$ of a $z=6$ galaxy from $S_{\rm 870\,\mu m}$ (or $S_{\rm 1.2 mm}$), one often needs an assumed SED function and an assumed $T_{\rm eqv}$ for the adopted function. Since it is extremely difficult to constrain the details of SED shape at this high redshift, often a simple OT-MBB or GP-MBB function is used by the observational studies \citep[\eg][]{C15, B16, C18a}. As an example, we fit the OT-MBB function to $S_{\rm 1.2 mm}$ of the $z=6$ \textsc{\small MassiveFIRE} galaxy with varying $T$. We show in the right panel of Figure~\ref{fig:SED} the OT-MBB function (with fixed $\beta=2.0$) that yields the simulated $L_{\rm IR}$ with the light blue line. The derived $T_{\rm eqv}$ for this function is 49.1 K. This is significantly higher than $T_{\rm mw}$, and as a result, the RJ side of the derived SED of this function appears to be much steeper than the simulated SED. It also poorly fits the simulated SED at wavelength close to $\lambda_{\rm peak}$. The derived $T_{\rm peak}$ is therefore very different from the true $T_{\rm peak}$ of the simulated SED. 

We also fit $S_{\rm 1.2 mm}$ of this galaxy by a GP-MBB function ($\lambda_1=100\,\rm \mu m$, $\beta=2.0$, $\alpha=2.5$). We show the result for $T=30.7$ K (purple line), $T=65.1$ (magenta line) and $T=80$ K (salmon line). For $T=T_{\rm mw}=30.7$ K, we use the same normalisation of the power-law component as for the $z=2$ galaxy (upper left panel), so that the SED shape is similar between these two galaxies. For $T=65.1$ K and $T=80.0$ K, we use the best-fitting normalisation factor derived based on the \textit{local} GOALS sample (see Table 1 of C12). We can see that the GP-MBB function appears to better describe the simulated SED shape compared with OT-MBB function, but in order to fit the simulated SED with reasonably good quality, a different choice of $N_{\rm pl}$ and $\lambda_1$ is needed.  With $T=T_{\rm mw}=30.7$ K, the GP-MBB function under-predicts the simulated $L_{\rm IR}$ ($3\times10^{11}\,L_{\odot}$) by $70\%$. Using $T_{\rm eqv,\,GP-MBB}=65.1$ K, this function leads to the right $L_{\rm IR}$. We also show the result for $T=80$ K, which over-predicts the $L_{\rm IR}$ by about a factor two.

In conclusion, we find that an OT-MBB function with a mass-weighted dust temperature well describes the long-wavelength ($\lambda\simgreat 200\rm \,\mu m$) part of the dust SED, but it does not well account for the Wien side of the SED and leads to significant under-estimate of $L_{\rm IR}$. A GP-MBB function can provide high-quality fitting to the simulated SED with good FIR+(sub)mm photometry of galaxy. Using single-band (sub)mm flux density of $z>4$ galaxies, $T_{\rm eqv}$ is very different from $T_{\rm mw}$ of the galaxy. We will discuss $T_{\rm eqv}$ for high-redshift galaxies, its evolution with redshift and its dependence on other galaxy properties in more details in Section~\ref{S4}.

\subsection{Comparing simulation to observation}
\label{S3c}

\begin{figure}
 \includegraphics[height=96mm, width=85mm]{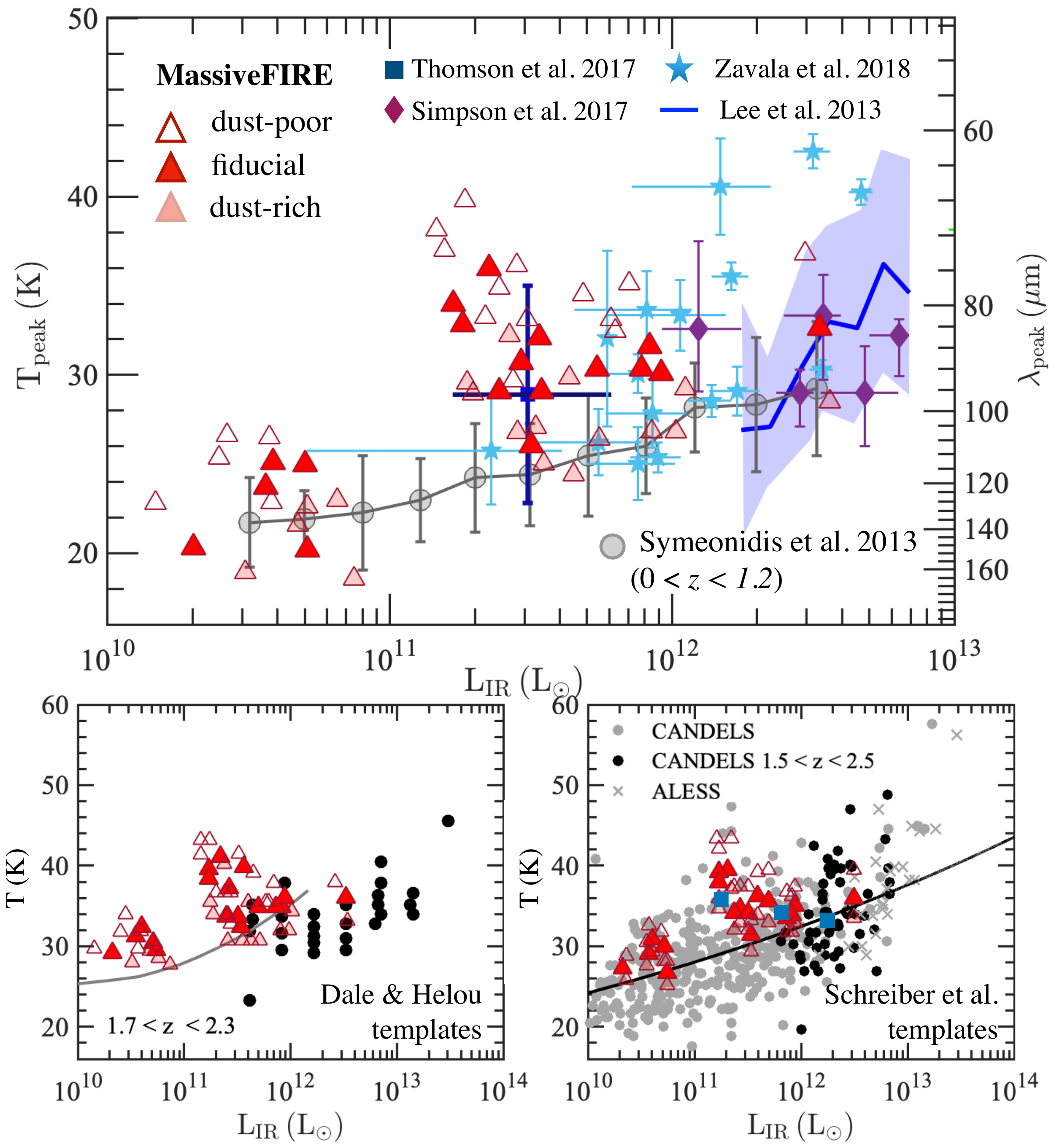}
 \vspace{-16pt}
 \caption{ Dust temperature vs. $L_{\rm IR}$ relation of the $z\sim2$ galaxies. The red triangles represent the simulated data of the \textsc{\small MassiveFIRE} sample at $z=2$. The unfilled, filled, semi-transparent symbols show the result for the dust-poor ($\delta_{\rm dzr}=0.2$), fiducial ($\delta_{\rm dzr}=0.4$) and dust-rich ($\delta_{\rm dzr}=0.8$) models, respectively. In the upper panel, we compare the simulated data to the observational results where dust temperature is derived using the SED fitting technique and with MBB-like functions (Eq.~\ref{eq.3}-\ref{eq.5}). The observation data by \citet[][hereafter S17]{SS17}, \citet[][hereafter Z18]{Z18} and the stacked result by \citet[][hereafter T17]{TS17} are represented by purple diamonds, cyan asterisks and blue square, respectively. The blue shaded area shows 1$\sigma$ distribution of the compilation of high-redshift COSMOS galaxies by \citet{L13}. The grey circles and error bars show the binned result and its 1$\sigma$ distribution of the \textit{Herschel}-selected sample at lower redshift ($z=0\sim1.2$) from \citet[][hereafter S13]{SV13}. To make fair comparison, we convert $T_{\rm eff}$ presented in S13, S17, T17, and Z18 to $T_{\rm peak}$. The relation between $T_{\rm peak}$ and $T_{\rm eff}$ for each study is shown in Figure~\ref{fig:TpTeff}. In the \textit{lower panels}, we show the observational data derived using empirical SED templates. The stacked result by \citet{M14} and the compilation by \citet{S18} are shown in the \textit{left and right panels}, respectively. The solid grey line in the \textit{lower left panel} represents a second-order polynomial fit to the data points of a lower-redshift bin ($0.2 < z < 0.5$). The solid black line in the \textit{lower right panel} represents the derived $T-L_{\rm IR}$ scaling relation by \citet{S18} using the combined HRS \citep{B10} + CANDELS \citep{G11,HK13} + ALESS \citep{HK13,SS14} samples from local to $z\sim4$. The blue squares show the stacked results for the three luminosity bins at $z\sim2$. The dust temperature in the \textit{lower panels} is defined using the same method as in \citet{M14} and \citet{S18}. \textbf{The dust temperature of the $z=2$ \textsc{\small MassiveFIRE} sample is in good agreement with the observational data.}}
    \label{fig:TL2}
     \vspace{-15pt}
\end{figure}

\begin{figure}
 \includegraphics[width=80mm]{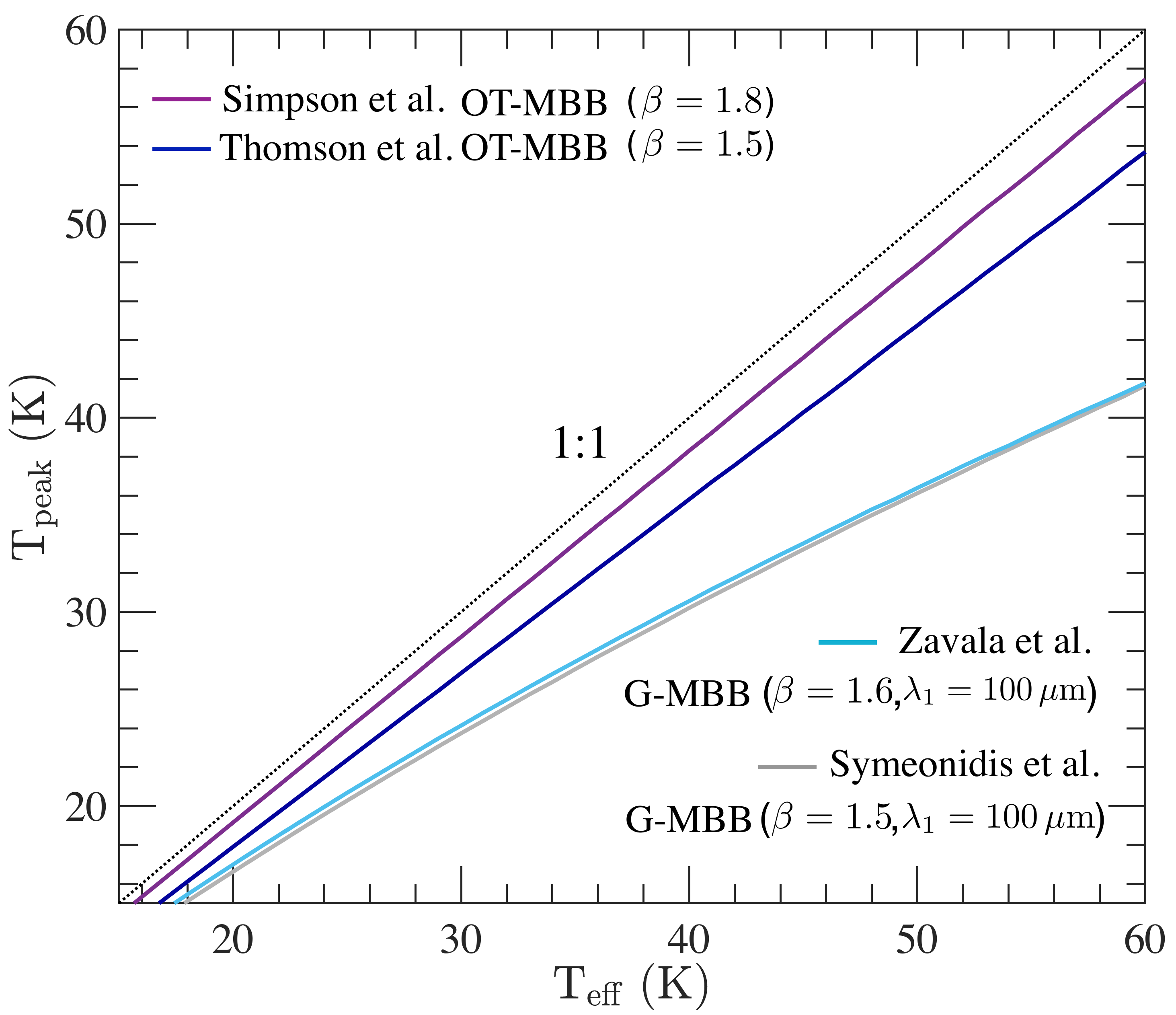}
 \caption{$T_{\rm peak}$ vs. $T_{\rm eff}$ relation of different MBB functions (see also Figure 20 of \citealt{C14}). The purple, blue, cyan and grey lines correspond to OT-MBB (Eq.~\ref{eq.5}) with $\beta=1.8$ \citep{SS17}, OT-MBB with $\beta=1.5$ \citep{TS17}, G-MBB (Eq.~\ref{eq.3}) with $\beta=1.6$ \citep{Z18} and G-MBB with $\beta=1.5$ \citep{SV13}. For the two G-MBB functions, $\lambda_1=100\,\rm \mu m$. We convert $T_{\rm eff}$ reported in the above papers to $T_{\rm peak}$ using the $T_{\rm peak}$ vs. $T_{\rm eff}$ relation and show the result in Figure~\ref{fig:TL2}. The black dashed line shows the 1-to-1 relationship.}
    \label{fig:TpTeff}
    \vspace{-10 pt}
\end{figure}

Due to the high confusion noise level of the \textit{Herschel} PACS/SPIRE cameras, most current observational studies on dust temperature at high-redshift are limited to the most IR-luminous galaxies in the Universe. For $z=2$, the observations are generally limited to $L_{\rm IR}\simgreat10^{12}\,L_{\odot}$. Applying the powerful stacking technique to the \textit{Herschel} images, it is also possible to probe the fainter regime of a few $10^{11}\,L_{\odot}$ at $z\sim2$ \citep[\eg][]{TS17, S18}. Yet another problem with the observational studies is the strong selection bias with flux-limited surveys, meaning that the selected galaxy sample is limited to increasing IR luminosity with redshift. It is therefore non-trivial to disentangle the dependence of dust temperature on redshift and that on other galaxy properties. Using simulated sample, we do not expect to have such problem. 

We start here by comparing the result of the \textsc{\small MassiveFIRE} sample at $z=2$ with the observational data from similar redshift. This is where the luminosity range of our simulated galaxies share the largest overlap with the current observational data. The selection methods of the quoted data are summarized in Table~\ref{T1}. At higher redshift, the observations are biased to higher $L_{\rm IR}$. In the following section, we will explicitly discuss the redshift evolution of dust temperatures with the \textsc{\small MassiveFIRE} sample. 

We present the result in Figure~\ref{fig:TL2}. In the upper panel, we compare the simulations with the observational data of which the (originally \textit{effective}) dust temperature is derived using SED fitting technique and with MBB functions (\ie~Eq.~\ref{eq.3}-\ref{eq.5}), while in the lower panels, we show examples where the dust temperature of both the simulated and observation data is derived using the SED template libraries. In order to make fair comparison among different observations and with the simulation data, we convert all different $T_{\rm eff}$ presented in the literature to $T_{\rm peak}$ in the upper panel. $T_{\rm peak}$ of the simulated galaxies are derived from the best-fitting GP-MBB function (Eq.~\ref{eq.4}, with $\rm \lambda_1 = 100\,\mu m$, $\beta=2.0$ and $\alpha=2.5$) to the FIR-to-mm photometry. 

In the upper panel, we show with the blue shaded block the data from the H-ATLAS survey \citep{L13}, that encompasses the high-redshift ($1.5<z<2.0$) \textit{Herschel}-selected galaxies in the COSMOS field. The height of the block represents $1\sigma$ dispersion. We also explicitly show the $z=1.5-2.5$ objects from \citet{SS17} (purple diamonds) and \citet{Z18} (cyan asterisks), which are selected at 850 $\rm \mu m$ from the deep SCUBA-2 Cosmology Legacy Survey \citep[S2CLS;][]{G17} probing the Ultra Deep Survey (UDS) and the Extended Groth Strip (EGS) field, respectively. And finally, we present the stacked result by \citet{TS17} (blue square), which is based on a high-redshift ($\langle{}{z}\rangle{}=2.23$) sample extracted from the High-redshift Emission Line Survey (HiZELS) \citep{SS13}, comprising 388 and 146 $\rm H_\alpha$-selected star-forming
galaxies in the COSMOS and UDS fields, respectively. And for purpose of reference, we show the binned data from \citet{SV13} by grey filled circles and error bars, which encompasses an IR-selected sample at $0.1<z<2$ selected from the COSMOS, GOODS-N and GOODS-S fields. We convert the \textit{effective} dust temperature $T_{\rm eff}$ presented in \citet{SV13}, \citet{SS17}, \citet{TS17} and \citet{Z18} to $T_{\rm peak}$. The relation between $T_{\rm peak}$ and $T_{\rm eff}$ for the fitting functions that are used by the four studies are plotted in Figure~\ref{fig:TpTeff}. \citet{SS17} (\citealt{TS17}) adopt an OT-MBB function (Eq.~\ref{eq.5}) with fixed $\beta=1.8$ ($\beta=1.5$), while \citet{SV13} (\citealt{Z18}) use a G-MBB function (Eq.~\ref{eq.3}) with fixed $\beta=1.5$ ($\beta=1.6$) and $\lambda_1=100\rm \,\mu m$. From Figure~\ref{fig:TpTeff}, we can see that $T_{\rm eff}$ presented in the four studies is higher than $T_{\rm peak}$. 

\begin{table*}
\caption{The selection methods for the observational data presented in Figure~\ref{fig:TL2}.}
\begin{tabular}{ p{2.7 cm} p{14 cm}  }
 \hline
\multicolumn{1}{}{} \; Data Source \;\;\; & Section Method \\
 \hline
 \citet{L13} &  \textit{Herschel}-selected galaxy sample in the COSMOS field with $\ge5\sigma$ detections in at least two of the five PACS+SPIRE bands and with photometric redshifts (hereafter photo-$z$) between 1.5 and 2.0. The $1\sigma$ sensitivity limits are 1.5, 3.3, 2.2, 2.9 and 3.2 mJy in the 100, 160, 250, 350, 500 $\rm \mu m$ bands, respectively.  Photo-$z$s are calculated using fluxes in 30 bands that cover the far-UV at 1550 $\angstrom$ to the mid-IR at 8.0 $\rm \mu m$.\\
  \hline 
  \citet{SV13} & A sample of IR-selected (\textit{Sptzer} MIPS 24 $\rm \mu m$ + \textit{Herschel} PACS/SPIRE) galaxies at $z=0-2$ in the COSMOS and GOODS N+S fields. The sample is confined to those 24 $\rm \mu m$-detected ($f_{24}>30\,\rm \mu Jy$ for GOODS N+S and $f_{24}>60\,\rm \mu Jy$ for COSMOS) sources that have at least two reliable photometric data points in the two \textit{Herschel} bands ($>3\sigma$). 1/3 of the sample have spectroscopic redshifts and the rest photometric redshits. \\
    \hline 
 \citet{SS17}  &  SCUBA-2-detected ($\sigma_{850} = \rm 2.0\,mJy$, at $\ge 4\sigma$) galaxies in the UKIDSS Ultra Deep Survey (UDS) field with photo-$z$ between 1.5 and 2.5. Photo-$z$s are determined using 11 bands, covering from $U$-band to near-IR at 4.5 $\rm \mu m$.\\
  \hline
 \citet{TS17} &  535 galaxies detected in the HiZELS at $K$-band, corresponding to the redshifted wavelength of $\rm H_\alpha$ line at $z=2.23$. The sample is confined to those with dust-corrected luminosities $L_{\rm H_\alpha}\ge 2.96\times10^{42}\rm \,erg\,s^{-1}$, corresponding to a SFR of $\approx4\,M_{\odot}\,\rm yr^{-1}$.\\
 \hline
  \citet{Z18} &  SCUBA-2-selected galaxies in the EGS field detected at $>3.75\sigma$ at 450 and/or 850 $\rm \mu m$ ($\sigma_{450}=1.9$ and $\sigma_{850}=0.46\,\rm mJy$ beam$^{-1}$). The PACS/SPIRE photometry is obtained from the PACS Evolutionary Probe \citep[PEP;][]{L11} and the \textit{Herschel} Multi-tiered Extragalactic Survey \citep[HerMES;][]{O12} programs. This sample consists of objects with optical spectroscopic redshift, optical photometric redshift and FIR photometric redshift estimates.\\
 \hline
   \citet{M14} & The sample consists of the NIR-selected galaxies in the GOODS-N ($K_{\rm s}<24.3$, down to a $3\sigma$ significance), GOODS-S ($K_{\rm s}<24.3$, down to a $5\sigma$ significance) and COSMOS ($K_{\rm i}<25$, down to a $3\sigma$ significance) fields. For each SFR-$M_*$ bin, the mean dust temperature of the galaxies in the bin is derived using their mean PACS+SPIRE flux densities .  $29\%$, $26\%$ and $3\%$ of these galaxies have spectroscopic redshift estimates, and the rest have photometric redshift estimates based on the available optical-to-NIR data.\\
 \hline
   \citet{S18} &  The sample consists of the $z=0.3-4$ NIR-selected (down to a $5\sigma$ significance) galaxies in the GOODS-N ($K_{\rm s}<24.5$), GOODS-S ($H<27.4-29.7$), UDS ($H<27.1-27.6$) and COSMOS ($H<27.4-27.8$ and $K_{\rm s}<23.4$ for the CANDELS and UVISTA-detected sources, respectively) fields. For $T$ measurement, the galaxies are required to have at least one detection at $\ge5\sigma$ significance at the \textit{Herschel} bands on both sides of the peak of the FIR SED. It is also complemented with the $z=0$ volume-limited sample from the HRS as well as the $z=2-4$ galaxies in the Extended Chandra Deep Field South (ECDFS) field as part of the ALESS program. The ALESS sample are selected at 870 $\rm \mu m$ in the single-dish LABOCA image. For the CANDELS and ECDFS-field galaxies, photo-$z$s are calculated using the available UV-to-NIR multi-wavelength data.\\
 \hline
\end{tabular}
\vspace{-13 pt}
\label{T1}
\end{table*}

In the lower panels, we compare the simulated result with the observational data from \citet{M14} (left) and \citet{S18} (right), both of which fit the galaxy photometry to the empirical SED template libraries. In particular, \citet{M14} adopt the \citet{DH02} SED template library and determine the temperature for each template by fitting their PACS+SPIRE flux densities with an OT-MBB function with fixed $\beta=1.5$ and then finding the $T_{\rm eff}$ for the best-fitting OT-MBB function. Their sample comprises of near-infrared (NIR)-selected galaxies in GOODS-N, GOODS-S and COSMOS fields with reliable SFR, $M_{\rm star}$ and redshift estimates.
The galaxies are binned in the SFR-$M_{\rm star}$-$z$ plane and dust temperatures are inferred using the stacked FIR ($100-500\;\rm \mu m$) flux densities of the SFR-$M_{\rm star}$-$z$ bins with least-$\chi^2$ method. We show the stacked result for their $1.7<z<2.3$ redshift bin with the black filled dots in the \textit{lower left panel}. For purpose of reference, we also show with the solid grey line the result of a lower-redshift bin ($0.2<z<0.5$) in the same panel.

In the \textit{lower right panel}, we also compare the simulation to the observational data of \citet{S18}, of which the galaxy catalogue is based on the CANDELS survey \citep{G11, K11}, a $z=2-4$ galaxy sample from the ALESS program \citep{HK13, SS14}, as well as the local \textit{Herschel} Reference Survey \citep[HRS,][]{B10}. The temperature is derived by fitting the PACS+SPIRE photometry to the \citet{S18} SED template library, which is constructed based on the \citet[][hereafter G11]{GH11} library of elementary templates with an assumed power-law distribution of $U$. The G11 templates are a set of MIR-to-mm spectra emitted by a uniform dust cloud of $1M_{\odot}$ when it is exposed to the \citet{M83} interstellar radiation field of a range of $U$. The temperature assigned to each \citet{S18} template of galaxy SED is the mass-weighted value of the G11 templates being used. We show in the \textit{lower right panel} the result for the CANDELS sample with the black and grey filled circles. The black circles explicitly represent the objects at $z=1.5-2.5$. We also show with blue squares the result of the stacked SEDs for $z=1.5-2.5$ derived based on the PACS/SPIRE photometry in the CANDELS sample. The result of the ALESS sample at higher redshift ($z=2-4$) is shown with grey crosses. The black curve shows the scaling relation $(T/{\rm K})=5.57\,(L_{\rm IR}/L_{\odot})^{0.0638}$ that is derived by \citet{S18} using the combination of the CANDELS, ALESS and HRS samples. 

For the simulated $z=2$ galaxies, we fit their PACS/SPIRE photometry to the \citet{DH02} (as \citealt{M14}) and \citet{S18} SED templates using least-$\chi^2$ method and find the temperature associated with the best-fitting template SED as defined in the literature. In other words, the temperature of the \textsc{\small MassiveFIRE} galaxies is not the same in each of the three panels. The temperature derived following the \citet{M14} and \citet{S18} methods are on average 5.2 and 4.2 K higher than $T_{\rm peak}$, respectively. Comparing the simulated with the observational data, we find an encouragingly good agreement over the common range of $L_{\rm IR}$, with either the observational data derived using SED fitting technique (upper panel), or using SED templates (lower panels). And apart from that, $T_{\rm peak}$ of the simulated $z=2$ galaxies appear to show no clear correlation with $L_{\rm IR}$ in all three panels, at least at $L_{\rm IR}\simgreat 10^{11}\,L_{\odot}$. This is consistent with the recent finding by \citet{S18} that the mean dust temperature derived from the stacked SEDs of the three $L_{\rm IR}$ bins of their $z\sim2$ sample shows almost no correlation over the range of $1.5\times10^{11}-1.5\times10^{12}\,L_{\odot}$ (blue squares) and lies systematically above the mean temperature of galaxies at lower redshift (black line). \textit{This suggests that high-redshift galaxies do not necessarily follow a single, \textit{fundamental} $L_{\rm IR}-T$ scaling relation}, which is typically derived using flux-limited observational data across a range of redshift but without much overlap of $L_{\rm IR}$ among different redshift bins. We will also show in Section~\ref{S3d3} that the dust temperatures of our \textsc{\small MassiveFIRE} sample increase with redshift at fixed $L_{\rm IR}$ from $z=2$ to $z=6$. \citet{M19} also report that the same redshift evolution extends to higher redshift (up to $z=10$) using a different suite of \textsc{\small FIRE} simulations.

\begin{figure*}
 \begin{center}
 \includegraphics[width=175mm]{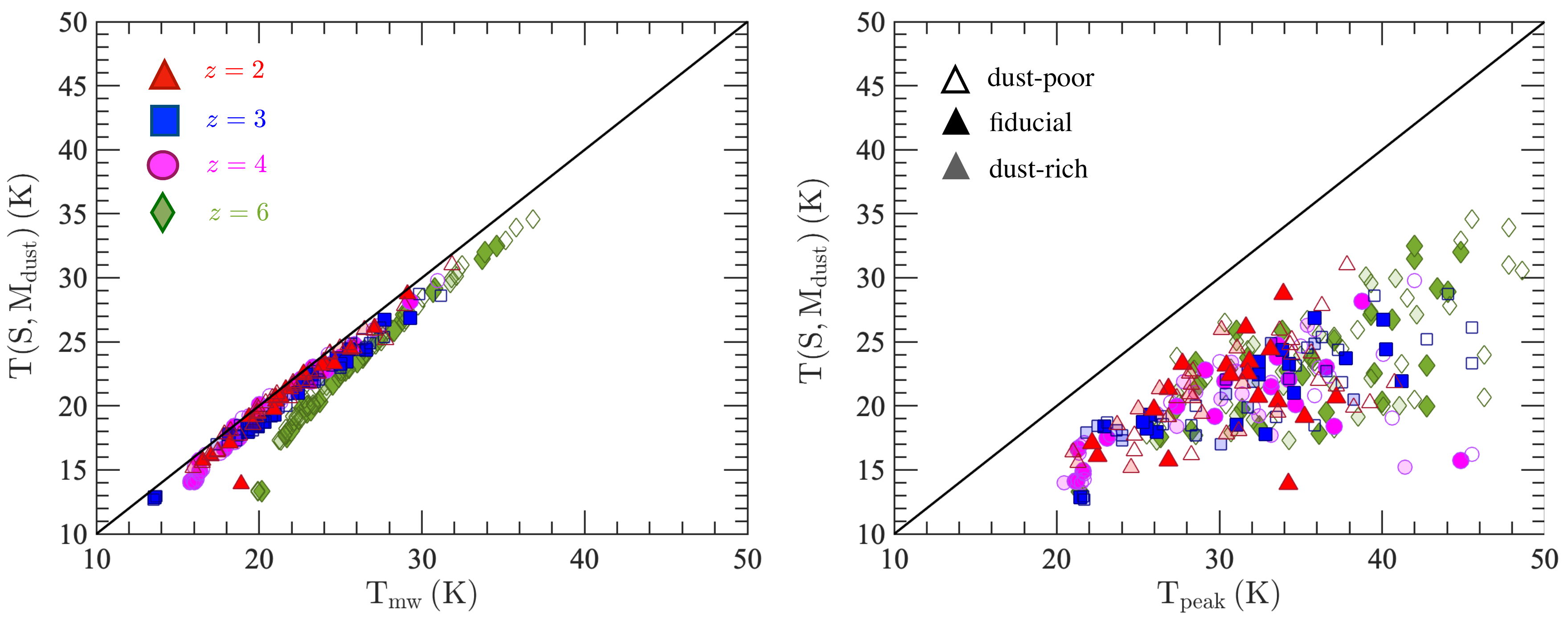}
 \caption{Relation of the temperature needed for dust mass estimate (calculated from Eq.~\ref{eq:Snu0OT}) against $T_{\rm mw}$ (left panel) and $T_{\rm peak}$ (right panel) of the \textsc{\small MassiveFIRE} sample at $z=2$ (red triangles), $z=3$ (blue squares), $z=4$ (magenta circles) and $z=6$ (green diamonds). For the $z=2-4$ galaxies, the flux density for mass estimate is measured at ALMA band 7 ($\lambda_{\rm o}=870\;\rm \mu m$), while for the $z=6$ galaxies (green), it is measured at ALMA band 6 ($\lambda_{\rm o}=1.2\,\rm mm$) so as ensure the rest-frame wavelength is on the optically-thin part of the RJ tail. The unfilled, filled and semi-transparent symbols represent the result for $\delta_{\rm dzr}=0.2$, 0.4 and 0.8, respectively. The solid diagonal line marks the 1-to-1 locus. \textbf{$T_{\rm mw}$ is the temperature needed for estimating dust mass using the RJ-tail approach. $T_{\rm peak}$ is a poor proxy for this temperature.}}
    \label{fig:RJ}
  \end{center}
\end{figure*}

The observational data shows nontrivial scatter, which is particularly clear in the upper and lower right panels. At $L_{\rm IR}\approx3\times10^{12}\,L_{\odot}$, for instance, $T_{\rm peak}$ (upper panel) is observed to be as low as $\sim25$ K and as high as $\sim45$ K. One possible reason is the intrinsic scatter of $\delta_{\rm dzr}$.  We show in Figure~\ref{fig:TL2} the result for the dust-poor ($\delta_{\rm dzr}=0.2$) and dust-rich ($\delta_{\rm dzr}=0.8$) models in each panel. The former (latter) show $\sim3$ K increase (decrease) of dust temperature(s) compared with the fiducial model ($\delta_{\rm dzr}$). This difference, however, still appears to be relatively smaller compared to the scatter of the observational data. A larger variance of $\delta_{\rm dzr}$ may lead to a larger scatter of temperature. Apart from that, another reason could be the variance of the conditions of the ISM structure on the unresolved scale (\eg~compactness and obscurity of the birth-clouds embedding the young stars) could also contribute to the scatter. We will discuss more about the impact of sub-grid models later in Section~\ref{S5}. And finally, given that the \textit{Herschel} cameras have fairly high confusion noise level, and it is rare that one galaxy has full reliable detection at every PACS/SPIRE+SCUBA band, we suggest that both factors can cause nontrivial uncertainty of observational result. Future infrared space telescope \citep[\eg~{\sc \small SPICA},][]{SA17,E18} spanning similar wavelength range and with higher sensitivity may help improve the constraint near emission peak and hence the observationally-derived dust temperatures.

We also note that $z=2$ {\sc \small MassiveFIRE} galaxies appear to show higher dust temperature compared to the lower-redshift counterparts in the observed sample, with either the temperature derived using SED fitting (\textit{upper panel}) technique or SED templates (\textit{lower panels}). Observationally, how dust temperature evolves at fixed $L_{\rm IR}$ (or $M_{\rm star}$) from $z=0$ to $z=2$ is still being debated \citep[\eg][]{H10,M12,MP13,L14,M14,B15,K17,S18}. Uncertainties can potentially arise from selection effects (surveys at certain wavelengths preferentially select galaxies of warmer/colder dust) \citep[\eg][]{ME10,HK11,MS19} and inconsistency in derivation of dust temperature. The dust temperature of galaxies in this redshift regime ($z<2$) is beyond the scope of this paper.

\begin{figure*}
 \begin{center} 
 \includegraphics[width=170mm]{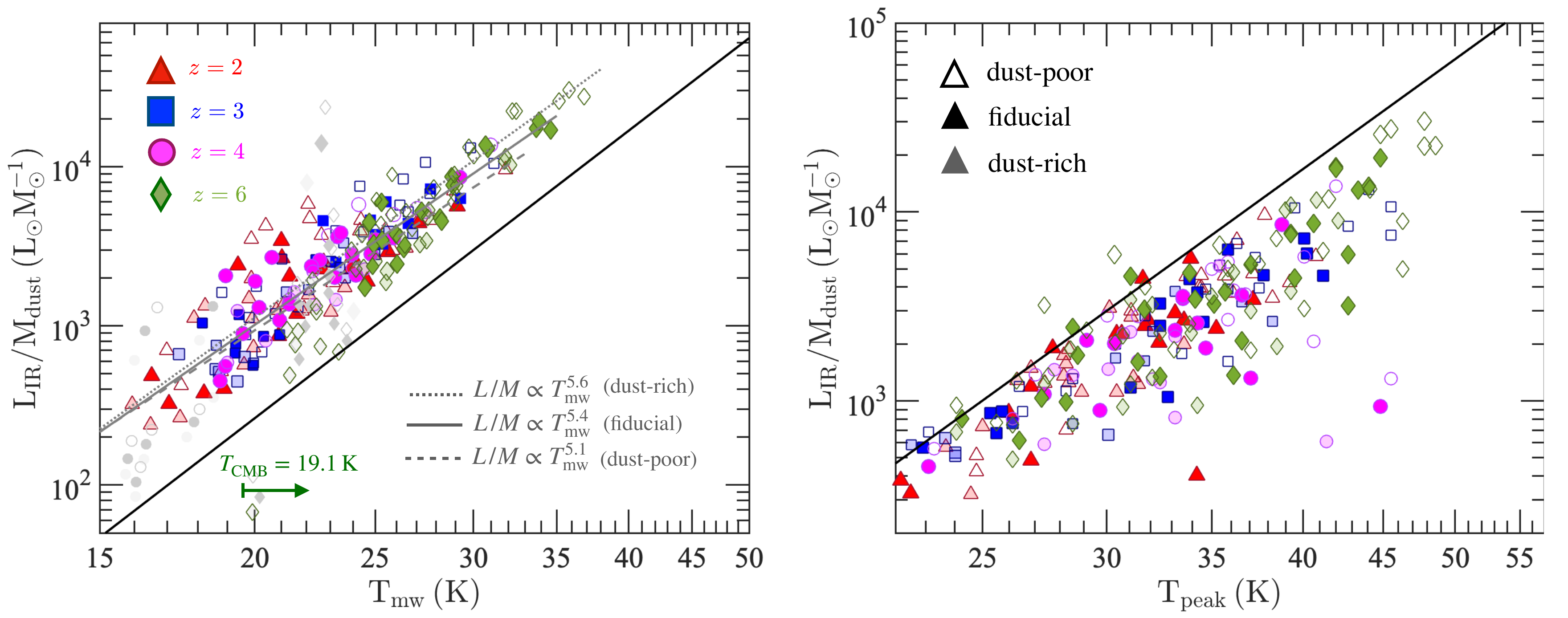}
 \caption{Relation of $L_{\rm IR}/M_{\rm dust}$ against $T_{\rm mw}$ (\textit{left panel)} and $T_{\rm peak}$ (\textit{right panel}) of the \textsc{\small MassiveFIRE} sample at $z=2-6$. The result for the fiducial, dust-poor and dust-rich cases are shown with unfilled, filled and semi-transparent symbols, respectively. In the left panel, the dotted, solid and dashed grey lines represent the best-fit power-law scaling relation for the dust-rich, fiducial and dust-poor cases, respectively. Those galaxies of which $T_{\rm mw}$ is strongly affected by CMB heating, \ie~$T_{\rm mw}-T_{\rm CMB}(z)<5\,\rm K$, are coloured by grey. They are excluded from the power-law fitting. The dust-rich (poor) case exhibits a flatter (steeper) $L_{\rm IR}/M_{\rm dust}$ vs. $T_{\rm mw}$ scaling relation compared with the fiducial model. The solid black line in each panel represents the expected analytic scaling using the optically-thin MBB function (Eq.~\ref{eq.5}), with the dust emissivity spectral index $\kappa_{870}=0.05\;\rm m^2\;kg^{-1}$.}
    \label{fig:Peak}
  \end{center}
\end{figure*}

\subsection{The role of dust temperature in scaling relationships} 
\label{S3d}

The scaling relationships of dust temperature against other dust/galaxy properties (such as total IR emission, sSFR and etc.) have been extensively studied in the past decade because of the significant boost of the number of detected high-redshift dusty star-forming galaxies by \textit{Herschel}, SCUBA and ALMA. We now have statistically large sample for revealing and studying the various scaling relationships of dust temperature. Here in this section, we show the result of the \textsc{\small MassiveFIRE} sample at $z=2-6$, discuss the physical interpretation of the scaling relations and specifically examine how each scaling relation differs by using different dust temperatures ($T_{\rm mw}$ vs. $T_{\rm peak}$). 

\subsubsection{$S\propto MT$ (optically-thin regime)}
\label{S3d1}

As mentioned above, the long-wavelength RJ tail can be well described by a single-$T$ OT-MBB function. This is a direct consequence of the rapid power-law decline of the dust opacity with wavelength as well as the fact that the coldest dust dominates the mass budget \citep[\eg][]{D01,HF13,LB14,U19}. At very long wavelength, the flux is only linearly dependent on $T$ in the RJ tail, and therefore the overall shape of the SED on the RJ side is largely set by the temperature of the mass-dominating cold dust. Hence, it has been proposed that the flux density originating from the optically-thin part of the RJ tail can be used as an efficient measure for estimating dust and gas mass (by assuming a dust-to-gas ratio) of massive high-redshift galaxies \citep[\eg][]{M12,S14,GS15,S16,H17,L18,P18,KS19}. Given the high uncertainties of the traditional CO methods and their long observing time, this approach represents an important alternative strategy for gas estimate \citep[\eg][]{SG16,S17,J18,HY18,W19,C19}. 

The RJ approach benefits from the effect of ``negative $K$-correction". Eq.~\ref{eq.5} can be re-written as \citep[\eg][]{S16}

\begin{align}
S_{\nu_{\rm o}} (T) &= \frac{(1+z)}{d^2_{\rm L}}2k_{\rm B}\kappa_{\nu}(\nu/c)^2\Gamma_{\rm RJ}(\nu_0,T,z) M_{\rm dust}T\notag\\
&= \psi(z) \Gamma_{\rm RJ} M_{\rm dust}T
\label{eq:Snu0OT}
\end{align}

\noindent where $\Gamma_{\rm RJ}$ is the RJ correction function that accounts for the departure of the Planck function from RJ approximate solution in the rest frame, and $\psi(z)$ has the unit of ${\rm mJy}\,M^{-1}_{\odot}\,{\rm K}^{-1}$. For given $\nu_{\rm o}$, $\kappa_\nu(\nu/c)^2$ scales as $(1+z)^4$ ($\beta=2.0$). On the other hand, $(1+z)\,d^{-2}_{\rm L}$ and $\Gamma_{\rm RJ}$ decline with redshift. The former term roughly scales as $(1+z)^{-2}$, while how $\Gamma_{\rm RJ}$ evolves with redshift depends on both $\nu_{\rm o}$ and $T$. The rise of $\kappa_\nu(\nu/c)^2$ with redshift can roughly cancel out or even reverse the decline of the other two components at $z\simgreat1$, with typical $T$ of galaxies and (sub)mm bands. For example, with $T=25$ K and ALMA band 6, $\psi\Gamma_{\rm RJ}$ stays about a constant from $z=2-6$, while with ALMA band 7, $\psi\Gamma_{\rm RJ}$ declines \textit{only} by less than a factor of two over the same redshift range (see Figure 2 of \citealt{S16}). (Sub)mm observations are therefore powerful for unveiling high-redshift dusty star-forming galaxies. In the RJ regime ($h\nu\ll k_{\rm B}T$), $\Gamma_{\rm RJ}\approx{}1$ and $S$ scales linearly to $M_{\rm dust}T$ at a given redshift.

The RJ approach relies on an assumed dust temperature. The proper temperature, $T$, needed for inferring dust (and gas) masses can be obtained from solving Eq.~\ref{eq:Snu0OT}, given $S_{\nu_{\rm o}}$, $M_{\rm dust}$ and $z$. This required $T$ value is close to the mass-weighted dust temperature, for galaxies from $z=2$ to $z=6$, and with varying $\delta_{\rm dzr}$, see Figure~\ref{fig:RJ}. The difference between these two temperatures is typically {less than} 0.03 dex. This again confirms that a single-$T$ OT-MBB function well describes the emission from the optically-thin RJ tail. 

However, using $T_{\rm peak}$ will lead to a poor constraint on $M_{\rm dust}$ and therefore gas mass of galaxy. First of all, it is systematically higher than $T_{\rm mw}$, and therefore can cause systematically underestimate of $M_{\rm dust}$. Secondly, there seems to be no strong correlation between $T_{\rm mw}$ and $T_{\rm peak}$ by comparing the \textit{left} and \textit{right panels}. So even by using $T_{\rm peak}$ to infer $T_{\rm mw}$ will produce systematic error. We will discuss the discrepancy between $T_{\rm peak}$ and $T_{\rm mw}$ in more details in the later sections. \textit{Using other effective temperatures that have strong correlation with $T_{\rm peak}$ will be problematic as well.}

\subsubsection{The $L_{\rm IR}$ vs. $MT^{4+\beta}$ relation}
\label{S3d2}

\begin{figure*}
   \vspace{-5pt}
 \begin{center}
 \includegraphics[width=165mm]{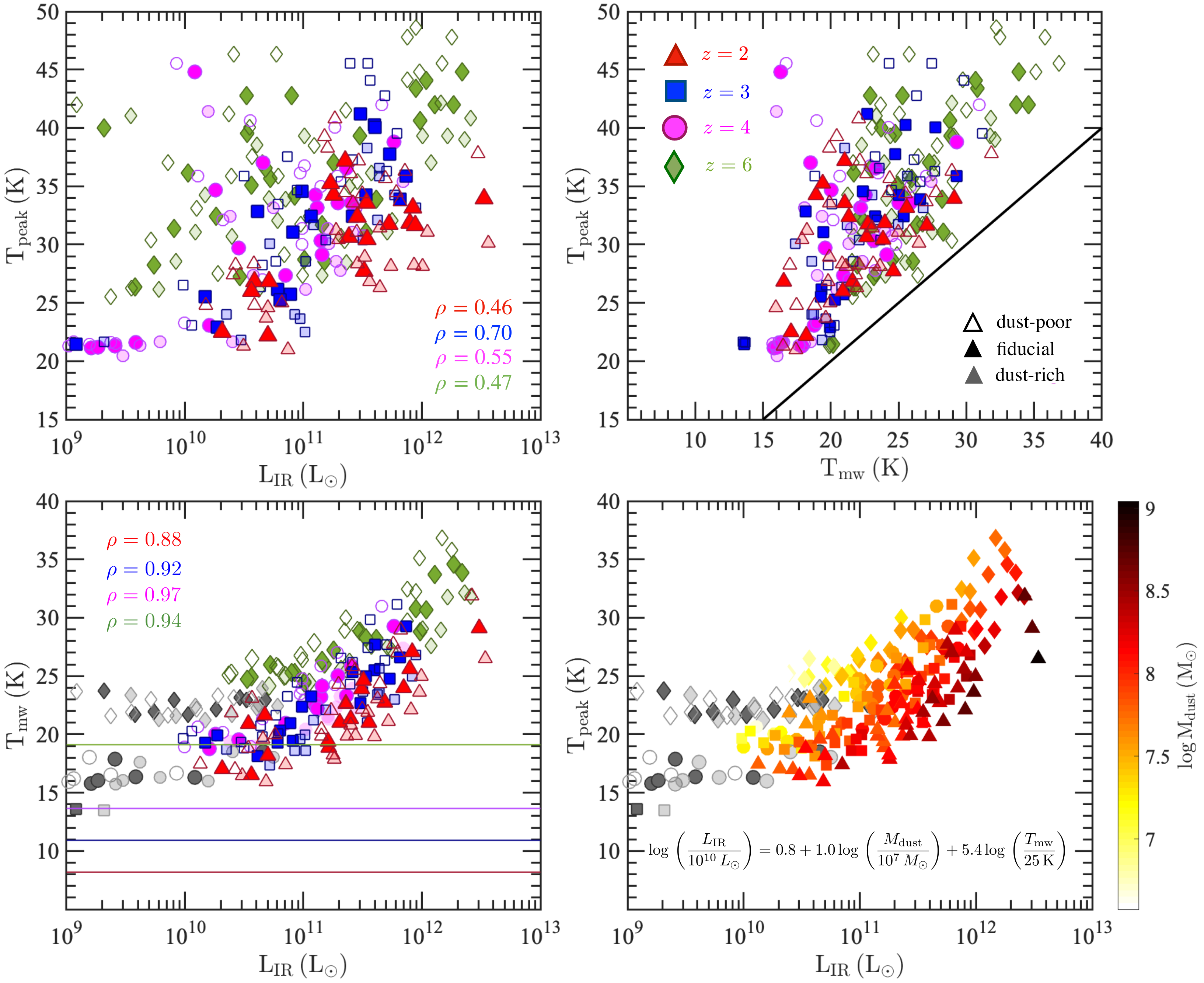}
 \caption{\textit{Upper left}: $T_{\rm peak}$ vs. $L_{\rm IR}$ relation of the \textsc{\small MassiveFIRE} galaxies at $z=2-6$. \textit{Upper right}: $T_{\rm peak}$ vs. $T_{\rm mw}$ relation. \textit{Lower panels}: $T_{\rm mw}$ vs. $L_{\rm IR}$ relation. In the \textit{left panel}, galaxies are coloured by their redshift, while in the \textit{right panel}, they are coloured by $M_{\rm dust}$. The galaxies that are strongly affected by the heating of the CMB background (\ie~$T_{\rm mw}\simless T_{\rm CMB}(z)+5$ K) are coloured by grey. The horizontal solid lines in the \textit{lower left panel} represent the CMB temperature at each redshift.  In the upper panels and the \textit{lower left panel}, the filled, unfilled and the semi-transparent symbols represent the fiducial, dust-poor and dust-rich models, respectively. The data from all three dust models are included in the lower right panel. }
    \label{fig:TLZ}
  \end{center}
\end{figure*} 

The scaling relation $L_{\rm IR}\propto M_{\rm dust}T^{(4+\beta)}$, which is frequently been adopted by many studies to probe and obtain useful physical insights for the star-forming conditions of the IR-luminous sources owing to its simplicity, is derived under the assumption of the optically-thin approximation (Eq.~\ref{eq:OTMBB}). 

The temperature in the above scaling relation is a measure of the luminosity per unit dust mass and often viewed as a proxy for the internal radiative intensity. Yet, it is \textit{not} obvious how this temperature parameter (\ie~$\sim(L_{\rm IR}/M_{\rm dust})^{1/6}$) is related to the physical, $T_{\rm mw}$ or the observationally accessible $T_{\rm peak}$.

We show in Figure~\ref{fig:Peak} the scaling relation of the light-to-mass ratio, $L_{\rm IR}/M_{\rm dust}$ against $T_{\rm mw}$ (left panel)  as well as $T_{\rm peak}$ (right panel) for the \textsc{\small MassiveFIRE} sample at $z=2-6$, and we explicitly present the result for the fiducial (filled symbols), dust-poor (unfilled symbols) and dust-rich (semi-transparent symbols) cases. 

In general, galaxy having higher dust temperature (both $T_{\rm mw}$ and $T_{\rm peak}$) emits more IR luminosity per unit dust mass. Focusing at first on $T_{\rm mw}$ (left panel), we see that $L_{\rm IR}/M_{\rm dust}$ of the \textsc{\small MassiveFIRE} galaxies appears to be systematically higher than from a simple single-$T$ OT-MBB function (Eq.~\ref{eq.6}), which is indicated by solid black line in both panels. The offset ($\sim0.3$ dex) between the simulated result and the analytic solution is due to the higher emissivity of the dense, warm dust in vicinity of the star-forming regions (see lower panels of Figure~\ref{fig:Image}), which accounts for a small fraction of the total dust mass but has strong emission, and shapes the Wien side of the overall SED of galaxy.

With all the galaxies from $z=2$ to $z=6$, we find that $L_{\rm IR}/M_{\rm dust}$ scales to $\approx{}T^{5.4}_{\rm mw}$. This is slightly flatter than the analytic solution derived using a single-temperature, optically-thin MBB function, \ie~$L_{\rm IR,\;OT}/M_{\rm dust}\propto T^6$ (Eq.~\ref{eq:OTMBB}, with $\beta=2.0$). We understand the shallower slope as an optical depth effect. In the optically-thin regime ($\tau\ll1$), $L/M\propto (1-e^{-\tau})/\tau\approx1$, while in the optically-thick regime ($\tau\gg1$), $L/M\propto\tau^{-1}$ (Eq.~\ref{eq.3x}). In the optically-thick regime, $L_{\rm IR}/M_{\rm dust}$ therefore decreases with increasing $\tau$. Galaxies of higher $T_{\rm mw}$ are more dust-rich (Section~\ref{S3d3}) and their star-forming regions tend to be more optically-thick, resulting in a flattening of the scaling relation.

Comparing the dust-poor (dust-rich) models with the fiducial case, the median of $T_{\rm mw}$ is higher (lower) by 0.84 (1.70) K. This is due to the optical depth effect. By reducing the amount of dust, the chance of receiving a short-wavelength photon increases because the optical depth from the emitting sources decreases. Therefore, dust is expected to be heated to higher temperature to balance the increased amount of absorption. Apart from that, $\delta_{\rm dzr}$ also mildly effects the normalisation of the $L_{\rm IR}/M_{\rm dust}$ vs. $T_{\rm mw}$ relation. The dust-poor (dust-rich) case shows about 0.13 (0.06) dex higher (lower) $L_{\rm IR}/M_{\rm dust}$, on the average, than the fiducial case, indicating a high (lower) luminosity emitted per unit dust mass. This is because a larger (reduced) mass \textit{fraction} of the total dust is heated by (can actually ``see") the hard UV photons {emitted} from the young stars due to the reduced optical depth \citep{S13,S16}. This dust component can be efficiently heated to a temperature much higher than the mass-weighted average of the bulk \citep{HF13, LB14, BM18}, and has a much higher $L/M$ ratio than the rest.

$T_{\rm peak}$ (right panel) also shows a positive correlation with $L_{\rm IR}/M_{\rm dust}$, although the strength of correlation is relatively weaker than that of $T_{\rm mw}$ ($\rho=0.81$ vs. 0.91, where $\rho$ is the Spearman rank correlation coefficient). Beside, $T_{\rm peak}$ also shows larger scatter than $T_{\rm mw}$. The $1\sigma$ dispersion of $L_{\rm IR}/M_{\rm dust}$ at fixed $T_{\rm peak}$ is 0.21 dex, which is higher than 0.14 dex at fixed $T_{\rm mw}$. This means that $T_{\rm peak}$ has relatively lower power to predict the luminosity-to-dust-mass ratio. Furthermore, $T_{\rm peak}$ is also more affected by a change of $\delta_{\rm dzr}$. The median $T_{\rm peak}$ of the dust-poor (dust-rich) case is 2.49 (2.63) K higher (lower) than the fiducial model, which is more than the change of $T_{\rm mw}$ with $\delta_{\rm dzr}$. This is because $T_{\rm peak}$ is more sensitive to the mass \textit{fraction} of ISM dust that is efficiently heated to high temperature by the hard UV photons emitted from young stars \citep[see also][]{FC17}.

\subsubsection{$L_{\rm IR}$ vs. $T$ relation}
\label{S3d3}

The dust temperature vs. total IR luminosity is one most extensively studied scaling relations. We have shown in Section~\ref{S3b} that our simulations have successfully produced the result at $z=2$ for galaxies that are in good agreement with the recent observational data at similar luminosity range. Here in this section, we focus on the evolution of dust temperature up to higher redshift. One major problem with the current observational studies on the $T-L$ scaling is the selection effects of the flux-limited FIR samples that have been used to probe such relation. Higher redshift sample is biased towards more luminous systems \citep{MD14}. How dust temperature evolves at fixed luminosity is still being routinely debated \citep[see \eg][]{M12,SV13,M14, B15, I16,C18b, S18}. We present the result using our sample with $L_{\rm IR}\approx 10^9-2\times10^{12}\,L_{\odot}$ from $z=2-6$. For $z=3-6$, there is no current data available that we can make direct comparison to at similar $L_{\rm IR}$ of our sample. Future generation of space infrared telescope, such as \textsc{\small SPICA}, can probe similar regime of IR luminosity at these epochs.

We present the temperature vs. luminosity relation of the \textsc{\small MassiveFIRE} galaxies at $z=2-6$ in Figure~\ref{fig:TLZ}. In the upper and lower left panels, we show $T_{\rm peak}$ vs. $L_{\rm IR}$ and $T_{\rm mw}$ vs. $L_{\rm IR}$ relation, respectively. 

Focusing at first on $T_{\rm peak}$ vs. $L_{\rm IR}$ relation (\textit{upper left}), we find a noticeable increase of $T_{\rm peak}$ with redshift at fixed $L_{\rm IR}$, albeit with large scatter at each redshift. Looking at the most luminous galaxy at each redshift, we see that $T_{\rm peak}$ increases from about 34 K at $z=2$ to $\sim43$ K at $z=6$ for the fiducial dust model ($\delta_{\rm dzr}=0.4$). With all the luminous galaxies with $L_{\rm IR}>10^{11}\,L_{\odot}$, we fit the evolution of $T_{\rm peak}$ with redshift as a power law and obtained

\begin{equation}
    {\rm log}\,\left(\frac{T_{\rm peak}(z)}{25\,{\rm K}}\right) = (-0.02\pm0.06)+(0.25\pm0.09)\,{\rm log}\,(1+z)
    \label{eq:Tevo}
\end{equation}

\noindent This result is in good quantitative agreement with the recent observational finding by \citet{I16} and \citet{S18}, although they use more IR-luminous sample at similar redshift range.

For each redshift, there is also a mild trend of declining $T_{\rm peak}$ with decreasing $L_{\rm IR}$ over the three orders of magnitude of $L_{\rm IR}$ being considered. For instance, $T_{\rm peak}$ of the $z=6$ galaxies at $L_{\rm IR}=10^{10}\;L_{\odot}$ is about 32 K, which is about 10 K lower than the value at $L_{\rm IR}=10^{12}\;L_{\odot}$, and is similar to the value of the brightest objects at $z=3$ and $z=4$. We find some faint objects at $\sim10^{10}\;L_{\odot}$ whose $T_{\rm peak}$ is as low as $\sim20$ K. We also note that the scatter of $T_{\rm peak}$ could be very large at the faint end even with the simple fiducial dust model. At $z=4$, some objects could be as hot as $\sim40$ K, while some could be as cold as $\sim20$ K. This large scatter is mainly driven by the difference of sSFR among those galaxies, which we will discuss in more details in the following section.

With such large scatter, the correlation between $T_{\rm peak}$ and $L_{\rm IR}$ appears to be fairly weak. The Spearman correlation coefficient ($\rho$) of the $T_{\rm peak}$ vs. $L_{\rm IR}$ relation at individual redshift ranges from 0.46 to 0.70 at the redshifts being considered. For the $z=2$ sample, there is no noticeable correlation at $L_{\rm IR}>10^{11}\,L_{\odot}$.

On the other hand, $T_{\rm mw}$ exhibits a tighter correlation with $L_{\rm IR}$ (lower left panel) ({$\rho$ ranging from 0.88 to 0.97}), with an increase of the normalisation of the $L_{\rm IR}$-$T_{\rm mw}$ relation with redshift. The increase of $T_{\rm mw}$ with redshift at fixed $L_{\rm IR}$ is clearly less prominent than $T_{\rm peak}$. At $L_{\rm IR}\approx10^{12}\,L_{\odot}$, for example, $T_{\rm mw}$ increases from $\sim27$ K at $z=2$ to only $\sim32$ K at $z=6$. The CMB heating sets a temperature floor for $T_{\rm mw}$ at the low luminosity end. 

The evolution of the $T_{\rm mw}$ vs. $L_{\rm IR}$ scaling is driven by $M_{\rm dust}$. At fixed $L_{\rm IR}$, galaxies at higher redshift have lower $M_{\rm dust}$. This can be clearly seen from the \textit{lower right panel}, where we colour the same data as in the \textit{lower left panel} by $M_{\rm dust}$ of galaxy. There is clear sign of anti-correlation between $T_{\rm mw}$ and $M_{\rm dust}$ at fixed $L_{\rm IR}$ \citep[see also][]{HJ12,B15,SH16,FC17,K17}. Applying multi-variable linear regression analysis to the $z=2-6$ galaxies, excluding those that are strongly affected by the heating of the CMB background (\ie~$T_{\rm mw}\simless T_{\rm CMB}(z)+5$ K), we obtain the scaling relation

\begin{align}
    {\rm log}\;\left(\frac{L_{\rm IR}}{10^{10}\,L_{\odot}}\right)&=\,(0.81\pm0.07) + (1.01\pm0.06)\,{\rm log}\,\left(\frac{M_{\rm dust}}{10^7\,M_{\odot}}\right)\notag \\ 
    &+(5.40\pm0.36)\,{\rm log}\,\left(\frac{T_{\rm mw}}{
    25\,\rm K}\right),\notag \\ 
    {\rm or} \,\,L_{\rm IR}&\propto M_{\rm dust}\,T^{5.4}_{\rm mw}\,\,(\beta=2.0).
    \label{eq.12}
\end{align}

\noindent It appears to be shallower than the classical $L_{\rm IR}\propto M_{\rm dust}T^{(4+\beta)}$ ($\beta=2.0$ for our adopted dust model, \cf~Fig.~\ref{fig:Dust}) relation derived based on the optically-thin approximation. We will discuss in Section~\ref{S5} about using this scaling relation to estimate $M_{\rm dust}$ and $T_{\rm mw}$ when only single data point is available at FIR-to-mm wavelengths.

\begin{figure*}
 \begin{center}
 \includegraphics[width=175mm]{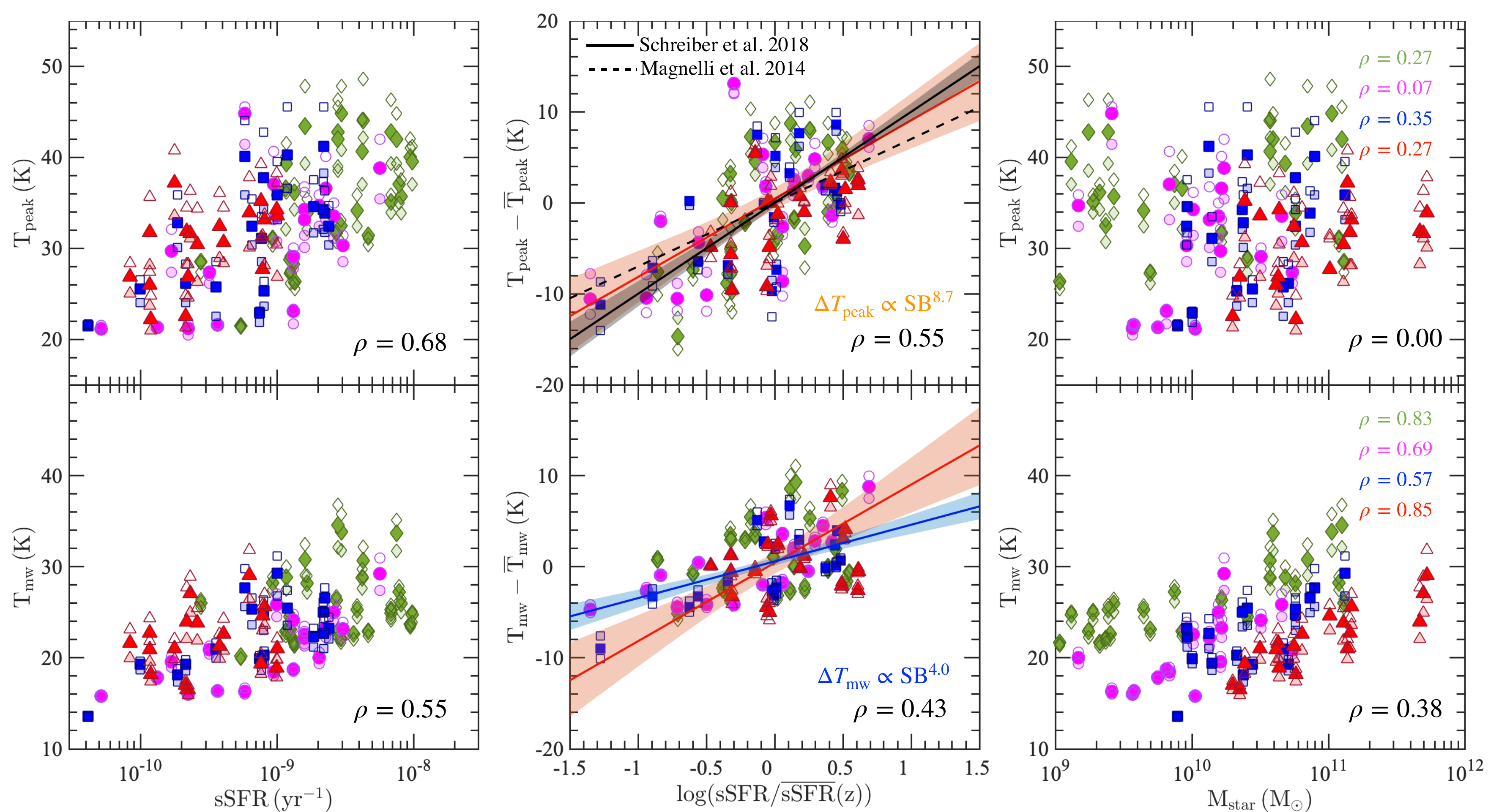}
 \caption{Relation of dust temperature against sSFR (left column), $\rm SB{}={}sSFR/\overline{sSFR}(z)$ (middle column) and $M_{\rm star}$ (right column) of the \textsc{\small MassiveFIRE} galaxies at $z=2$ (red triangles), $z=3$ (blue squares), $z=4$ (magenta circles) and $z=6$ (green diamonds). The result of $\delta_{\rm dzr}=0.4$, $\delta_{\rm dzr}=0.8$, and $\delta_{\rm dzr}=0.2$ are shown with filled, semi-transparent and unfilled symbols, respectively. We show the result with $T_{\rm peak}$ and $T_{\rm mw}$ in the upper and lower panels, respectively. The solid and dashed lines in the upper middle panel represent the observed scaling relation by \citet{S18} and \citet{M14}, respectively. The orange and blue lines in the middle panels show the best-fitting line for the {\sc \small MassiveFIRE} sample. The former and latter correspond to $T_{\rm peak}$ and $T_{\rm mw}$, respectively. The shaded areas represent the $95\%$ (\ie~$2\sigma$) confidence interval of each scaling relation. \textbf{$T_{\rm peak}$ exhibits relatively stronger correlation with sSFR than $T_{\rm mw}$, but weaker correlation with $M_{\rm star}$. } }
    \label{fig:Tssfr}
  \end{center}
\vspace{-10 pt}
\end{figure*}

\subsubsection{sSFR vs. T relation}
\label{S3d4}

The sSFR vs. dust temperature relation is one other frequently studied scaling relation which provide useful physical insights to dust temperature and is complementary to the $L_{\rm IR}$ vs. temperature relation. 

In Figure~\ref{fig:Tssfr}, we show the relation of dust temperature against ${\rm sSFR}={\rm SFR_{\rm 20\,Myrs}}/M_{\rm star}$ for the \textsc{\small MassiveFIRE} sample at $z=2-6$ in the \textit{left panels}. We present the result for $T_{\rm peak}$ and $T_{\rm mw}$ in the upper and lower \textit{left panels}, respectively. 

The dust temperatures are positively correlated with sSFR ($\rho=0.68$ for the sSFR vs. $T_{\rm peak}$ relation and $\rho=0.55$ for the sSFR vs. $T_{\rm mw}$ relation). Galaxies at higher redshift have, on average, higher sSFR, which is a direct consequence of the evolution of the star-formation main sequence. SFR is a proxy for the internal radiative intensity (most UV emission originates from the young stellar populations in the galaxies), and $M_{\rm dust}$ is about linearly scaled to $M_{\rm star}$ in the \textsc{\small MassiveFIRE} galaxies, the sSFR ($\sim$SFR/$M_{\rm dust}$) can be viewed as a proxy for the total energy input rate per unit dust mass. It is therefore expected that to first order, dust temperature is positively correlated with sSFR of galaxies. This is indeed what we can see from both of the \textit{left panels} of Figure~\ref{fig:Tssfr}. For instance, the $z=2$ galaxies (red) have a median sSFR of $3\times10^{-9}\;\rm yr^{-1}$ and median $T_{\rm mw}=20$ K ($T_{\rm peak}=30$ K). Both sSFR and dust temperature (both $T_{\rm peak}$ and $T_{\rm mw}$), on average, increases with redshift. The $z=6$ sample (green) have a median sSFR of $2\times10^{-8}\;\rm yr^{-1}$ and median $T_{\rm mw}=26$ K ($T_{\rm peak}=37$ K). 

The correlation persists when focusing on each individual redshift.  In the middle panels, we show the result when both temperature and sSFR are normalised by the median value of the whole sample ($\overline{T}_{\rm mw}$ or $\overline{T}_{\rm peak}$, $\overline{\rm sSFR}$) at each different redshift. With $T_{\rm peak}$ (upper middle panel), the simulated galaxies, including all objects at $z=2-6$, exhibit a positive correlation ($\rho=0.55$) between starburstiness\footnote{The median sSFR at $z=2$, $z=3$, $z=4$ and $z=6$ of the \textsc{\small MassiveFIRE} sample are $2.1\times10^{-10}$, $5.8\times10^{-10}$, $8.7\times10^{-10}$ and $3.3\times10^{-9}\,{\rm yr^{-1}}$, respectively. SFRs are averaged over the past 20 Myrs.} (\ie~$\rm SB{}={}sSFR/\overline{sSFR}(z)$) and normalised $T_{\rm peak}$. The derived scaling relation (solid orange line) is in good qualitative agreement with the recent observations by \citet{M14} (dotted black line) and \citet{S18} (solid black line), despite that both studies include samples at lower redshifts ($z<2$) which our simulations do not probe. We also find that compared to $L_{\rm IR}$, $T_{\rm peak}$ is more strongly correlated with sSFR at each given redshift, which is in agreement with the previous finding by \citet{M14} (see also \citealt{L14}).

However, due to the inhomogeneity of dust distribution in galaxies and the complexity in star-dust geometry, the radiative energy emitted from the young stellar populations is not expected to evenly heat the ISM dust in the galaxy. Most of the UV photons are absorbed by the dense dust cloud in vicinity of the young star-forming regions, while the majority of the dust in the ISM is heated by the old stellar populations with more extended distribution, as well as the secondary photons re-emitted from the dust cloud near the young star clusters. For such reason, $T_{\rm peak}$ is expected to be more sensitive to the emission from the warm dust component, which is more closely tied to the young star clusters, while $T_{\rm mw}$ is determined by the cold dust component and therefore can be relatively less sensitive to the sSFR of galaxy than $T_{\rm peak}$.

This indeed can be seen from comparing the \textit{upper} and \textit{lower middle panels} of Figure~\ref{fig:Tssfr}. First of all, $\Delta T_{\rm peak}$ ($T_{\rm peak}-\overline{T}_{\rm peak}$) shows a relatively stronger correlation with SB than $\Delta T_{\rm mw}$ ($T_{\rm mw}-\overline{T}_{\rm mw}$). With all the $z=2-6$ \textsc{\small MassiveFIRE} galaxies, the Spearman correlation coefficient of the $\Delta T_{\rm peak}$ vs. SB scaling is $\rho=0.55$, while that of the $\Delta T_{\rm mw}$ vs. SB scaling is $\rho=0.43$. Secondly, over about two orders of magnitude of SB ($\sim0.1-10$), the scaling relation with $\Delta T_{\rm peak}$ is relatively steeper, 

\begin{equation}
\Delta T_{\rm peak} \propto {\rm SB}^{\,8.61\pm 1.38}\;{\rm vs.}\;\Delta T_{\rm mw} \propto {\rm SB}^{\,4.03\pm 0.48}.
\label{eq.13}
\end{equation}

\begin{figure*}
 \begin{center}
 \includegraphics[width=160mm]{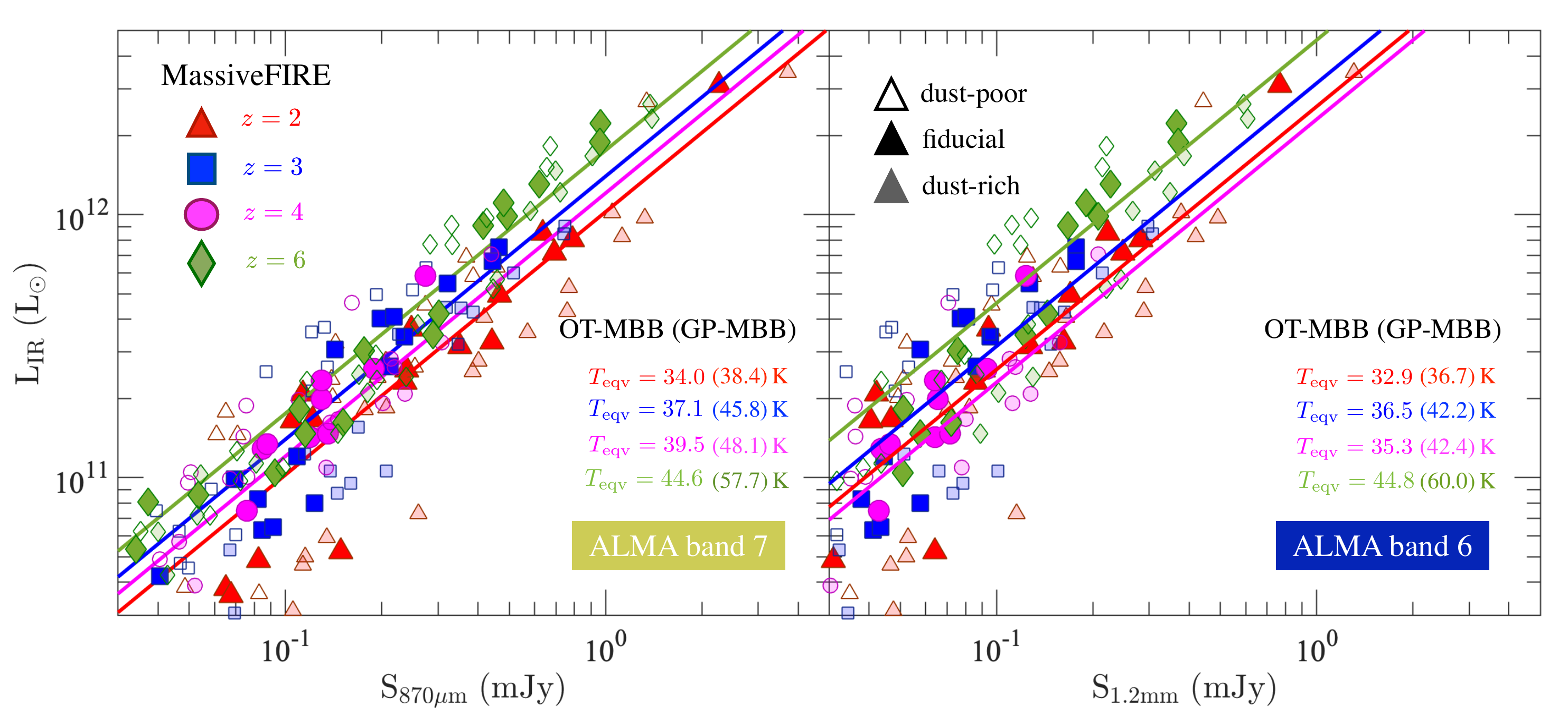}
 \caption{Relation of $L_{\rm IR}$ vs. $S_{\rm 870\mu m}$ (\textit{left panel}) and $S_{\rm 1.2 mm}$ (\textit{right panel}) of our {\sc\small MassiveFIRE} galaxy sample at $z=2-6$. The unfilled, filled and semi-transparent symbols represent the result for a range of dust-to-metal ratios $\delta_{\rm dzr}=0.2$, $\delta_{\rm dzr}=0.4$ and $\delta_{\rm dzr}=0.8$, respectively. The coloured lines show the $L_{\rm IR}$ vs. $S_{\rm 870\mu m}$ (and $S_{\rm 1.2 mm}$) relation expected from an OT-MBB function (Eq.~\ref{eq.5}, with fixed $\beta=2.0$) with an equivalent temperature ($T_{\rm eqv,\,OT-MBB}$) that yields the $L_{\rm IR}$ of the {\sc\small MassiveFIRE} sample at each redshift. The sample-average value of $T_{\rm eqv}$ for each redshift, ALMA band, and SED fitting function is labeled in the figure. {\bf Overall, $T_{\rm eqv}$ increases with redshift for the galaxies in our sample}.}
    \label{fig:LS}
  \end{center}
  \vspace{-20pt}
\end{figure*}

\noindent This is because the UV photons from the young star clusters preferentially heat the dense dust cloud in the neighbourhood to high temperature, which boosts the MIR emission and helps shift the SED peak to shorter wavelength. However, the heating of the bulk of the dust is inefficient. The reason is that once the UV photons get absorbed and re-emit as FIR photons, the chance of them being absorbed by dust again becomes much lower as a consequence of the declining opacity with wavelength ($\kappa_{\lambda}\propto\lambda^{-2}$) \citep{S13}. It is also interesting to note that both $T_{\rm peak}$ and $T_{\rm mw}$ are less correlated with SB when $\rm sSFR$ is averaged over longer period of time \citep{SHF16,F17b,F18}. By averaging sSFR over a period of 100 Myrs instead of 20 Myrs, for example, {$\rho$} of the $T_{\rm peak}$ ($T_{\rm mw}$) vs. SB relation declines from {0.55 (0.43) to 0.13 (0.22)}. 

We note that comparing to the recent observations of the $z=2-4$ star-forming galaxies \citep{SP15}, the median sSFR of the \textsc{\small MassiveFIRE} galaxies is about $0.2-0.3$ dex lower, but is still within the lower $1\sigma$ limit of the observational data \citep[see][]{F16,F17}. This discrepancy of sSFR is commonly seen in the current cosmological galaxy simulations. A systematic increase of SFR will lead to more heating to the ISM dust and hence higher simulated dust temperatures, but will not affect the slope of the sSFR vs. $T$ relation. The increment of $T_{\rm peak}$ is estimated to be about $1.7-2.6$ K according to the sSFR vs. $T$ relation (Figure~\ref{fig:Tssfr}), which appears to be insignificant compared to the scatter of the observational data (Figure~\ref{fig:TL2}). It should also be noted that the impact of an increased sSFR on $T_{\rm peak}$ can easily be offset by an increase of dust mass, which can potentially be driven by an increased $\delta_{\rm dzr}$, dust opacity, or gas metallicities - all these properties are currently uncertain at high redshifts.

Finally, we show the relation between dust temperatures and $M_{\rm star}$ in the \textit{right panels}. Looking at the upper panel, it is clear that $T_{\rm peak}$ has very weak correlation with $M_{\rm star}$. This again shows that $T_{\rm peak}$ is strongly influenced by the emission from the warm dust that is associated with the recently formed young stars and does not have as strong correlation with the total stellar mass of a galaxy. In contrast, $T_{\rm mw}$ is less sensitive to the variance of recent star-forming conditions and therefore shows relatively small scatter at given $M_{\rm star}$ at each redshift. The normalisation of the $T_{\rm mw}$ vs. $M_{\rm star}$ relation increases with redshift, which is driven by the rise of ${\rm SFR}/M_{\rm dust}$ (\ie~energy injection rate per unit dust mass). We also notice a slight increase of $T_{\rm mw}$ with $M_{\rm star}$. This is owing to the decrease of $M_{\rm star}/M_{\rm dust}$ with $M_{\rm star}$ of the $\textsc{\small MassiveFIRE}$ sample. As a result, ${\rm SFR}/M_{\rm dust}$ slightly increases with $M_{\rm star}$ (\ie~${\rm SFR}/M_{\rm dust}\propto {\rm sSFR}(M_{\rm star}/M_{\rm dust}) \propto M^{0.3}_{\rm star}$) at given redshift.

\vspace{-10pt}
\section{(Sub)Millimetre broadband fluxes}
\label{S4}

A major problem for probing the dust properties in the high-redshift ($z>4$) is that most observations of dust emission at such high redshift are limited to a single broadband flux detected by ALMA band 7 or 6. Deriving infrared luminosities and hence SFRs of these $z\simgreat4$ objects is very challenging without FIR constraints and depends highly on the assumed \textit{equivalent} dust temperature for the flux-to-luminosity conversion. The same problem also applies to many faint (\ie~below a few mJy) submm-selected objects at lower redshift ($2<z<4$) that do not have \textit{Herschel} FIR coverage. Therefore, an accurate estimate of $T_{\rm eqv}$ of the adopted SED function for different redshifts is critical.

In this section, we will analyse the $T_{\rm eqv}$ distribution of galaxies at $z=2-6$ with the help of the \textsc{\small MassiveFIRE} sample. Specifically, in Section~\ref{S4a}, we will examine the redshift evolution of $T_{\rm eqv}$ and its dependence on $\delta_{\rm dzr}$, offering a `cookbook' for converting between (sub)mm and $L_{\rm IR}$ observations. In Section~\ref{S4b}, we will compare $T_{\rm eqv}$ with $T_{\rm mw}$ and $T_{\rm peak}$, and provide a physical interpretation of this dust temperature.
 
 \begin{figure*}
 \begin{center}
 \includegraphics[width=160mm]{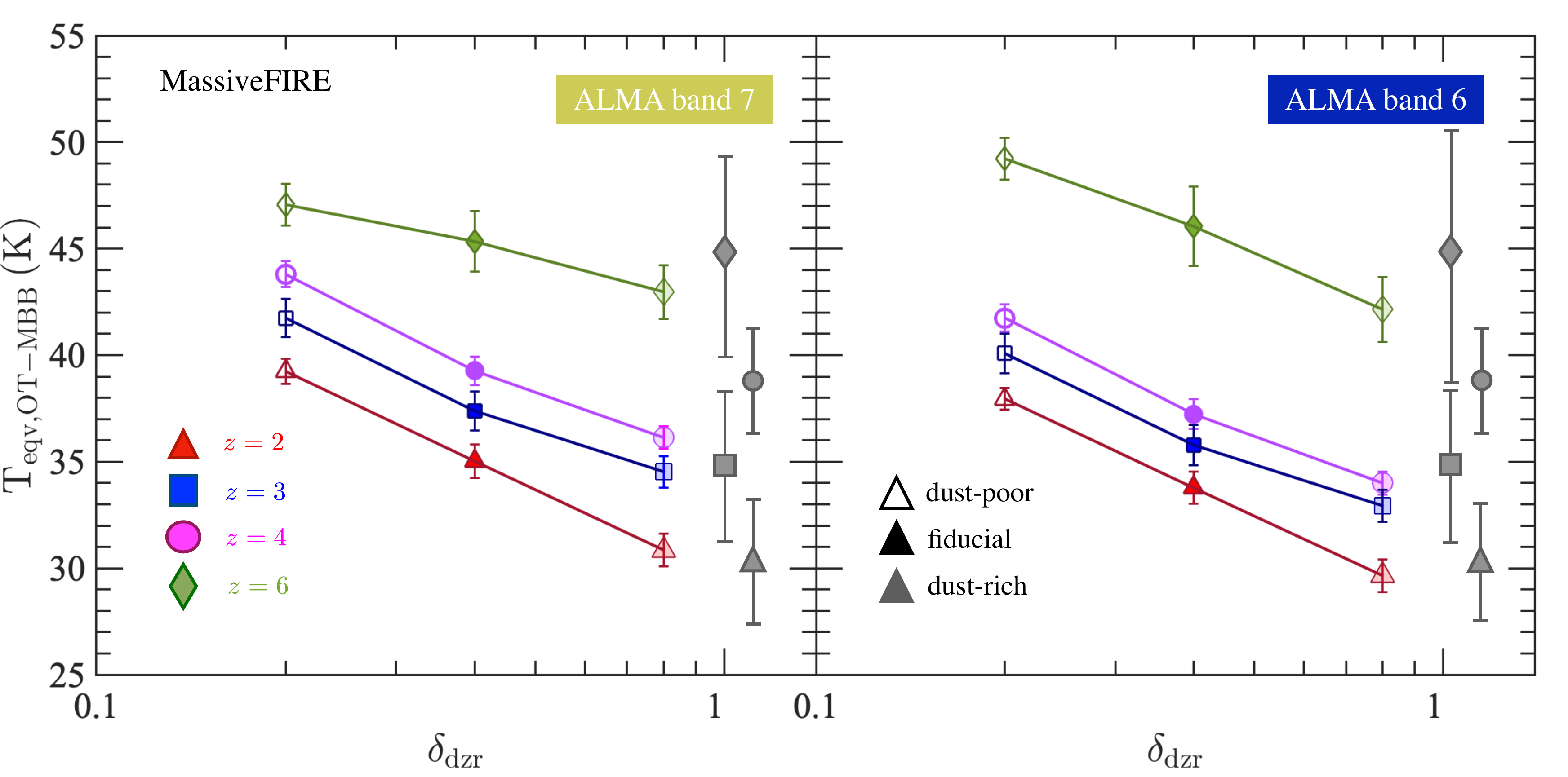}
 \caption{Relation of equivalent dust temperature ($T_{\rm eqv}$) vs. dust-to-metal ratio ($\delta_{\rm dzr}$) of the $z=2-6$ {\sc\small MassiveFIRE} sample. $T_{\rm eqv}$ is the effective dust temperature in the OT-MBB function (Eq.~\ref{eq.5}, with $\beta=2.0$) that yields the true $L_{\rm IR}$ of the galaxy from the flux densities at ALMA band 7 at 870 $\rm \mu m$ (\textit{left panel}) and band 6 at 1.2 mm (\textit{right panel}). The result of $\delta_{\rm dzr}=0.4$, $\delta_{\rm dzr}=0.8$, and $\delta_{\rm dzr}=0.2$ are shown with filled, semi-transparent and unfilled symbols, respectively. The grey error bars in both panels represent the $1\sigma$ dispersion of $T_{\rm eqv}$ for each redshift. \textbf{$T_{\rm eqv}$ increases with redshift, and at the same redshift, $T_{\rm eqv}$ shows negative correlation with $\delta_{\rm dzr}$.}}
    \label{fig.DZR}
  \end{center}
  \vspace{- 20 pt}
\end{figure*} 

\subsection{The flux-to-luminosity conversion}
\label{S4a}

$L_{\rm IR}$ is often extrapolated from a single broadband (sub)mm flux given the lack of additional submm or FIR constraints. A typical approach is to assume that the SED has an OT-MBB (or G-MBB) shape with a chosen value of the dust temperature parameter. However, as we have shown in Figure~\ref{fig:SED} and discussed in Section~\ref{S3a}, choosing a dust temperature parameter that is not compatible with the assumed functional shape of SED can result in significant biases for estimating $L_{\rm IR}$ of a galaxy. By definition, this problem is avoided if the adopted dust temperature is chosen to be $T_{\rm eqv}$.

With the {\sc\small MassiveFIRE} sample, we are able to predict the full dust SED for the high-redshift ($z=2-6$) objects covering over two orders of magnitude of IR luminosity ($L_{\rm IR}\approx10^{10}-10^{12}\;L_{\odot}$). We predict the observed flux densities at ALMA band 7 ($S_{870\rm \mu m}$) and band 6 ($S_{1.2\rm mm}$) given the SED and redshift as well as $L_{\rm IR}$. Many of these objects have $S_{\rm 870\mu m}\,(S_{\rm 1.2 mm})\simgreat0.1$ mJy, which are over the $3\sigma$ detection limit of ALMA band 6 and 7 using a typical integration time of 1 hour. With the calculated $S_{870\rm \mu m}$ (and $S_{1.2\rm mm}$) of each galaxy, we find the OT-MBB (with $\beta=2.0$) and GP-MBB functions (with $\beta=2.0$, $\lambda_1=100\rm \;\mu m$, $\alpha=2.5$ and the suggested value of $N_{\rm pl}$ by C12), normalised to match their observed flux densities at both ALMA bands, that can predict their true $L_{\rm IR}$. By adjusting the temperature parameter in the fitting function to match both observed submm flux density and true $L_{\rm IR}$, we obtain $T_{\rm eqv}$, \ie, \emph{the value of $T$ necessary for obtaining an accurate estimate of $L_{\rm IR}$ from the measured (sub)mm flux densities for each galaxy}.

In Figure~\ref{fig:LS}, we show the relation of $L_{\rm IR}$ against $S_{870\rm \mu m}$ (\textit{left panel}) and $S_{1.2\rm mm}$ (\textit{right panel}) for the $z=2-6$ {\sc\small MassiveFIRE} galaxies. For each redshift, we also show the expected $L_{\rm IR}$ vs. $S_{\rm 870\rm \mu m}$ (and $S_{\rm 1.2\rm mm}$) relation using the mean $T_{\rm eqv}$ for galaxies above 0.1 mJy. The latter temperature is provided for the two different ALMA bands and for redshifts $z=2-6$. We present the results for OT-MBB and GP-MBB functional shapes.

There appears to be a clear trend of increasing $T_{\rm eqv}$ with redshift, with either forms of fitting function (GP or OT-MBB) and with either ALMA band 6 or 7. This shows that a higher $T_{\rm eqv}$ is typically needed for deriving $L_{\rm IR}$ of galaxies at higher redshift. Using OT-MBB function, for example, the mean $T_{\rm eqv}$ increases from 34.0 K at $z=2$ (red triangles) to 44.6 K at $z=6$ (green diamonds) for ALMA band 7. Applying the typical $T_{\rm eqv}$ for $z=2$ to a $z=6$ galaxy will therefore lead to a significant underestimate of $L_{\rm IR}$.

For the same redshift, the normalisation of the $L_{\rm IR}$ vs. $S_{\rm 870\,\mu m}$ ($S_{\rm 1.2mm}$) relation depends on dust mass. We explicitly show in Figure~\ref{fig:LS} the result for dust-rich and dust-poor models. At fixed observed broadband flux density, the $L_{\rm IR}$ of dust-rich galaxies lies systematically below the fiducial model (vice versa for dust-poor galaxies). This result indicates that a galaxy of given observed (sub)mm flux density tends to have lower (higher) $L_{\rm IR}$ if it contains more (less) amount of dust.

This finding can be understood as follows. By increasing the dust mass, both $L_{\rm IR}$ and $S_{\rm 870\rm \mu m}$ ($S_{\rm 1.2 mm}$) increase but the latter changes by a larger degree. Hence, the normalisation of the relation declines. The increase of $S_{\rm 870\rm \mu m}$ ($S_{\rm 1.2 mm}$) is mainly driven by dust mass, as $S_{\rm 870\rm \mu m}$ ($S_{\rm 1.2 mm}$) is linearly scaled to $M_{\rm dust}$ (Eq.~\ref{eq:Snu0OT}). On the other hand, the increase of $L_{\rm IR}$ is due to enhanced optical depth --- a larger fraction of UV photons gets absorbed by dust and re-emitted in the infrared/submm. A lower $T_{\rm eqv}$ is therefore needed to account for the decrease of the normalisation of the $L_{\rm IR}$ vs. $S_{\rm 870\rm \mu m}$ ($S_{\rm 1.2 mm}$) relation with increasing dust mass. This anti-correlation of $T_{\rm eqv}$ with $\delta_{\rm dzr}$ is more clearly shown in Figure~\ref{fig.DZR}.

We therefore provide a two-parameter fit for $T_{\rm eqv}$ with $\delta_{\rm dzr}$ and redshift as predictor variables. Using all the $z=2-6$ objects with $S_{\rm 870 \rm \mu m}>0.1\,{\rm mJy}$, including the data for $\delta_{\rm dzr}=0.2-0.8$, we perform a multiple linear regression analysis

\begin{equation}
{\rm log}\,(T_{\rm eqv}/25\,{\rm K})=a+b\,{\rm log\,(\delta_{\rm dzr}/0.4)}+c\,{\rm log}\,(1+z).
\label{Eq:Tfit}
\end{equation}

\noindent We present the best-fit regression parameters $a$, $b$ and $c$ for ALMA band 6 and 7, and for OT-MBB and GP-MBB functions in Table~\ref{T2}. These derived scaling relations are useful for converting a measured (sub)mm flux density into $L_{\rm IR}$, provided the redshift and dust-to-metal ratio of galaxy can be constrained.

The photometric redshift of the (sub)mm-detected galaxies can be determined when multi-band optical and NIR data are available, and the more accurate spectroscopic redshift can subsequently be determined if several atomic/molecular emission lines (\eg~CO, CII, NII, OIII) are identified \citep[\eg][]{L17,HL18,P19}. In contrast, $\delta_{\rm dzr}$ is more difficult to constrain from direct observation and is not yet well understood. Recent studies have reported differing results on how $\delta_{\rm dzr}$ depends on redshift and other galaxy properties \citep{I03,MT16,WS17,D19}. We will discuss in more detail about the recent observations of $\delta_{\rm dzr}$ and the implication of the reduced $\delta_{\rm dzr}$ at high redshifts in Section~\ref{S5c}.

\begin{table}
\caption{Scaling relations between $T_{\rm eqv}$,  $\delta_{\rm dzr}$ and redshift. $z=a+b\times x +c\times y$, where $z={\rm log}\,(T_{\rm eqv}/25\,\rm K)$, $x={\rm log}\,(\delta_{\rm dzr}/0.4$) and $y={\rm log}\,(1+z)$.}
\begin{tabular}{c c c c  c }
 \hline
\multicolumn{1}{}{} &  OT$^{i)}$ (band 7) & OT$^{i)}$ (band 6)  & GP$^{ii)}$ (band 7) & GP$^{ii)}$ (band 6)  \\
 \hline
 a &  $-0.01\pm0.03$ & $-0.05\pm0.04$ & $0.00\pm0.05$ & $-0.08\pm0.05$ \\
  \hline
 b &  $-0.13\pm0.03$ & $-0.15\pm0.03$ & $-0.22\pm0.04$ & $-0.24\pm0.05$ \\
  \hline
 c &  $0.31\pm0.05$ & $0.36\pm0.06$ & $0.45\pm0.07$ & $0.54\pm0.08$ \\
 \hline
\end{tabular}
\footnotesize{$^{i)}$ With fixed $\beta=2.0$.}\\ \footnotesize{$^{ii)}$ With $\lambda_1=100\rm \,\mu m$, $\beta=2.0$, $\alpha=2.5$ and the fiducial $N_{\rm pl}$ by \citet{C12}.}
\label{T2}
\end{table}

\subsection{The equivalent dust temperature}
\label{S4b}

\begin{figure}
  \vspace{-10pt}
 \begin{center}
 \includegraphics[width=86mm]{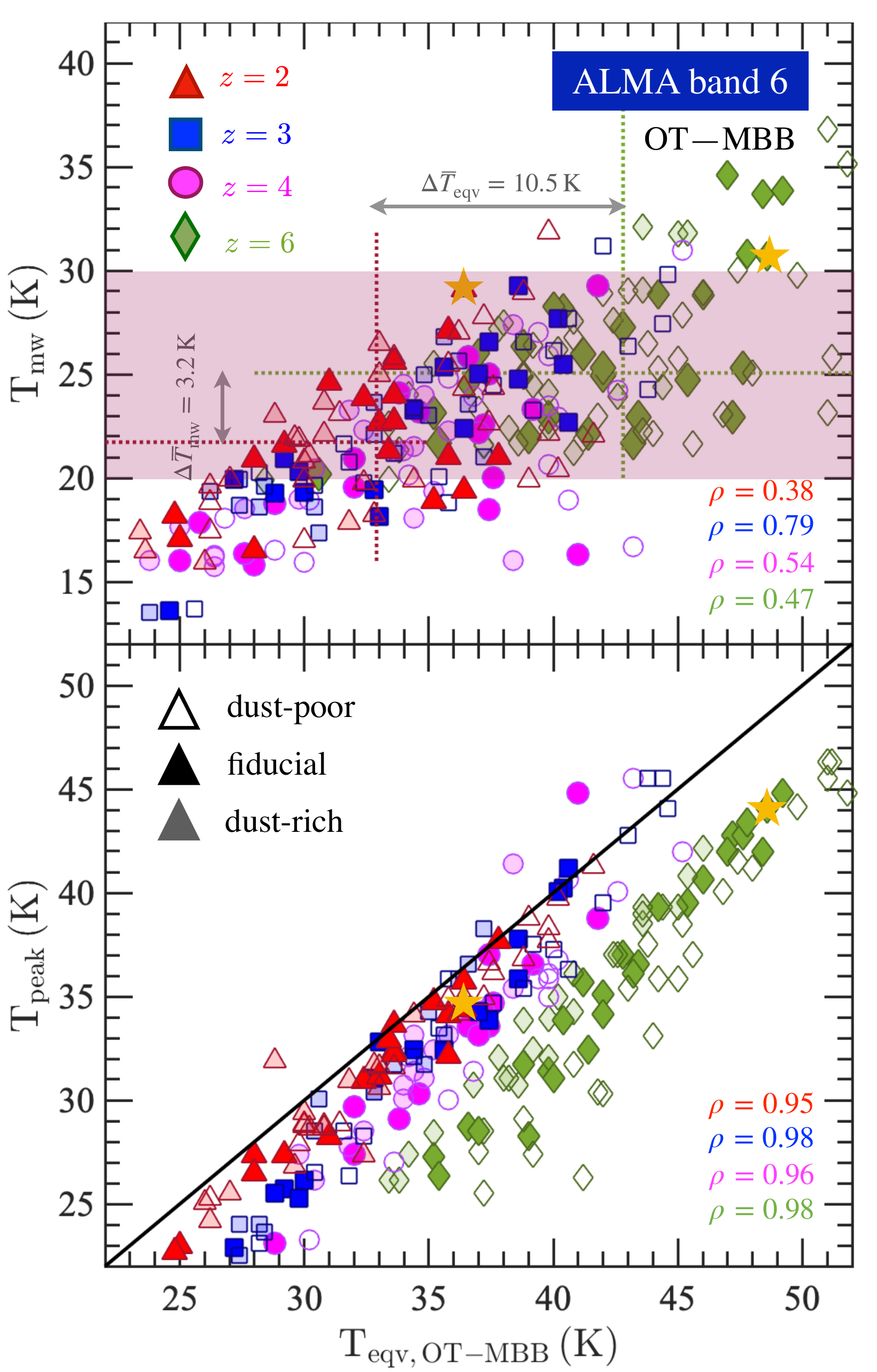}
 \vspace{-10pt}
 \caption{Relation of $T_{\rm mw}$ (upper panel) vs. $T_{\rm eqv,\,OT-MBB}$ and $T_{\rm peak}$ (lower panel) vs. $T_{\rm eqv,\,OT-MBB}$ of the $z=2-6$ \textsc{\small MassiveFIRE} galaxies, where $T_{\rm eqv,\,OT-MBB}$ is the equivalent dust temperature for the adopted OT-MBB function (Eq.~\ref{eq.5}, with fixed $\beta=2.0$) that yields the right $L_{\rm IR}$ from $S_{\rm 1.2 mm}$. In the \textit{upper panel}, the two horizontal dotted lines mark the median $T_{\rm mw}$ of the $z=2$ (red) and $z=6$ (green) samples, while the two vertical dotted lines mark their mean $T_{\rm eqv,\,OT-MBB}$. The purple shaded box shows $T_{\rm mw}=25\pm5$ K, where $T_{\rm mw}=25$ K is the suggested dust temperature for estimating dust/gas mass using the RJ approach by \citet{S16} (Section~\ref{S3d1}). The two yellow asterisks in each panel mark the selected $z=2$ (left) and $z=6$ (right) galaxies. Their SEDs are shown in Figure~\ref{fig:SED}. The two galaxies have similar $T_{\rm mw}$, but very different $T_{\rm peak}$ and $T_{\rm eqv}$. {\bf $T_{\rm eqv}$ exhibits stronger correlation with $T_{\rm peak}$ than $T_{\rm mw}$}.}
    \label{fig:Teff}
      \end{center}
\end{figure}

\begin{figure*}
 \begin{center}
 \includegraphics[height=105mm,width=175mm]{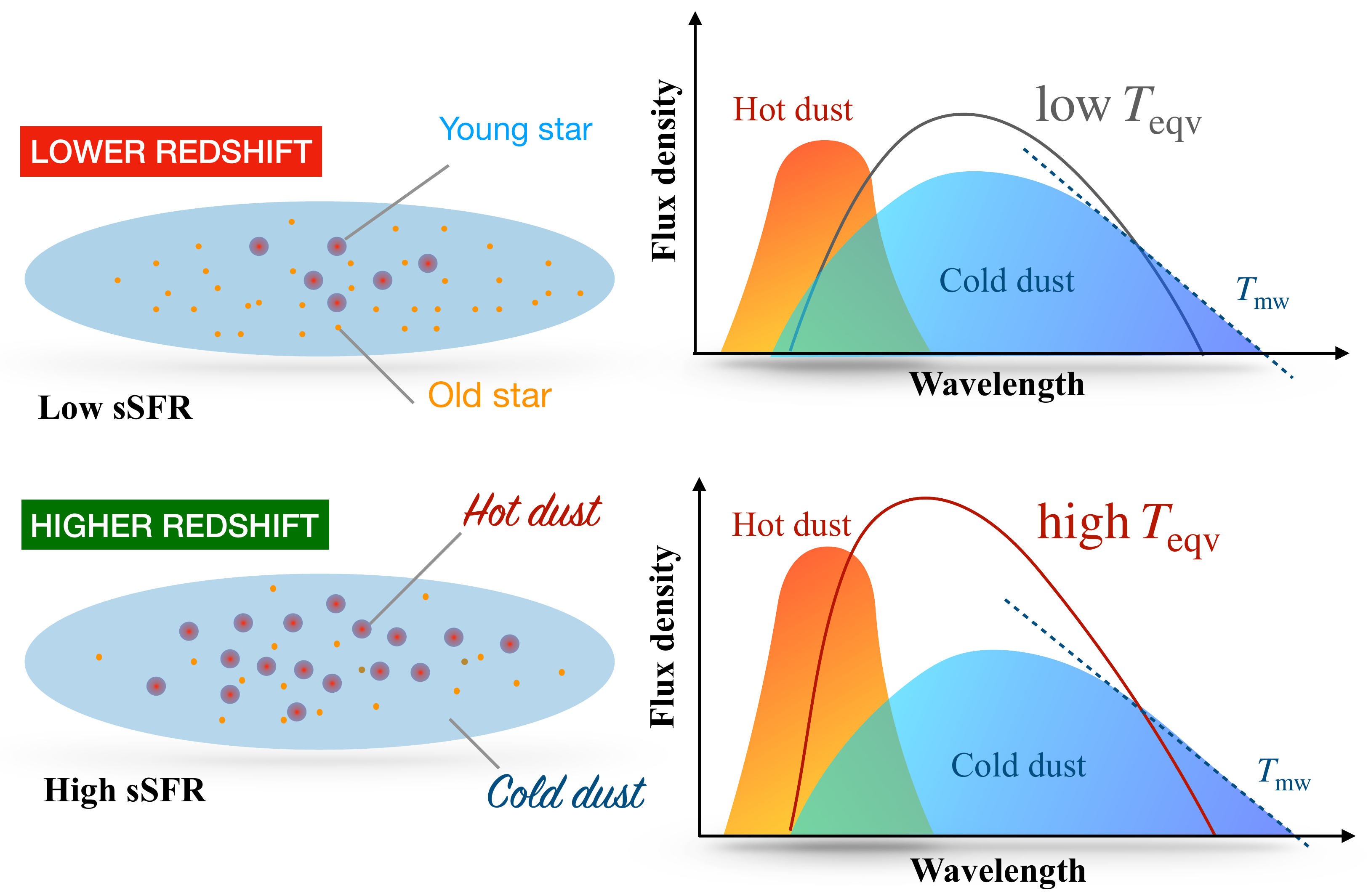}
  \vspace{-10pt}
  \caption{Schematic figure for the `two-phase' model of ISM dust and the implication on the dust SED. Higher-redshift galaxies have higher sSFR and more young ($t_{\rm age}\simless10\,\rm Myrs$) star clusters efficiently heat the dense dust in vicinity of the star-forming regions to high temperature. This hot dust component boosts the overall SED of galaxy at MIR. A higher equivalent temperature ($T_{\rm eqv}$) is thus needed to account for the more prominent MIR emission of galaxies at higher redshift. $T_{\rm eqv}$ is not well correlated with the mass weighted temperature ($T_{\rm mw}$) of galaxy. $T_{\rm mw}$ is determined by the cold dust component and it sets the slope of the RJ tail.}
  \label{fig:Scheme} 
  \end{center}
\end{figure*}

$T_{\rm eqv}$ depends on redshift and $\delta_{\rm dzr}$ in a clear and systematic manner, see Table~\ref{T2}. For the OT-MBB functional shape, for example, $T_{\rm eqv}$ scales as $\propto{}(1+z)^{0.31}\delta_{\rm dzr}^{-0.13}$ for ALMA band 7. This means that by applying a typical $T_{\rm eqv}$ for $z=2$ to a $z=6$ galaxy would lead to an underestimate of $L_{\rm IR}$ by a factor of $\sim4$ (Eq.~\ref{eq:LSequiv}). Also, at a given redshift, an order-of-magnitude increase of $\delta_{\rm dzr}$ corresponds to a $\sim0.13$ dex decrease of the best-fitting $T_{\rm eqv}$. This corresponds to a decrease of $L_{\rm IR}$ by a factor of $\sim4$ (Eq.~\ref{eq:LSequiv}). Therefore, not taking the correlation of $T_{\rm eqv}$ with redshift and $\delta_{\rm dzr}$ into account can potentially lead to significant biases in the $L_{\rm IR}$ (and hence SFR) estimates.

The scaling $T_{\rm eqv}\propto(1+z)^{0.31}$ (for band 7 and OT-MBB) is quantitatively similar to the one for $T_{\rm peak}$ (Eq.~\ref{eq:Tevo}), meaning that $T_{\rm eqv}$ also evolves more quickly with redshift compared to $T_{\rm mw}$ (see \textit{left panels} of Figure~\ref{fig:TLZ}). A natural question arises --- what drives the evolution of $T_{\rm eqv}$ with redshift?

To answer this question, we show in Figure~\ref{fig:Teff} the $T_{\rm eqv}$ vs. $T_{\rm mw}$ (\textit{upper panel}) and $T_{\rm eqv}$ vs. $T_{\rm peak}$ (\textit{lower panel}) relations of the \textsc{\small MassiveFIRE} sample at $z=2-6$. In this figure, $T_{\rm eqv}$  is calculated using an OT-MBB functional form (with fixed $\beta=2.0$) given a flux density at ALMA band 6. Using ALMA band 7 or a different form of MBB function results in qualitatively similar results and thus does not affect our conclusions.

It is clear from Figure~\ref{fig:Teff} that $T_{\rm eqv}$ is more strongly correlated with $T_{\rm peak}$ than $T_{\rm mw}$, either by looking at the $z=2-6$ sample as a whole, or each individual redshift. For each redshift, $T_{\rm peak}$ scales approximately linearly with $T_{\rm eqv}$, with a high {Spearman} correlation coefficient {$\rho\simgreat0.95$}. In contrast, the relation between $T_{\rm mw}$ and $T_{\rm eqv}$ is sub-linear and shows large scatter. As shown in the \textit{upper panel}, galaxies with similar $T_{\rm mw}$ can have very different $T_{\rm eqv}$ ($\Delta T_{\rm eqv}>10$ K) and thus a large range of $L_{\rm IR}/S$ ratios (Eq.~\ref{eq:LSequiv}).

To understand the origin of the scatter in $T_{\rm eqv}$ and fixed $T_{\rm mw}$, we selected two galaxies from the \textsc{\small MassiveFIRE} sample with similar $T_{\rm mw}(\approx 30\,{\rm K})$, one from $z=2$ and the other from $z=6$, and study their SEDs and their $T_{\rm eqv}$ in more detail. The two galaxies are marked in both panels of Figure~\ref{fig:Teff} by yellow asterisks, and their SEDs are presented in Figure~\ref{fig:SED}. The $z=6$ galaxy has $T_{\rm eqv,\,OT-MBB}=49.1$ K which is about 14 K higher than the $z=2$ galaxy.

Figure~\ref{fig:SED} shows that the two galaxies have different SED shape at short wavelengths. The $z=6$ galaxy shows more prominent MIR emission due to its more active recent star formation. Its sSFR ($=5.0\times10^{-9}\,{\rm yr^{-1}}$) is about one order of magnitude higher than that of the $z=2$ galaxy. Young star clusters in this high-redshift galaxy efficiently heat the dense, surrounding dust, which boosts the MIR emission and thus leads to a relatively high $T_{\rm peak}$ ($=44.6$ K) to account for the more prominent MIR emission of this galaxy. Furthermore, the $z=6$ galaxy is less dust-enriched than the $z=2$ galaxy (having only $1/7$ of dust mass), and its SFR/$M_{\rm dust}$ ratio is roughly 4 times higher.

The increased SFR/$M_{\rm dust}$ ratio would leave an imprint on the temperature of the diffuse dust if the heat budget of the young stars were evenly distributed in the ISM dust. However, the bulk of the diffuse cold dust is clearly not heated efficiently as the two galaxies have almost the same $T_{\rm mw}$ (29.1 K vs. 30.7 K). A number of factors can influence how efficiently the bulk of the dust is heated, such as the spatial distribution of dust in galaxy and the optical depth in vicinity of the star-forming cores \citep[\cf][]{N18,K19}. These conditions can be significantly different among galaxies and therefore $T_{\rm mw}$ is not expected to be well correlated with $T_{\rm peak}$ (see the \textit{upper right panel} of Figure~\ref{fig:TLZ}).  This example strongly indicates that a `two-phase' picture of ISM dust is needed to account for the discrepancy between $T_{\rm mw}$ and $T_{\rm peak}$, see Figure~\ref{fig:Scheme}.

Clearly, $T_{\rm eqv}$ depends on the exact form of the MBB function and the observing frequency band. As is shown in Table~\ref{T2}, $T_{\rm eqv}$ is higher at $z=2$ by 0.07 dex and it increases faster with redshift at $z=2-6$ when a GP-MBB function is used. Using the same MBB function, $T_{\rm eqv}$ also appears to be slightly higher (by $\sim0.05$ dex at $z=2-4$ for an OT-MBB function) when a flux density is measured at ALMA band 7 than band 6. As $T_{\rm eqv}$ depends both on the specific form of MBB function and the observing wavelength, $T_{\rm eqv}$ should \textit{not} be interpreted as a physical temperature but rather understood as a parametrisation of SED shape.  

It may appear reasonable to use sSFR as a predictor variable instead of $(1+z)$, given that the former depends strongly on redshift (Figure~\ref{fig:Tssfr}) and is physically linked to the amount of hot dust in galaxies. The reasons for adopting $(1+z)$ are two-folds. First of all, observationally, redshift of the (sub)mm-selected galaxies can be accurately determined through atomic/molecular emission lines, as discussed in Section~\ref{S4a}. sSFR estimates, however, are uncertain because SFR derived based on the non-$L_{\rm IR}$ indicators (\eg~UV continuum and $\rm H_\alpha$ flux) are uncertain due to the variation of the dust attenuation laws \citep{WG12,CC13,N18}. Secondly, the mapping between observed (sub)mm flux and rest-frame SED introduces an explicit redshift dependence on $T_{\rm eqv}$ --- as is shown in the lower panel of Figure~\ref{fig:Teff}, the normalisation of the $T_{\rm peak}$ vs. $T_{\rm eqv}$ relation declines with redshift using the same functional form and the observing frequency, indicating that a higher $T_{\rm eqv}$ (\ie~a steeper MBB function) is needed to derive $L_{\rm IR}$ when the rest-frame observing wavelength gets closer to the emission peak. Therefore, the $(1+z)$ term in Eq.~\ref{Eq:Tfit} accounts both (indirectly) for the cosmic time dependence of the sSFR and (directly) for the redshift of electromagnetic radiation.

Finally, adding $M_{\rm star}$ as a predictor variable results in a regression coefficient for the $M_{\rm star}$ term being consistent with zero. This means that our obtained fitting functions for $T_{\rm eqv}$ do not depend on the selection function of $M_{\rm star}$ of the {\sc\small MassiveFIRE} sample, which can be different from that of the observations. Replacing the dependence on $\delta_{\rm dzr}$ by a dependence on $M_{\rm star}$ or $Z_{\rm gas}$\footnote{$Z_{\rm gas}$ is calculated using gas particles with temperature between $7,000-15,000$ K and density above 0.5 $\rm cm^{-3}$, which represent the nebular gas where the strong nebular emission lines originate \citep{M16}.} leads to a decreased goodness-of-fit for $T_{\rm eqv}$. 

\section{Discussion}
\label{S5}

\subsection{Deriving $M_{\rm dust}$}
\label{S5a}

Many dust-enshrouded galaxies at high redshift ($z>2$) have been detected at (sub)mm wavelengths in the past years, thanks to the unprecedented sensitivity of ALMA. These (sub)mm-detected objects often lack a reliable measure of FIR photometry and many are extremely faint at UV/optical wavelengths \citep[\eg][]{D09,W12,R15,FE18}. A reliable estimate of their dust mass from full SED fitting is often not possible \citep[\cf][]{B18}.
 
In the optically-thin regime, the flux density in the RJ tail has a simple analytic form (Eq.~\ref{eq:Snu0OT}), and $M_{\rm dust}$ can be derived from the flux density given $T_{\rm mw}$ (Section~\ref{S3d1}). However, it is difficult to constrain $T_{\rm mw}$ of high-redshift galaxies when individual star-forming regions are not resolved.

Fortunately, we find that $T_{\rm mw}$ does not strongly vary from galaxy to galaxy. This is noteworthy, given that our sample spans a wide range of cosmic time ($z=2-6$), stellar mass ($M_{\rm star} = 10^9-10^{12}\,M_{\odot}$), sSFR ($10^{-10}-10^{-8} \, \rm yr^{-1}$), and IR luminosities ($L_{\rm IR}=10^9-3\times10^{12}\,L_{\odot}$). In particular, $68\%\,(\ie~1\sigma)$ of the galaxies in our sample have mass-weighted dust temperatures $T_{\rm mw} = 25\pm5$ K,
corresponding to a $20\%$ uncertainty of estimating the dust mass as the mass estimates scale only {\it linearly} 
with $T_{\rm mw}$, while $90\%$ of our sample lies within $T_{\rm mw} = 25\pm8$ K ($32\%$ uncertainty of $M_{\rm dust}$). Our findings support the empirical approach of adopting a constant $T_{\rm mw} = 25$ K to estimate the ISM mass of high redshift galaxies via Eq.~\ref{eq:Snu0OT} and $\delta_{\rm dgr}$ \citep{S16}.

While adopting a constant $T_{\rm mw}$ is a good assumption to first order, and the only option if the (sub)mm flux density is measured at only a single wavelength, additional constraints on the SED may help to determine $T_{\rm mw}$ and improve the accuracy of measuring ISM masses. Specifically, in Section~\ref{S3d3}, we show that $T_{\rm mw}$ is well correlated with $L_{\rm IR}$ and that the redshift evolution of the $L_{\rm IR}$ vs. $T_{\rm mw}$ relation is driven by the evolving dust mass. In fact, $L_{\rm IR}$, $M_{\rm dust}$ and $T_{\rm mw}$ follow a tight scaling relation (Eq.~\ref{eq.12}) for $T_{\rm mw}{}\gg{}T_{\rm CMB}$. Hence, given $S\propto M_{\rm dust}T_{\rm mw}$, it should be possible to simultaneously infer $M_{\rm dust}$ and $T_{\rm mw}$ from a combined measurement of $S$ and $L_{\rm IR}$.

Recent studies have shown that the broadband rest-frame $8\,\rm \mu m$ luminosity, $L_8$, can be a rough tracer of $L_{\rm IR}$ over a range of galaxies \citep[see][]{E11,M13, MM14,S18}. One important practical interest for using $L_8$ is that it will be easily accessible by the upcoming James Webb Space Telescope (JWST) for galaxies up to $z\sim3$. The unprecedented sensitivity of the Mid-Infrared Instrument (MIRI) on board the JWST, covering the wavelength range of 5 to 28 $\rm \mu m$, will significantly enlarge the sample size of distant galaxies with measured MIR broadband spectroscopy. We thus propose to use $L_8$ to infer $L_{\rm IR}$ for the (sub)mm-detected galaxies at $z\simless3$ that have no constraint on SED shape near the emission peak \citep[\cf][]{CE01,E02,R06,R09,S09,RE10,RR13,SP16,A18}.

Hence, we propose to derive $M_{\rm dust}$ (as well as $T_{\rm mw}$) of high-redshift galaxies by combining mid-infrared (e.g., from JWST) and far-infrared/submm (e.g., ALMA) data sets. Specifically, by combining Eq.~\ref{eq:Snu0OT} and~\ref{eq.12}, we obtain

\begin{align}
    {\rm log}\,\left(\frac{M_{\rm dust}}{M_{\odot}}\right)&=1.23\,{\rm log}\,\left(\frac{S}{{\rm mJy}}\right)-0.23\,{\rm log}\,\left(\frac{L_{\rm IR}}{L_{\odot}}\right)+\mathcal{F}(z)\nonumber \\ \label{eq:Mdust1}
    {\rm or}\,\,\,\,  M_{\rm dust}&\propto \left (\frac{S}{L_{\rm IR}}\right)^{0.23}S,
\end{align}

\noindent where $\mathcal{F}(z)=-0.85+1.23\,{\rm log}\,(\psi(z)\Gamma_{\rm RJ})$ and $\psi(z)$ has the unit of ${\rm mJy}\,M^{-1}_{\odot}\,{\rm K}^{-1}$. Assuming that $L_{\rm IR}=\alpha\,L_8$ \citep{M13,S18}, we can rewrite the above equation as

\begin{equation}
    {\rm log}\,\left(\frac{M_{\rm dust}}{M_{\odot}}\right)=1.23\,{\rm log}\,\left(\frac{S}{{\rm mJy}}\right)-0.23\,{\rm log}\,\left(\frac{L_8}{L_{\odot}}\right)+\mathcal{G}(z) 
    \label{eq:Mdust2}
\end{equation}

\noindent where $\mathcal{G}(z)=-0.23\log\alpha + \mathcal{F}(z)$. In general, $\Gamma_{\rm RJ}$ is a function of $T$ and Eq.~\ref{eq:Snu0OT} \& ~\ref{eq:Mdust2} need to be solved numerically.

It is important to note that $\alpha$ can depend on the variation in the detailed conditions of the star-forming regions. Recent observational evidence has shown that scatter in $\alpha$ can be driven by certain galaxy properties, such as sSFR, $Z_{\rm gas}$ and compactness of IR-emitting regions \citep[\eg][]{N12,S18,EL18}. These dependences on intrinsic galaxy properties can then translate to an apparent dependence of $\alpha$ on redshift and starburstiness (SB, see Section~\ref{S3d4} for definition) of galaxy. Therefore, to improve the accuracy of $L_{\rm IR}$ estimates from $L_8$, one needs to rely on either a direct measurement or an observational proxy of these properties and/or galaxy redshift/SB. Furthermore, one should also caution the contribution of PAH molecules and AGN activity to the 8 $\rm \mu m$ features \citep[\eg][]{S04,P08,K12,M13,SA14,KP15,R16,L19}. This topic is beyond the scope of the present paper and we leave it to a future study.

According to Eq.~\ref{eq:Mdust1}, a factor of 2 uncertainty in $L_{\rm IR}$ translates into $\sim{}20\%$ uncertainty in the derived dust mass, i.e., matches the intrinsic level of error of the constant $T_{\rm mw}=25$ K method \citep{S16}. Therefore, increasing complexity by deriving $T_{\rm mw}$ and $M_{\rm dust}$ from $L_{\rm IR}$ and $S$ will only be beneficial if $L_{\rm IR}$ can be constrained to within a factor of 2 or better. 

Finally, we note that our sample does not include the most luminous submm galaxies that can have $S_{\rm 870 \mu m}$ fluxes of much higher than a few mJy \citep[\eg][]{O17,O18} and, hence, we cannot rule out that $T_{\rm mw}$ significantly exceeds 20-30 K in such objects. While submm-luminous galaxies are typically interpreted as having high SFRs, Eq.~\ref{eq:Snu0OT} shows that submm fluxes are simply the product of $M_{\rm dust}$ and $T_{\rm mw}$ ($\Gamma_{\rm RJ}$ is a weak function of $T$). Hence, as long as $T_{\rm mw}$ is not significantly higher in these objects, a straightforward interpretation is that the most submm-luminous galaxies are those with the highest $M_{\rm dust}$. In fact, $S_{\rm 870\mu m}$ ($S_{\rm 1.2mm}$) is nearly doubled as $M_{\rm dust}$ gets doubled (by comparing the dust-rich and fiducial cases in Figure~\ref{fig:LS}), while $T_{\rm mw}$ decreases by only $\sim1$ K (\ie~$\simless5\%$). \textit{This example also suggests that caution needs to be taken when directly converting (sub)mm flux densities to $L_{\rm IR}$  (and SFR), without taking into account how dust mass (and optical depth) alters the SED shape of galaxy} (\cf~\citealt{S76,HK11,S13,SH16}, and see the lower panel of Figure~\ref{fig:Dust}, where we show how SED is altered by $M_{\rm dust}$ without having a different SFR of galaxy).

\subsection{The increase of $T_{\rm eqv}$ with redshift and its observational evidence}
\label{S5b}

Adopting $T_{\rm eqv}$ and an SED shape is another way to estimate the IR luminosity from submm fluxes, see Section~\ref{S4b}. Hence, if $T_{\rm eqv}$ is known, it is possible to use the approach described in the previous section to infer dust masses and mass-weighted temperatures. This could be a particularly useful approach at $z\simgreat3$, where the potential MIR diagnostics redshift out of the wavelengths accessible by JWST.

Using the simplified functional forms of SED (an OT-MBB or GP-MBB), the obtained $T_{\rm eqv}$ increases monotonically with redshift (Table~\ref{T2}). The typical $T_{\rm eqv}$ of a $z=6$ galaxy is as high as $45-50$ K for an OT-MBB function, which is significantly higher than the mean $T_{\rm mw}$ ($\sim25$ K) at this redshift. This result is consistent with what has been implied by some recent observational findings, including the unusual relationship between the IR excess (IRX$\equiv L_{\rm IR}/L_{\rm UV}$) and the UV spectral slope ($\beta$) of the Lyman break galaxies (LBGs) at high redshifts.

Empirically, IRX is used as a proxy of total dust mass and $\beta$ is a measure of dust column density. It has been found that galaxies between $0<z<4$ follow a well-defined sequence on the IRX-$\beta$ diagram, although there exists nontrivial scatter that depends on the stellar populations and the detailed dust and ISM properties \citep{M95,M99,SS09,T12,FC17,P17,ND18,M19}. However, recent observations of higher-redshift LBGs show evident IRX deficit \citep[\eg][]{C15,B16}. A significant fraction of the $z>4$ UV-selected galaxies have no observed dust continuum at ALMA bands. For the ALMA-detected objects, their estimated IRX appears to be significantly lower than the value inferred from their measured $\beta$ using the canonical IRX-$\beta$ relations found by the local samples. The IRX deficit of the selected high-redshift LBGs is challenging to explain with the current dust attenuation models \citep{FH17,FC17,ND18}.

Instead of the high-redshift populations having significantly different dust properties, an alternative solution is that they have a higher $T_{\rm eqv}$, which results in a higher derived $L_{\rm IR}$ with a given observed (sub)mm flux density. For example, \citet{B16} report that among 330 LBGs in the Hubble Ultra Deep Field spanning the redshift range of $z=2-10$, only 6 were detected at $\simgreat2\sigma$ at 1.2mm (band 6) by the ALMA Spectroscopic Survey (ASPECS). This is significantly lower than the number (35) extrapolated from the $z=0$ IRX-$\beta$ relation based on their UV properties and by assuming a constant (equivalent) dust temperature of 35 K. The authors suggest that using a monotonic increase of (equivalent) dust temperature with redshift, \ie~$T\propto (1+z)^{0.32}$ (an OT-MBB function is assumed), the number of detected sources can be consistent with the SMC IRX-$\beta$ relation, which is derived based on the local metal-poor populations \citep{SS09}. Encouragingly, this suggested redshift dependence of temperature well agrees with that of $T_{\rm eqv}$ found by the {\sc\small MassiveFIRE} sample (see Table~\ref{T2}). This indicates that a significant deviation of the dust properties of the high-redshift UV-selected populations is not needed for explaining their \textit{observed} IRX deficit \citep[\cf][]{C18a}.

Finally, $T_{\rm eqv}$ should not be deemed equivalent to the \textit{mean intensity of the radiation field}, <$U$>${}\propto{}L_{\rm IR}/M_{\rm dust}$ \citep[\eg][]{DL07}, despite that the latter has also been found to increase with redshift \citep[\eg][]{M12,B15,MR17,S18}. This is because $T_{\rm eqv}$ is a parametrisation of SED shape, and it depends on both the assumed functional form of SED as well as the observing wavelength (Section~\ref{S4b}), while <$U$> ($\propto T^{5.4}_{\rm mw}$, Eq.~\ref{eq.12}) represents the physical dust temperature. It is also important to note that since $T_{\rm mw}$ not only evolves with redshift, but also shows a clear dependence on $L_{\rm IR}$ at fixed redshfit (see the \textit{lower left panel} of Fig.~\ref{fig:TLZ}). The observed redshift evolution of <$U$>  therefore can potentially depend on the selection function of $L_{\rm IR}$. \textit{The selection bias caused by using a flux-limited sample (galaxies at higher redshift are confined to higher $L_{\rm IR}$) can lead to a steeper increase of <$U$>  with redshift than is measured at fixed $L_{\rm IR}$.}

\subsection{The dependence of $T_{\rm eqv}$ on $\delta_{\rm dzr}$}
\label{S5c}

Fig.~\ref{fig.DZR} shows that $T_{\rm eqv}$ is anti-correlated with $\delta_{\rm dzr}$ at fixed redshift. Specifically, an order-of-magnitude decrease in $\delta_{\rm dzr}$ translates to $\sim{}0.13$ dex increase of required $T_{\rm eqv}$, corresponding to a factor of $\sim4$ increase of $L_{\rm IR}$ (Eq.~\ref{eq:LSequiv}). As noted before (Section~\ref{S4a}, \ref{S5a}), $\delta_{\rm dzr}$ directly affects the total dust mass (and optical depth) of a galaxy, thereby altering its SED shape. Hence, it can be one source of uncertainty in estimating $L_{\rm IR}$ through $T_{\rm eqv}$.

Observationally, while $\delta_{\rm dzr}$ has been found to be fairly constant across a wide range of galaxies at low redshifts by different studies \citep{I90,LF98,J02,G05,WD11}, there is also evidence of a reduced $\delta_{\rm dzr}$ in the low-metallicity environments \citep{HF12,R14,MT16,DL16,D17,CS18,D19}. This can imply a lower $\delta_{\rm dzr}$ in higher-redshift galaxies since they are known to have lower metallicities than the low-redshift galaxies \citep{E06,M08,D08,M10,D11,LC13,o16,M16}. A direct implication of the decrease of $\delta_{\rm dzr}$ with redshift is that it can further mitigate the IRX deficit problem of the high-redshift LBGs, see Section~\ref{S5b}. A reduced $\delta_{\rm dzr}$ leads to an additional increase of $T_{\rm eqv}$ and hence a higher upper confidence limit of the IRX of the undetected objects, meaning that the dust properties of these high-redshift LBGs are more probable to be consistent with the canonical dust attenuation laws that are derived based on the low-redshift observations.

However, $\delta_{\rm dzr}$ of high-redshift galaxies is not yet well understood. Recent studies based on foreground absorbers towards $\gamma$-ray burst (GRB) afterglows and quasars (QSOs) as well as distant lensed galaxies have shown different trends of how $\delta_{\rm dzr}$ depends on redshift \citep{DK09,CD13,D13,Z13,DL16,WS17}, which can be due to selection effects and observational uncertainties \citep{MDA14}, as well as the different choice of $\delta_{\rm dzr}$ measures \citep{WS17}. Moreover, because these studies are often limited to only a handful of galaxies, it becomes difficult to distinguish the explicit redshift dependence of $\delta_{\rm dzr}$ from the intrinsic correlation with the galaxy properties, although some studies based on the QSO damped Lyman-$\alpha$ absorbers (QSO-DLAs) find that the QSO-DLAs over a range of redshifts ($z=2-6$) follow a similar $\delta_{\rm dzr}$-$Z_{\rm gas}$ relation \citep{D13,DL16,WS17}.

Observationally, gas-phase galaxy metallicities ($Z_{\rm gas}$) can be derived using the ratios between (rest-frame) optical auroral and nebular line fluxes and with calibration on theoretical models \citep[\eg][]{K02,KE08,M08,Z11,SR14}. However, this method can only be used for galaxies up to $z\sim3$, above which the emission lines redshift out of the wavelengths of the current ground-based NIR spectrographs. An alternative method is to use the equivalent width of (rest-frame) UV absorption features, which has been used for galaxies up to $z\sim5$, but is still limited because of the faintness of the features \citep{H98,ES12,FC16}.

To overcome these shortcomings, \citet{RP18} have recently proposed a new method of using the (rest-frame) FIR [$\rm O_{III}$] 88 $\rm \mu m$\,/\,[$\rm N_{II}$] 122 $\rm \mu m$ line ratio for probing the gas metallicities of galaxies at $z>4$, where both characteristic lines shift to the submm range that is accessible with ALMA. Using the previously reported FIR line measurements of a sample of local normal and star-forming galaxies by the \textit{Infrared Space Observatory} \citep[ISO,][]{K96}, \citet{RP18} find that the derived galaxy mass-metallicity relation is consistent with the result derived using optical emission lines \citep{T04}. The gas metallicities of three $z=2\sim3$ submm-luminous galaxies derived using \textit{Herschel} measurements are also in good agreement with the high-redshift relationships previously derived by \citet{M08} and \citet{M10}, despite that the  stellar mass estimates of these obscured dusty high-redshift galaxies have large uncertainties. These results suggest that FIR emission lines could be promising tool for estimating $Z_{\rm gas}$ of $z>4$ galaxies.

Therefore, if $Z_{\rm gas}$ could be used as a predictor for $\delta_{\rm dzr}$ (which is currently uncertain), it will further help improve the accuracy of $T_{\rm eqv}$ (and hence $L_{\rm IR}$) estimates at $z>4$, which is important since the MIR diagnostics for $L_{\rm IR}$ are inaccessible by JWST at this epoch (Section~\ref{S5a}). This will in turn help improve our constraints on the total obscured SFR in the early Universe, where currently only UV-based SFR estimates are available \citep{C18b,C18a}.

\subsection{The sub-resolution structure of the birth-clouds}
\label{S5d}

\begin{figure}
 \centering
 \hspace*{-10 pt}
 \includegraphics[width=85mm]{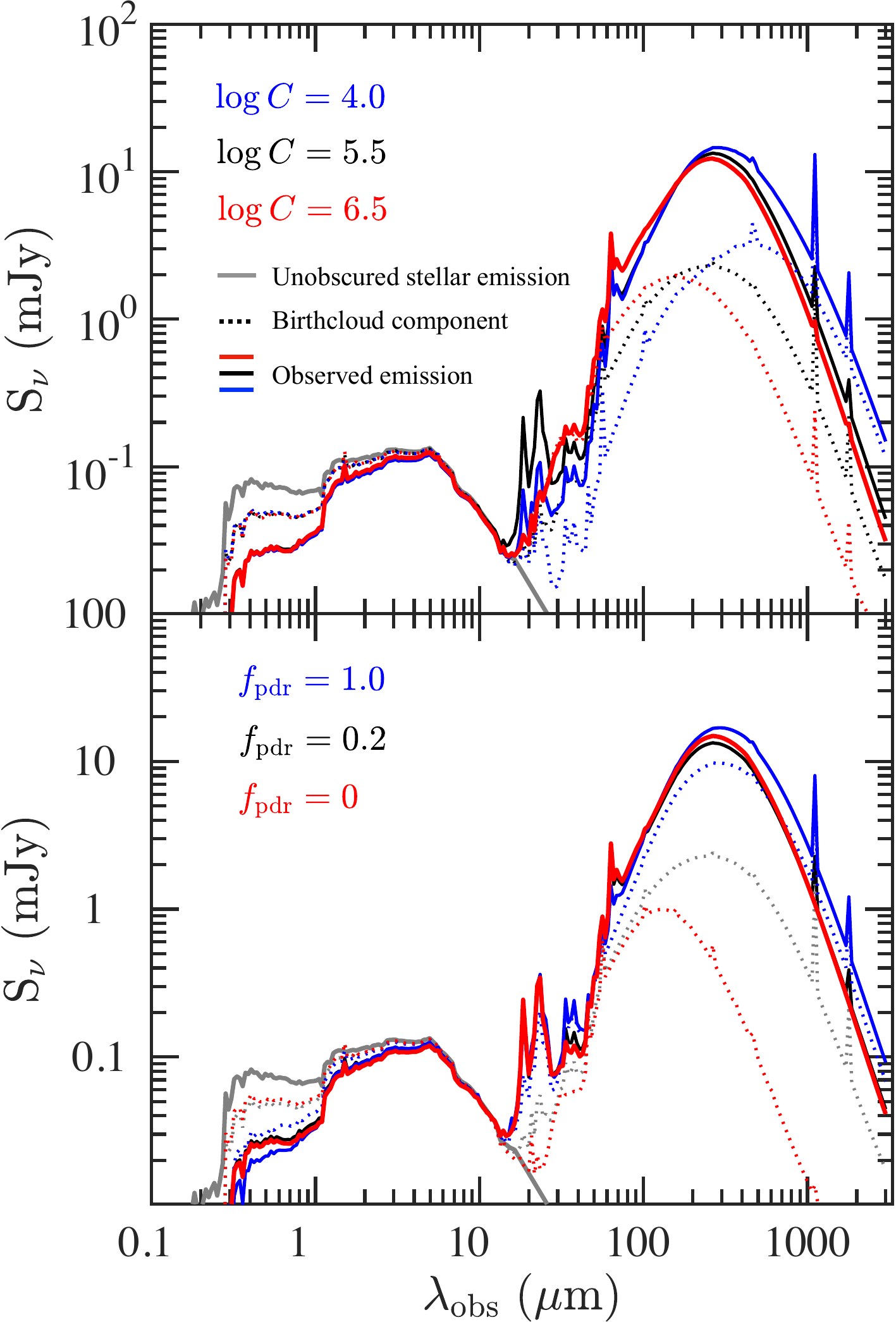}
 \vspace{-5 pt}
  \caption{SEDs of a $z=2$ {\sc\small MassiveFIRE} galaxy generated by different dust models. In the left panel, we show the observed SEDs for ${\rm log}\;C=6.5$ (red), $5.5$ (black) and $4.0$ (blue) with fixed $f_{\rm pdr}\, (=0.2)$. In the right panel, we show the result for $f_{\rm pdr}=0$ (red), $f_{\rm pdr}=0.2$ (black) and $f_{\rm pdr}=1.0$ (blue) with fixed ${\rm log}\;C\,(=5.5)$. In each panel, the grey curve shows the intrinsic stellar emission, while the solid red, black and blue curves show the observed SEDs, each corresponding to a different dust model. Source SEDs from birth-clouds associated with the star forming regions are shown with dotted lines with the corresponding colour for each model.} 
  \vspace{-15pt}
  \label{fig:Maps} 
\end{figure}

\begin{figure}
 \centering
 \includegraphics[width=85mm]{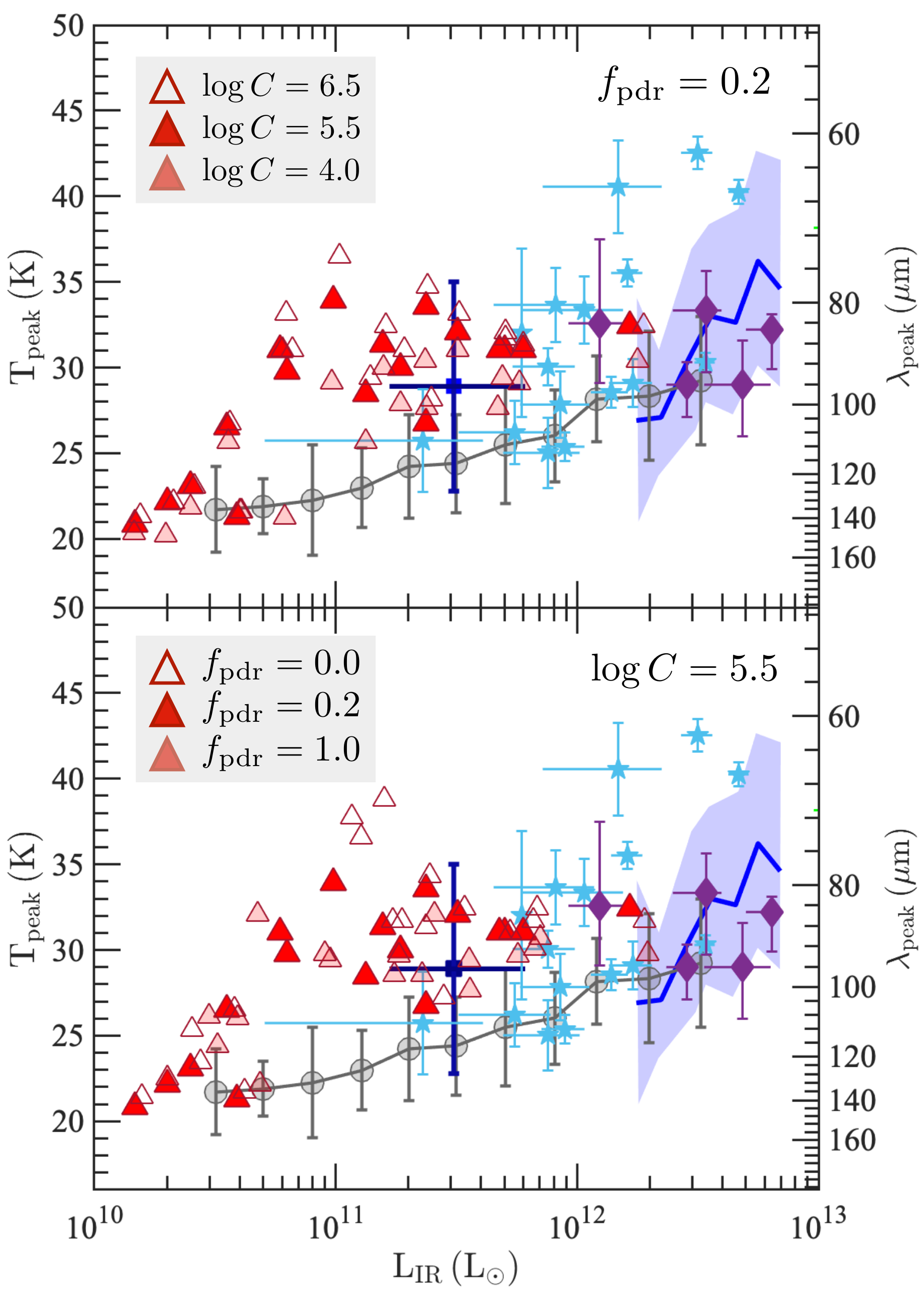}
 \vspace{-10 pt}
  \caption{The same as the {\it upper panel} of Figure~\ref{fig:TL2} except that the data of the $z=2$ \textsc{\small MassiveFIRE} sample are produced by including the \textsc{\small MappingsIII} birth-cloud model (Section~\ref{S5d}). In the \textit{upper panel}, we show the result for ${\rm log}\;C=6.5$ (unfilled), $5.5$ (filled) and $4.0$ (semi-transparent) with fixed $f_{\rm pdr}\,(=0.2)$, while in the \textit{lower panel}, we show the result for $f_{\rm pdr}=0$ (unfilled), $f_{\rm pdr}=0.2$ (filled) and $f_{\rm pdr}=1.0$ (semi-transparent) with fixed ${\rm log}\;C\,(=5.5)$. For all different models, $\delta_{\rm dzr}=0.4$. } 
  \vspace{-15pt}
  \label{fig:TLC} 
\end{figure}

Observational evidence has indicated that young star clusters reside in dense dusty birth-clouds \citep[\eg][]{C97, TP04, WC11,P14,K15}. To check the uncertainty arising from potentially unresolved small-scale ISM structure, we have repeated the analysis presented in this paper with additional RT analysis by \textsc{\small SKIRT} as \citet{L18}, where we include a sub-grid model for birth-clouds embedding the young stars (our `alternative' RT model). We summarise the detail of this sub-grid model and the main results from this model in this subsection.

In brief, all the young star particle of a galaxy that has formed less than 10 Myrs ago is assigned a {\sc\small MappingsIII} source SED \citep{G08}. {\sc\small MappingsIII} SED templates are parameterised by the SFR and the metallicity of the star-forming regions, the pressure of the ambient ISM, the HII region compactness (${\rm log}\;C$), and the covering fraction of the associated PDR ($f_{\rm pdr}$). 

To explore how our results depend on this choice, the \textit{upper} and \textit{lower} panels in Figure~\ref{fig:Maps} show the overall SED of one of our galaxies for different values of ${\rm log}\;C$ and $f_{\rm pdr}$, respectively. As ${\rm log}\;C$ increases, the birth-clouds become more compact and the dust associated with the clouds attain higher mean temperature because of the stronger incident radiation onto dust grains. The source SED of this dust component (shown with dashed lines) shifts to shorter wavelength, and so does the overall SED of the galaxy. $f_{\rm pdr}$ is a measure of the survival timescale of birth-clouds \citep{J10}. Increasing $f_{\rm pdr}$ results in a larger fraction of the stellar emission being absorbed by dust in the birth-clouds, which results in more energy being re-emitted as IR light. The mean dust temperature of the birth-clouds, however, decreases. For the total emission of galaxy, a higher $f_{\rm pdr}$ typically leads to higher $L_{\rm IR}$. Whether the emission peak of the \textit{overall} SED shifts to shorter or longer wavelengths with $f_{\rm pdr}$, however, depends on the value of ${\rm log}\;C$. 

The sub-grid model has minor impact on $T_{\rm mw}$ of galaxies. $T_{\rm mw}$ increases with ${\rm log}\,C$ at fixed $f_{\rm pdr}$, and decreases with $f_{\rm pdr}$ at fixed ${\rm log}\,C$. The reason is that the photons emitted from the birth-clouds are more energetic if the birth-clouds are more compact (higher ${\rm log}\,C$) and less dust-obscured (low $f_{\rm pdr}$). But the resulting difference of $T_{\rm mw}$ is typically no more than $\pm1$ K ($\pm5\%$) by exploring the parameter space of the {\sc\small MappingsIII} model.

$T_{\rm peak}$, however, is more sensitive to the uncertainty of the small-scale ISM structure. In some cases, especially for strongly star-forming galaxies, $T_{\rm peak}$ can differ by much as 10 K when the {\sc\small MappingsIII} parameters are varied.  $T_{\rm peak}$ is typically higher with increasing ${\rm log}\,C$, and for low/intermediate ($\sim4.0-5.5$) value of ${\rm log}\,C$, decreases with  $f_{\rm pdr}$. For the $z=2$ MassiveFIRE sample, ${\rm log}\,C=6.5$ (max) leads to a median $T_{\rm peak}$ higher than ${\rm log}\,C=4.0$ (min) by about 4 K with fixed $f_{\rm pdr}=0.2$, and $f_{\rm pdr}=1.0$ (max) yields a median $T_{\rm peak}$ lower than $f_{\rm pdr}=0$ (min) by about 2.5 K with fixed ${\rm log}\,C=5.5$. Uncertainty of the small-scale ISM conditions could introduce scatter in the observed $T_{\rm peak}$ vs. $L_{\rm IR}$ relation in addition to galaxy-by-galaxy variations of $\delta_{\rm dzr}$ (Figure~\ref{fig:TLC}).

Including the sub-grid birth-cloud model strengthens the correlation between $T_{\rm peak}$ (and $T_{\rm eqv}$) and sSFRs of galaxies. By pre-processing starlight in birth-clouds, the range of the physical conditions surrounding star-forming regions is reduced. We note, however, that none of the trends reported in this paper change on a qualitative level by including or excluding the {\sc\small MappingsIII} birth-cloud model. 

\section{Summary and Conclusion}
\label{S6}

In this paper, we study dust temperatures of high-redshift galaxies and their scaling relationships with the help of cosmological zoom-in simulations and dust RT modelling. Our sample consists of massive ($M_{\rm star}>10^{10}\;M_{\odot}$) $z=2-6$ galaxies extracted from the {\sc\small MassiveFIRE} suite \citep{F16, F17}, a set of cosmological hydrodynamic zoom-in simulations from the {\sc\small FIRE} project \citep{H14}. The sample encompasses 18 central galaxies at $z=2$ and their most massive progenitors up to $z=6$, together with a disjoint set of 11 central galaxies at $z=6$. We generate FIR-to-mm broadband fluxes and spectra for our galaxy sample with {\sc\small SKIRT}. 

We explicitly define and discuss four different dust temperatures that are commonly used in the literature, $T_{\rm mw}$, $T_{\rm peak}$, $T_{\rm eff}$ and $T_{\rm eqv}$. $T_{\rm mw}$ is the physical, mass-weighted temperature that can be extracted from RT analysis, but is often not easily accessible to observations. $T_{\rm eff}$ and $T_{\rm peak}$ are derived from SED fitting: $T_{\rm eff}$ is the $T$ parameter in the best-fit modified blackbody function and $T_{\rm peak}$ is the inverse of emission peak wavelength. These two are the temperatures that are often adopted for analysing large statistical sample of galaxies by observational studies. And finally, $T_{\rm eqv}$ is the temperature one needs to convert single (sub)mm data to total IR luminosity based on an assumed SED shape. \\

The main findings of this paper are:

\begin{itemize}[leftmargin=0.5cm]
 \item \textsc{\small FIRE} simulations together with RT processing successfully reproduce $T_{\rm peak}$ of $z=2$, $L_{\rm IR}\simgreat10^{11}\,L_{\odot}$ galaxies, in good agreement with recent observations (Figure~\ref{fig:TL2}). The observational data shows large scatter, which may be driven by galaxy-to-galaxy variations of $\delta_{\rm dzr}$ as well as local variations in the physical conditions of unresolved birth-clouds embedding young star clusters (Section~\ref{S3c},~\ref{S5d}).
 
 \item $T_{\rm mw}$ is only weakly correlated with $T_{\rm peak}$ over $z=2-6$ (Figure~\ref{fig:TLZ}). The former sets the slope of the RJ tail (Figure~\ref{fig:SED}), and is the temperature needed for estimating dust and gas mass of distant galaxies (Figure~\ref{fig:RJ}). Using $T_{\rm peak}$, or $T_{\rm eff}$ (\eg~derived from full SED fitting), which is strongly correlated with $T_{\rm peak}$, can lead to a systematic bias/error of the derived dust/gas mass, and may lead to an inaccurate interpretation of the star-forming conditions in high-redshift galaxies (Section~\ref{S3d1}). 
 
 \item $T_{\rm peak}$ is well correlated with sSFR ($\rho\sim0.7$) (Figure~\ref{fig:TLZ}). Recently formed stars efficiently heat the dense, warm dust in the close vicinity of star-forming regions. The emission from this warm dust component boosts the overall dust SED at MIR, and helps to shift the emission peak to shorter wavelength (Figure~\ref{fig:Scheme}). $T_{\rm mw}$ is less well correlated with sSFR ($\rho\sim0.55$) and the scaling relation shows a flatter slope ($\Delta T_{\rm mw}\propto {\rm SB}^{4.0}$ vs. $\Delta T_{\rm peak}\propto {\rm SB}^{8.7}$). The bulk of the cold diffuse dust is not as effectively heated as the warm dust component (Section~\ref{S3d4}).
 
 \item $T_{\rm peak}$ scales as $(1+z)^{0.25}$ at fixed $L_{\rm IR}$ between $z=2-6$ driven by the increasing sSFR at higher redshift, which is consistent with recent observations (Section~\ref{S3d3}). $T_{\rm mw}$ evolves only weakly with redshift at fixed $L_{\rm IR}$ at $z=2-6$ (Figure~\ref{fig:TLZ}).  
 
 \item Of the galaxies in our sample, $68\%$ have mass-weighted dust temperatures $T_{\rm mw} = 25\pm5$ K (Figure~\ref{fig:Teff}). This temperature range corresponds to an uncertainty of $20\%$ in estimating $M_{
\rm dust}$ from a single submm band since the mass estimates scale only {\it linearly} 
with $T_{\rm mw}$. Furthermore, $90\%$ of our sample lies within $T_{\rm mw} = 25\pm8$ K. Our findings support the empirical approach of adopting a constant $T_{\rm mw} = 25$ K to estimate the ISM mass of high redshift galaxies \citep{S16}.
 
 \item $T_{\rm mw}$ is well correlated with $L_{\rm IR}$ at $T_{\rm mw}\gg{}T_{\rm CMB}$ at a given redshift (Figure~\ref{fig:TLZ}). The normalisation of this relation evolves weakly with redshift but the slope does not change. At higher redshift, galaxies of the same $L_{\rm IR}$ have higher $T_{\rm mw}$ but lower $M_{\rm dust}$. Using the $z=2-6$ sample, we derive the scaling relation $L_{\rm IR}\propto{} M^{1.0}_{\rm dust}T^{5.4}_{\rm mw}$, which appears to be shallower than the classical $L_{\rm IR}\propto M_{\rm dust}T^{4+\beta}$ ($\beta\approx 2.0$ for our adopted dust model) relation expected from the optically-thin assumption (Section~\ref{S3d3}).
 
 \item We propose to use this scaling relation to derive $M_{\rm dust}$ (and $T_{\rm mw}$) of high-redshift (sub)mm-detected galaxies, assuming that their $L_{\rm IR}$ can be constrained, for example, via the mid-IR luminosity probed by the \textit{Spitzer} telescope and the upcoming JWST. We showed that this method improves over the $T_{
 \rm mw}=25$ K approach if $L_{\rm IR}$ can be constrained to within a factor of 2 or better (Section~\ref{S5a}).
 
 \item $T_{\rm eqv}$ increases with redshift, meaning that a higher temperature is needed to convert observed (sub)mm broadband fluxes to $L_{\rm IR}$ (and hence SFRs) of galaxies at higher redshift. $T_{\rm eqv}$ is tightly correlated ($\rho\simgreat0.95$) with $T_{\rm peak}$, a much stronger correlation than with $T_{\rm mw}$ (Figure~\ref{fig:Teff}). In particular, two galaxies at different redshifts can have very different $T_{\rm eqv}$ ($\Delta T_{\rm eqv}>10$ K) but similar $T_{\rm mw}$ (Section~\ref{S4}).
 
 \item We find an anti-correlation between $T_{\rm eqv}$ and the dust-to-gas ratio, $\delta_{\rm dzr}$. Hence, at a given redshift, dust-poorer galaxies need, on the average, a higher $T_{\rm eqv}$ for the (sub)mm-flux-to-IR-luminosity conversion. We express $T_{\rm eqv}$ as a power-law function of $\delta_{\rm dzr}$ and $(1+z)$, and perform linear regression analysis using the \textsc{\small MassiveFIRE} sample at $z=2-6$. The best-fit parameters of the scaling relation are provided in Table~\ref{T2}. We present the result for both ALMA band 6 and 7. We propose to apply the scaling relation to more accurately convert between (sub)mm flux and IR luminosity (and SFR) of high-redshift galaxies (Section~\ref{S4}).
 
\end{itemize}

To summarise our results, we find that the observationally-derived temperatures, in particular, $T_{\rm peak}$, generally differ from $T_{\rm mw}$. $T_{\rm peak}$ shows a steeper slope and a stronger correlation with sSFR, and evolves more quickly with redshift compared with $T_{\rm mw}$. We also find that $T_{\rm eqv}$ is more strongly correlated with $T_{\rm peak}$ than with $T_{\rm mw}$.

The difference between $T_{\rm peak}$ and $T_{\rm mw}$ may be understood by a `two-phase' picture of ISM dust. $T_{\rm mw}$ is set by the diffuse, cold dust component which dominates the total dust mass, while $T_{\rm peak}$ is also influenced by the dense, warm dust component in the close vicinity of young star clusters. The former component is typically heated less effectively by young stars than the latter so that $T_{\rm peak}$ and $T_{\rm mw}$ are not well correlated with each other.

The increase of $T_{\rm eqv}$ with redshift is consistent with recent observational evidence, including low number counts of (sub)mm sources in ALMA blind surveys \citep[][and references therein]{B16,C18a} and the unusual IRX-$\beta$ relation of high-redshift galaxies \citep[\cf][]{C15,M19}. However, as we argue in this paper, the rise of $T_{\rm eqv}$ with redshift is not simply a sign of dust being hotter at higher redshift, but it reflects the change in SED shape. In particular, higher $T_{\rm eqv}$ is often a consequence of a more prominent MIR emission of galaxies at higher redshift, resulting from more active star formation. However, as $T_{\rm mw}$ evolves only weakly between $z=2$ and $z=6$, the temperature of the majority of the dust component ($\sim{}T_{\rm mw}$) does not significantly change despite the change in $T_{\rm eqv}$. In this sense, dust in galaxies with higher $T_{\rm eqv}$ is not necessarily \textit{physically} hotter.

In conclusion, dust temperature is important for estimating and probing key physical properties (\eg~dust/gas mass, IR luminosity) and ISM conditions of high-redshift galaxies. A proper interpretation of dust temperatures and their scaling relationships requires taking into account the differences between temperatures derived from the SED shape and the physical, mass-weighted dust temperature. Upcoming facilities, such as JWST, \textsc{\small SPICA} and CCAT-prime, will significantly improve our capability of constraining key dust properties of galaxies in the distant Universe.

\section*{Acknowledgements}

We are grateful to the referee for the careful review and insightful suggestions that have helped improve the quality of this manuscript. We acknowledge the comments from Andreas Faisst, Rob Ivison, Allison Kirkpatrick, Georgios Magdis and Ian Smail to an early version of this manuscript. We thank Caitlin Casey for a discussion on \citealt{C18b,C18a} before they were published on arXiv, which had helped the writing of this paper.  We are grateful to the technical guidance by Peter Camps and Maarten Baes. LL would like to acknowledge the stimulating atmosphere during the Munich Institute for Astro- and Particle Physics (MIAPP) 2018 programme \textit{the Interstellar Medium of High Redshift Galaxies}, and especially thank the conversation with the programme coordinators. This research was supported by the MIAPP of the Deutsche Forschungsgemeinschaft (DFG) cluster of excellence ``Origin and Structure of the Universe". RF acknowledges financial support from the Swiss National Science Foundation (grant no 157591). Simulations were run with resources provided by the NASA High-End Computing (HEC) Program through the NASA Advanced Supercomputing (NAS) Division at Ames Research centre, proposal SMD-14-5492. Additional computing support was provided by HEC allocations SMD-14-5189, SMD-15-5950, by NSF XSEDE allocations AST120025, AST150045, by allocations s697, s698 at the Swiss National Supercomputing Centre (CSCS), and by S3IT resources at the University of Zurich. DK acknowledges support from the NSF grant AST-1715101 and the Cottrell Scholar Award from the Research Corporation for Science Advancement. CAFG was supported by NSF through grants AST-1517491, AST-1715216, and CAREER award AST-1652522; by NASA through grant 17-ATP17-0067; by STScI through grant HST-AR-14562.001; and by a Cottrell Scholar Award from the Research Corporation for Science Advancement. PFH was supported by an Alfred P. Sloan Research Fellowship,
NASA ATP Grant NNX14AH35G, and NSF Collaborative Research Grant \#1411920 and CAREER grant \#1455342. EQ was supported in part by a Simons Investigator Award from the Simons Foundation and by NSF grant AST-1715070. The Flatiron Institute is supported by the Simons Foundation.

\bibliographystyle{mnras}
\bibliography{Liangetal2019}

\bsp
\label{lastpage}
\end{document}